\documentclass[twocolumn]{aastex631}

\usepackage{hyperref}

\newcommand{\ha}{${\rm H\alpha}$}
\newcommand{\hb}{${\rm H\beta}$}

\newcommand{\nii}{\hbox{[N\,{\sc ii}]}}
\usepackage{threeparttable}
\newcommand{\oii}{[\textrm{O}~\textsc{ii}]}

\begin{document}

\title{Environmental Regulation of Dust and Star Formation Unveiled by Subaru Dual Narrow-band Imaging: Degree-scale Balmer Decrement Mapping across a $z = 0.9$ Supercluster}

\correspondingauthor{Zhaoran Liu}
\email{zhaoran.liu@astr.tohoku.ac.jp}
\author[0009-0002-8965-1303]{Zhaoran Liu}
\affiliation{Astronomical Institute, Graduate School of Science, Tohoku University, 6–3 Aoba, Sendai 980-8578, Japan}

\author[0000-0002-2993-1576]{Tadayuki Kodama}
\affiliation{Astronomical Institute, Graduate School of Science, Tohoku University, 6–3 Aoba, Sendai 980-8578, Japan}

\author[0000-0002-1428-7036]{Brian C. Lemaux}
\affiliation{Gemini Observatory, NSF NOIRLab, 670 N. A'ohoku Place, Hilo, Hawai'i, 96720, USA}
\affiliation{Department of Physics \& Astronomy, University of California, Davis, One Shields Ave., Davis, CA 95616, USA}

\author[0000-0002-7598-5292]{Mariko Kubo}
\affiliation{Astronomical Institute, Graduate School of Science, Tohoku University, 6–3 Aoba, Sendai 980-8578, Japan}
\affiliation{School of Science, Kwansei Gakuin University, Sanda, Hyogo 669-1337, Japan}

\author[0000-0002-5963-6850]{Jose Manuel P\'erez-Mart\'inez}
\affiliation{Instituto de Astrof\'isica de Canarias (IAC), E-38205, La Laguna, Tenerife, Spain.}
\affiliation{Universidad de La Laguna, Dpto. Astrof\'isica, E-38206, La Laguna, Tenerife, Spain.}
\affiliation{Astronomical Institute, Graduate School of Science, Tohoku University, 6–3 Aoba, Sendai 980-8578, Japan}

\author[0000-0002-0479-3699]{Yusei Koyama}
\affiliation{Department of Astronomical Science, The Graduate University for Advanced Studies, 2-21-1 Osawa, Mitaka, Tokyo 181-8588, Japan}
\affiliation{National Astronomical Observatory of Japan, 2-21-1 Osawa, Mitaka, Tokyo 181-8588, Japan}

\author[0000-0002-4937-4738]{Ichi Tanaka}
\affiliation{Subaru Telescope, National Astronomical Observatory of Japan, National Institutes of Natural Sciences \\
650 North A'ohoku Place, Hilo, HI 96720, USA}

\author[0000-0002-9509-2774]{Kazuki Daikuhara}
\affiliation{Astronomical Institute, Graduate School of Science, Tohoku University, 6–3 Aoba, Sendai 980-8578, Japan}
\affiliation{Institute of Space and Astronautical Science, Japan Aerospace Exploration Agency, 3-1-1, Yoshinodai, Chuou-ku, Sagamihara,
Kanagawa 252-5210, Japan}
\author[0000-0001-8255-6560]{Roy R. Gal}
\affiliation{Institute for Astronomy, University of Hawai’i, 2680 Woodlawn Drive, Honolulu, HI 96822, USA}

\author[0000-0001-7523-140X]{Denise Hung}
\affiliation{Gemini Observatory, NSF NOIRLab, 670 N. A'ohoku Place, Hilo, Hawai'i, 96720, USA}

\author[0000-0003-4907-1734]{Masahiro Konishi}
\affiliation{Institute of Astronomy, Graduate School of Science, The University of Tokyo, 2-21-1 Osawa, Mitaka, Tokyo 181-0015, Japan}

\author[0009-0000-5163-2305]{Kosuke Kushibiki}
\affiliation{Advanced Technology Center, National Astronomical Observatory of Japan, 2-21-1 Osawa, Mitaka, Tokyo 181-8588, Japan}
\affiliation{Department of Astronomical Science, The Graduate University for Advanced Studies, 2-21-1 Osawa, Mitaka, Tokyo 181-8588, Japan}

\author[0000-0002-0322-6131]{Ronaldo Laishram}
\affiliation{National Astronomical Observatory of Japan, 2-21-1 Osawa, Mitaka, Tokyo 181-8588, Japan}
\affiliation{Astronomical Institute, Graduate School of Science, Tohoku University, 6–3 Aoba, Sendai 980-8578, Japan}

\author[0000-0003-2119-8151]{Lori M. Lubin}
\affiliation{Department of Physics \& Astronomy, University of California, Davis, One Shields Ave., Davis, CA 95616, USA}

\author[0000-0002-0724-9146]{Kentaro Motohara}
\affiliation{Institute of Astronomy, Graduate School of Science, The University of Tokyo, 2-21-1 Osawa, Mitaka, Tokyo 181-0015, Japan}
\affiliation{National Astronomical Observatory of Japan, 2-21-1 Osawa, Mitaka, Tokyo 181-8588, Japan}

\author{Hidenori Takahashi}
\affiliation{Institute of Astronomy, Graduate School of Science, The University of Tokyo, 2-21-1 Osawa, Mitaka, Tokyo 181-0015, Japan}



\begin{abstract}

We present results from a dual narrow-band imaging survey targeting the CL1604 supercluster at $z = 0.9$  using the Subaru Telescope. By combining the \textit{NB921} filter on HSC and the \textit{NB1244} filter on SWIMS, we can detect redshifted \ha\ and \hb\ emission lines from the supercluster. This unique technique allows us to measure both star formation rates and dust extinction for a sample of 94 emission-line galaxies across the supercluster. We find that dust extinction, estimated from the Balmer decrement (\ha/\hb\ ratio), increases with stellar mass in star-forming galaxies, whereas relatively quiescent systems exhibit comparatively low extinction. Among galaxies with intermediate masses ($10^{8.5}<M_* < 10^{10.5}\,M_\odot$), the dust-corrected \ha-based star formation rates align with the main sequence at this epoch. More massive galaxies, however, deviate from this relation, exhibit redder colors, and reside predominantly in higher-density environments. Although stellar mass, SFR, and galaxy color are clearly influenced by environment, we detect no strong, systematic environmental dependence of dust extinction for the whole sample.



\end{abstract}




\section{Introduction} \label{sec:intro}

Distant galaxy clusters play a critical role in hierarchical structure formation, profoundly influencing the growth and transformation of galaxies within them. In this framework, galaxies form and evolve within large-scale structures, and over cosmic time, groups and clusters develop due to the gravitational accumulation of galaxies along these structures \citep[e.g.,][]{Bond96, Springel05}. As galaxies assemble in denser regions, their evolution is regulated not only by internal processes but also by external factors such as interactions with neighboring galaxies and the intra-cluster or intra-group gas \citep[e.g.,][]{Abadi99, Mori2000, Boselli06}. These processes are believed to be responsible for the strong environmental dependence of galaxy properties, such as star-forming activities and morphologies, observed in the $z<1$ Universe \citep[e.g.,][]{Dressler80, Kodama01, Lewis02, thomas05, Muzzin12}.

It is first noted by \citet{Dressler80} that the galaxies in nearby rich clusters tend to have redder colors and lower star formation activities compared to their field counterparts. These differences between cluster and field galaxies were later confirmed by subsequent works \citep[e.g.,][]{Postman84, Dressler97}. Observational evidence from higher redshift clusters reveals that the fraction of blue, star-forming galaxies increases with redshift, known as the Butcher–Oemler effect \citep{Butcher78, Butcher84}, suggesting an evolution in galaxy populations within dense environments over cosmic time. Studies have suggested that a combination of bursty star formation and galaxy mergers in early overdensities may be responsible for the observed differences between cluster and field galaxies \citep[e.g.,][]{Bower92, Bell04, Hayashi12, Tomczak17, Shimakawa18, Pelliccia19, Laishram24, Morishita25}. Numerical simulations further support this scenario, predicting that galaxies in massive protoclusters already experience a peak in star formation activity at early epochs (\( z \gtrsim 2 \)) \citep{Chiang17}.

To reveal the physical mechanisms behind the co-evolution of galaxies and their environments, it is crucial to observe distant clusters at various redshifts. By studying their star formation activities and physical properties, we can deepen our understanding of how environmental effects on galaxies change throughout cosmic time. At intermediate redshifts ($0.5 < z < 1.2$), large spectroscopic and imaging surveys of mature clusters and infalling galaxy populations --- such as RCS-1 \citep{Gladders05}, the Panoramic Imaging and Spectroscopy of Cluster Evolution with Subaru \citep{Kodama05}, ORELSE \citep{Gal08, Lubin09}, RCS-2 \citep{Gilbank11}, GCLASS \citep{Muzzin12}, and GOGREEN \citep{Balogh17} --- have significantly advanced our knowledge of where and when galaxies are quenched as they enter dense environments. An important finding from these cluster surveys is the ``inside-out'' transformation of galaxy star formation. \citet{Koyama08} observed enhanced star formation activities in medium-density regions --- such as cluster outskirts and groups --- and suppression of star formation in cluster cores, suggesting that quenching (and possible boosting prior to that) occurs from the inside out. This pattern of enhanced star-forming activity in intermediate-density regions and suppression in the densest environments is also supported by other observations \citep{Poggianti08, Sobral11, Muzzin12}.

At higher redshifts ($z > 1.5$), observations of protoclusters reveal a more complex picture. In some cases, star formation is elevated in dense regions relative to lower-density surroundings \citep{Dannerbauer14, Umehata15, Shimakawa18, miller18, jose24uss, Morishita24b}, further supporting the idea that environmental effects on galaxies are redshift-dependent and evolve over time, possibly reflecting the early stages of cluster assembly.

However, to further understand the physical mechanisms driving such transformations, quantitatively measuring the physical properties of galaxies in different environments remains essential. One of the biggest challenges in obtaining fundamental quantities, such as star formation rates (SFRs), is correcting for dust extinction. Among the various SFR tracers, \ha\ emission has been widely used to probe star formation through both spectroscopy and narrow-band imaging \citep[e.g.,][]{Kennicutt94, kennicutt98_2, Kodama04, sobral12, Steidel14}, owing to its origin as a hydrogen recombination line. This makes \ha\ a direct tracer of star-forming regions, in contrast to another widely used tracer, \oii, which is influenced by metallicity and ionization conditions \citep{Kewley02}, and can arise from sources unrelated to star formation. Moreover, \ha\ has a longer rest-frame wavelength than ultraviolet and \oii\ emission lines, making it less sensitive to dust attenuation. Yet, dust extinction can still significantly attenuate H$\alpha$ flux \citep[e.g.,][]{Schreiber09, Kashino13,Koyama19, Liu24}, and accurately assessing this attenuation remains challenging, especially when relying on the stellar continuum, where age and metallicity can introduce degeneracies. Moreover, dust extinction itself provides key information about the modes of star formation and the physical processes triggering these activities. Furthermore, dust extinction is known to correlate with stellar mass in star-forming galaxies, introducing systematic uncertainties when studying galaxies across different environments and evolutionary stages \citep[e.g.,][]{Garn10, Zahid13, Shapley22}.

Several approaches are commonly used to probe dust extinction in galaxies: (i) Multiwavelength Photometry-based Spectral Energy Distribution (SED) Fitting: This method relies primarily on broadband photometric data to model the observed SED, offering a cost-effective means of analysis. However, it suffers from significant uncertainties due to intrinsic degeneracies between dust extinction and stellar age \citep[e.g.,][]{Buat12,Cunha15,Hirashita17}. (ii) Far-Infrared (FIR) Dust Emission Measurements: In this approach, dust re-emits absorbed ultraviolet and optical energy at longer wavelengths in the FIR range \citep[e.g.,][]{Buat96,Takagi99,Berta16,Faisst20}. However, probing the dust content from FIR observations requires assumptions about dust temperature and grain size, introducing substantial uncertainties. Multiple FIR data points can constrain the dust emission spectrum but are costly to obtain. Moreover, FIR continuum is challenging to detect for typical star-forming galaxies at high redshift due to their low dust masses and the sensitivity limits of current facilities \citep[e.g.,][]{Fudamoto17, Pozzi21}. (iii) Hydrogen Recombination Line Ratios: This method compares observed hydrogen emission line ratios to theoretical predictions under Case B conditions, leveraging the stronger dust attenuation at shorter wavelengths. Combined with a suitable dust extinction law \citep[e.g.,][]{Dong08, Garn10, Groves12, 2013ApJ...763..145D, shapley23, Reddy23, Liu24, Jose24}, it provides precise measurements of dust extinction.

Although hydrogen recombination line ratios are considered the most precise and affordable indicators of dust extinction, applying them to the study of galaxy properties in large-scale structures at high redshifts is challenging due to the limitations of current near-infrared spectroscopic facilities and the selection biases inherent in target pre-selection. To address these issues, we innovatively utilize narrow-band imaging with the Subaru Telescope to capture \ha\ and \hb\ emissions at  $z = 0.9$ . This approach allows us to trace star formation activities and dust extinction across the structures in the supercluster CL1604 in a certain redshift slice hosting the two cluster/groups. This study marks a pioneering effort to investigate the Balmer decrement solely through narrow-band imaging, thereby overcoming the limitations of spectroscopic methods. By leveraging Subaru's large field of view and specialized narrow-band filters, we can effectively study star formation and dust extinction across the supercluster. Situated at a critical cosmic epoch where clusters transition from actively forming protoclusters to quiescent phases in the local universe, this dedicated dual narrow-band survey enables us to gain deeper insights into galaxy transformations without the complications typically introduced by dust extinction. 

We introduce our target supercluster in Section~\ref{sec:cl1604}. Observations and data reduction procedures are described in Section~\ref{sec:data}. Our analyses of photometry, emission line fluxes, sample selection, and physical property estimation are detailed in Section~\ref{sec:analysis}. We present our results in Section~\ref{result} and compare them with existing literature in Section~\ref{sec:discussion}. Finally, we summarize our conclusions in Section~\ref{sec:summary}. Throughout the paper, we adopt the AB magnitude system \citep{oke83,Fukugita96}, cosmological parameters of $\Omega_m=0.3$, $\Omega_\Lambda=0.7$, $H_0=70$\,km\,s$^{-1}\,{\rm Mpc}^{-1}$, and the \citet{Chabrier03} initial mass function (IMF).


\section{CL1604: A Large Scale Structure at z $\sim$ 0.9}
\label{sec:cl1604}
The CL1604 supercluster is one of the most thoroughly studied large-scale structures at $z \sim 0.9$. Initially identified as two separate clusters by \citet{Gunn1986}, the supercluster's redshifts and preliminary velocity dispersions were first measured by \citet{Postman98}.  More detailed follow-up came through the Observations of Redshift Evolution in Large-Scale Environments (ORELSE) survey \citep{Gal08, Lubin09}, which targeted CL1604 as one of the distant overdensities to investigate galaxy evolution on large scales. Since then, CL1604 has been observed across multiple wavelengths, including optical, near-infrared, and mid-infrared imaging, optical spectroscopy, Hubble Space Telescope (HST) imaging, X-ray and radio data \citep[e.g.,][]{Gal08, Kocevski11, Lemaux10, Lemaux12, Shen17}. These efforts have revealed eight distinct clusters and groups within the supercluster. Some are large and contain mostly red early-type galaxies (ETGs) and hot intracluster gas. Others are smaller groups or filaments with active star formation and bright AGN.


The Subaru Telescope has played an important role in studying the CL1604 supercluster with its wide field of view and diverse narrow-band filters. Using data from the Hyper Suprime-Cam Subaru Strategic Program (HSC-SSP), \citet{Hayashi19} reported that the supercluster extends farther than previously mapped, particularly in the north–south direction, reaching over 50 comoving Mpc. Their study also identified new candidate clusters and groups at $z \sim 0.9$ in CL1604’s northern and southern regions, later confirmed through follow-up spectroscopy with the Subaru Faint Object Camera and Spectrograph (FOCAS) and the Gemini Multi-Object Spectrographs (GMOS) on the Gemini North Telescope. \citet{Asano20} then took a closer look at CL1604-D, a cluster at $z = 0.923$ in the western-central region of CL1604. They utilized Subaru’s Simultaneous-color Wide-field Infrared Multi-object Spectrograph (SWIMS) with the \textit{NB1261} filter to detect \ha\ emission from star-forming galaxies. Their observations suggested that merging and interacting galaxies are more common in group environments, while starburst galaxies appear to be absent from the cluster core. Together, these findings highlight the diverse and dynamic nature of CL1604, making it an ideal laboratory to study galaxy evolution across different environments at $z \sim 1.0$.

\section{Observations} \label{sec:data}

\begin{figure}[ht]
    \centering
    \includegraphics[width=0.45\textwidth]{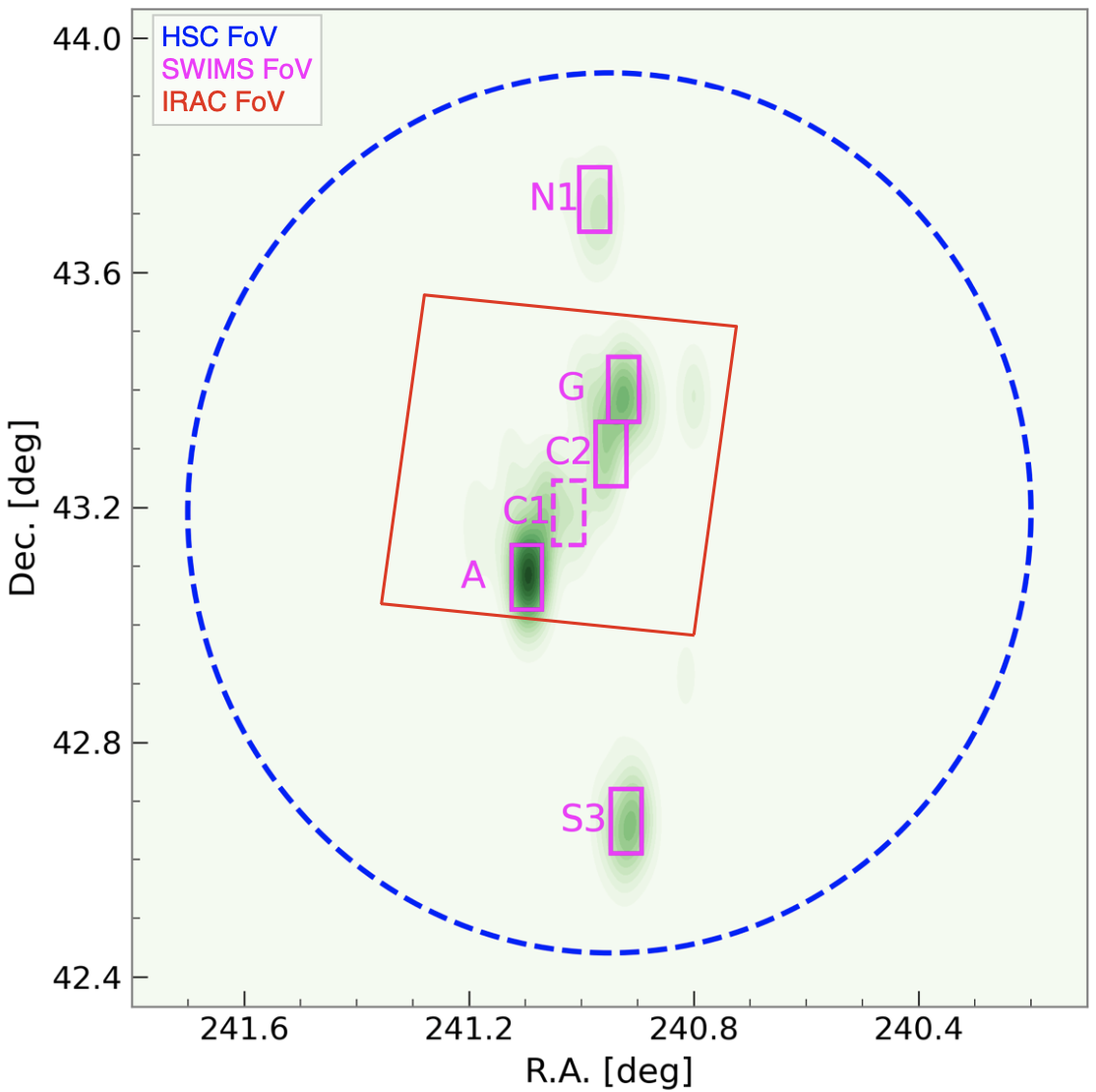}  
   \caption{Field of view for narrow-band imaging with SWIMS \textit{NB1244} (magenta rectangle) and HSC \textit{NB921} (blue circle). The \textit{J}-band imaging (the corresponding broad-band of SWIMS \textit{NB1244}) was conducted with MOIRCS, except for field C1, which was observed using SWIMS (dashed rectangle). The area with archival IRAC coverage is indicated by the red rectangle (see Section \ref{sec:spitzer}). The background green contours represent galaxy surface density derived from sources within the redshift slice $z=0.88$–$0.91$, including both spectroscopically confirmed galaxies (Section \ref{sec:spec}) and the dual emitters identified in this study. Galaxy density is computed using Gaussian kernel density estimation. We note that \hb\ emitters selected solely based on HSC \textit{NB921} imaging are not included in the density estimation.}
    \label{fig:fov}
\end{figure}

\begin{figure}[ht]
    \centering
    \includegraphics[width=0.46\textwidth]{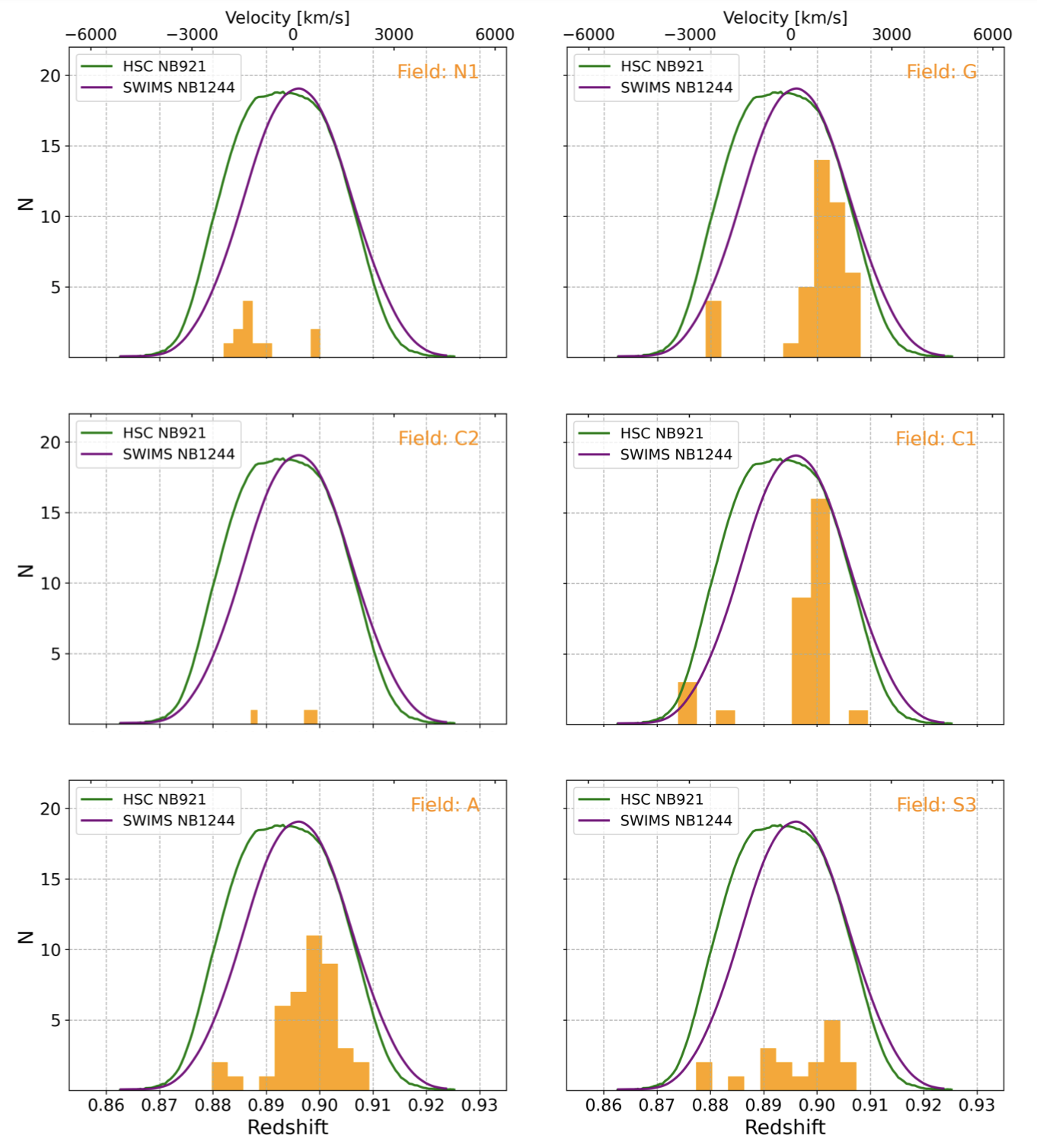}
    \caption{Redshift distribution of spectroscopically confirmed galaxies (see Section \ref{sec:spec}) within the filter coverage of six Subaru pointings, overlaid with the transmission curves of the \textit{NB921} and \textit{NB1244} narrow-band filters (arbitrarily scaled). The corresponding velocity distribution, centered at $z=0.895$, is also shown. The narrow-band filters effectively capture the redshift peak at $z=0.895$ over a considerable velocity range.}
    \label{fig:filter}
\end{figure}

Building on the extensive existing datasets, this work aims to accurately characterize dust extinction and star formation activities of galaxies within the CL1604 supercluster using Subaru observations. Our analysis is based on two key imaging datasets: (1) \hb\ imaging, using the Hyper Suprime-Cam (HSC; \citealt{Miyazaki12, Miyazaki18}) \textit{NB921} filter along with the corresponding \textit{z}-band, and (2) \ha\ imaging, conducted with the SWIMS \textit{NB1244} filter and the corresponding broadband \textit{J}-band. The broadband imaging was conducted with SWIMS for one pointing, and with the Multi-Object Infrared Camera and Spectrograph (MOIRCS; \citealt{Ichikawa06, Suzuki08}) for the rest of the pointings due to facility availability. The layout of our pointings was designed to cover the regions with the highest density of spectroscopically confirmed sources whose redshifts place their \ha\ and \hb\ emission lines within the passbands of our narrow-band filters. Our Subaru observations are summarized in Figure~\ref{fig:fov} and Table~\ref{tab:subaru-obs}, while the response functions of the \textit{NB921} and \textit{NB1244} filters are illustrated in Figure~\ref{fig:filter}. The astrometry and photometry of the Subaru observations are calibrated using the Panoramic Survey Telescope and Rapid Response System (Pan-STARRS1; \citealt{Schlafly12, Tonry12, Magnier13, Chambers16}). For the narrow-band observations, the photometric zero points are determined using the corresponding broad-band filters, namely the \textit{J}-band for \textit{NB1244} and the \textit{z}-band for \textit{NB921}.

In addition to the Subaru observations, our analysis incorporates archival Spitzer data and optical spectroscopy. The subsequent sections provide a detailed description of the observations, data reduction procedures, and the utilization of archival datasets.

\subsection{SWIMS NB1244, J, H band}\label{sec:SWIMS}

We use unique \textit{NB1244} filter onborad SWIMS which can capture \ha\ emission line at  $z = 0.9$ , as shown in Figure \ref{fig:filter}. Originally designed for the University of Tokyo Atacama Observatory (TAO) 6.5 m telescope \citep{Yoshii16}, currently under construction on the summit of Co. Chajnantor in Chile, SWIMS is capable of simultaneous imaging or spectroscopy in both blue (0.9 – 1.4 $\mu$m) and red (1.4 – 2.5 $\mu$m) channels. An overview of the instrument is provided by \citet{motohara16} and \citet{konishi18}. 

Before its commissioning on the TAO, SWIMS was temporarily installed on the Subaru Telescope for observational use. Our \textit{NB1244} observations were carried out during this period, specifically in the S22A semester (Program ID: S22A-006; PI: T. Kodama). The primary goal of our observations was to cover as many fields as possible with the \textit{NB1244} filter, given its unique availability on SWIMS, as the S22A semester represented the last opportunity to maximize \textit{NB1244} observations of CL1604 in the northern sky before the instrument was shipped to TAO in Chile. As a result, we observed six fields with \textit{NB1244} and one field with the \textit{J}-band, while the \textit{H}-band was obtained simultaneously with the \textit{J}-band in the same field.

The data reduction is performed using the \texttt{SWSRED}\footnote{\href{http://133.11.160.242/TAO/swims/?Data_Reduction/Imaging_Data_Reduction}{SWSRED - A Data Reduction Pipeline for SWIMS}} software developed by Masahiro Konishi, which is built upon the \texttt{MCSRED} framework. After image cleaning, astrometric matching, and frame co-adding by \texttt{SWSRED} for the two individual chips of the SWIMS detectors, the two chips are mosaicked to obtain the final combined image. 

\subsection{MOIRCS J, H band}\label{sec:MOIRCS}
We obtain \textit{J, H} bands for rest of the five fields that was observed by \textit{NB1244} and lack broad-band observations. The observations were conducted during the S23A semester (S23A-011; PI: Z. Liu) with Subaru MOIRCS. We reduce the images with \texttt{MCSRED}\footnote{\href{https://www.naoj.org/staff/ichi/MCSRED/mcsred_e.html}{MCSRED - A Data Reduction Pipeline for MOIRCS}}, a pipeline originally developed by Ichi Tanaka for reducing Subaru MOIRCS data, following the procedure described in \citet{Tanaka11}. 


\subsection{HSC NB921 and z band}\label{sec:HSC}
The HSC \textit{NB921} observations were conducted during S23A (S23A-011; PI: Z. Liu) and the data reduction was done with \texttt{hscPipe}\footnote{\href{https://hsc.mtk.nao.ac.jp/pipedoc/pipedoc_8_e/index.html}{hscPipe - A Data Reduction Pipeline for HSC}} \citep{Bosch18}.



Given that the \textit{z}-band serves as the corresponding broad-band for \textit{NB921}, applying a consistent photometry process to both bands is important for obtaining accurate color differences. Rather than using the available photometry from the HSC-SSP catalog to calculate the narrow-band color excess, we choose to retrieve the reduced HSC $g$, $r$, $i$, $z$, $y$ images from the HSC-SSP survey PDR2 \citep{Aihara18, Aihara19} in our field of interest and stitch patches from different tracts together to create a large image. We then perform our own photometry on both \textit{NB921} and \textit{z} band using a consistent methodology, which is further detailed in Section \ref{sec:photometry}.

\begin{table*}[ht]
    \centering
    \caption{Summaries of Subaru Observations}
    \begin{threeparttable}
    \begin{tabular}{llllll}
        \hline
        Instrument & Filter  & Limiting Magnitude (mag) & Seeing (arcsec) & Field Coverage & Reference \\
        \hline
        SWIMS & \textit{NB1244} & 23.2 & 0.62 & N1, G, C2, C1, A, S3  & S22A-006\\
        SWIMS & \textit{J} & 24.5 & 0.57 & C1 & S22A-006\\
        SWIMS & \textit{H} & 24.1 & 0.51 & C1 & S22A-006\\
        MORICS & \textit{J} & 24.7 & 0.64 & N1, G, C2, A, S3  & S23A-011\\
        MORICS & \textit{H} & 24.4 & 0.76 & N1, G, C2, A, S3  & S23A-011\\
        HSC & \textit{NB921} & 25.1 & 0.68 & N1, G, C2, C1, A, S3  & S23A-011\\
        HSC & \textit{g} & 26.6 & 0.77 & N1, G, C2, C1, A, S3 & HSC-SSP PDR2 \tnote{a}\\
        HSC & \textit{r} & 26.2 & 0.76 & N1, G, C2, C1, A, S3 & HSC-SSP PDR2\\
        HSC & \textit{i} & 26.2 & 0.58 & N1, G, C2, C1, A, S3 & HSC-SSP PDR2\\
        HSC & \textit{z} & 25.3 & 0.68 & N1, G, C2, C1, A, S3 & HSC-SSP PDR2\\
        HSC & \textit{y} & 24.5 & 0.68 & N1, G, C2, C1, A, S3 & HSC-SSP PDR2\\
        \hline 
    \end{tabular}
    \begin{tablenotes}
      \item[a]Wide-field imaging survey with Subaru HSC \citep{Aihara18, Aihara19}.
    \end{tablenotes}
    \end{threeparttable}
    \label{tab:subaru-obs}
\end{table*}

\subsection{Archival Data}\label{sec:archival}
CL1604 is equipped with a wealth of archival photometric and spectroscopic data from previous studies. It has been partially observed with the Spitzer Space Telescope and benefits from extensive optical spectroscopy conducted with KECK/DEIMOS, Subaru/FOCAS, and Gemini-N/GMOS. In the following sections, we will briefly describe these existing datasets and how they are incorporated into our analysis.
\begin{figure}[ht]
    \centering
    \includegraphics[width=0.49\textwidth]{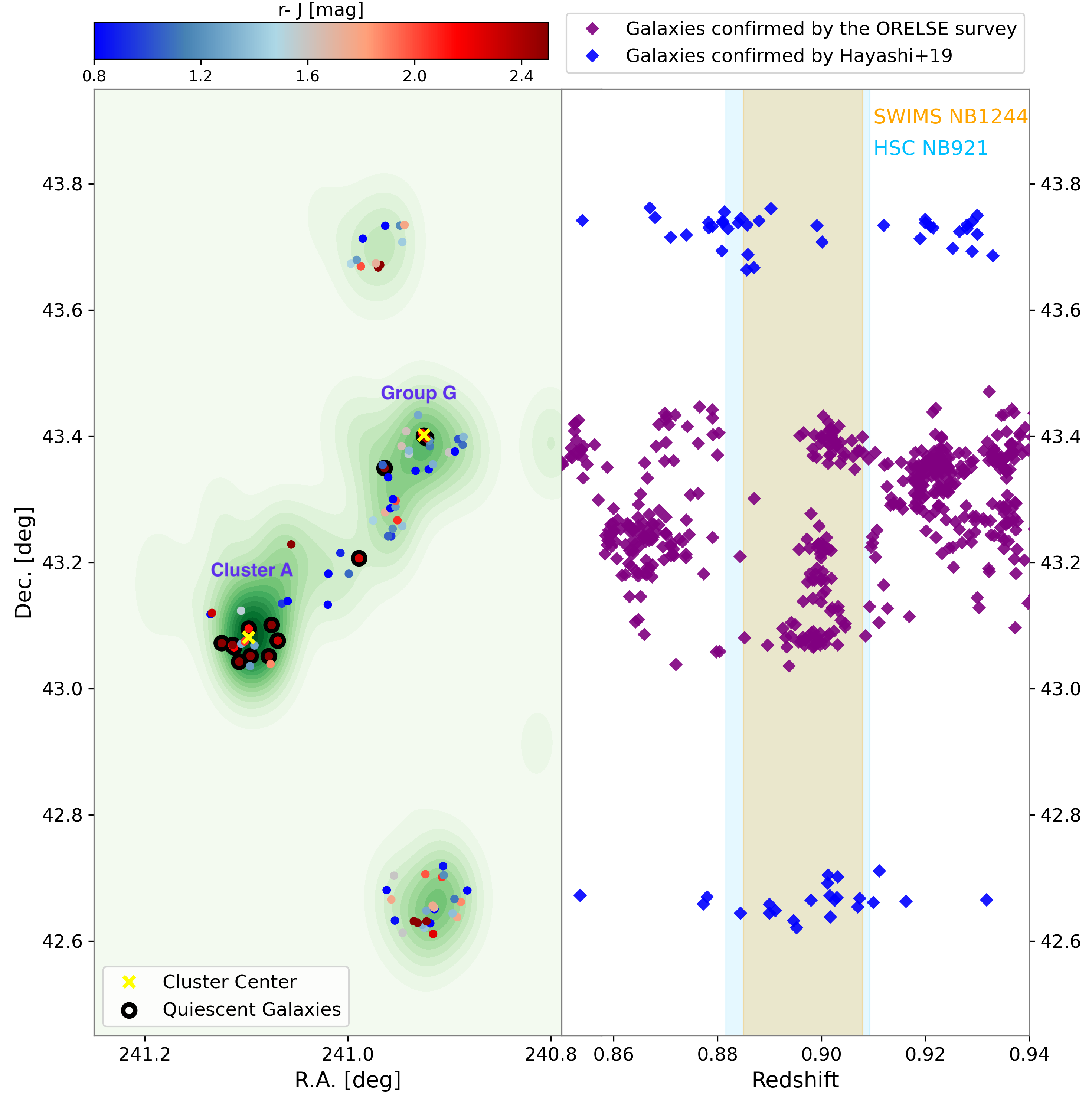}  
    \caption{(Left) Spatial distribution of emission-line galaxies within our narrow-band redshift coverage, color-coded according to their $r-J$ colors. The quiescent galaxies are defined by the rest-frame $UVJ$ diagram and selected from the subsample with IRAC coverage (Section \ref{sec:spitzer}). The yellow crosses mark the cluster/group centers as defined by \citet{Lemaux12}. (Right) Spectroscopically confirmed galaxies from the ORELSE survey \citep{Lemaux12} and \citet{Hayashi19}. The shaded regions correspond to the filter effective width coverage for \textit{NB1244} (orange) and \textit{NB921} (blue).}
    \label{fig:spatial}
\end{figure}

\subsubsection{Spitzer Imaging}\label{sec:spitzer}
The CL1604 supercluster was partially observed through the Spitzer program GO-30455 (PI: L. M. Lubin), with imaging conducted at 3.6 -- 8 $\mu$m using the Infrared Array Camera (IRAC) and at 24 $\mu$m using the Multiband Imaging Photometer for Spitzer (MIPS). Since only the central region is covered by Spitzer (Figure \ref{fig:fov}), incorporating these observations into our analysis presents challenges, as it may introduce systematic differences between fields with and without Spitzer coverage. This issue is particularly problematic when examining the environmental dependence of galaxy properties, therefore, we do not include the Spitzer observations for the SED fitting that will be detailed in Section \ref{physical properties}. However, in order to investigate whether galaxies are quenched, we employ the rest-frame $UVJ$ color-color diagram \citep{williams2009}. While the Subaru data extend up to the \textit{H}-band, allowing us to constrain the rest-frame \textit{U} and \textit{V} bands, estimating the rest-frame \textit{J}-band (corresponding to the observed \textit{K}-band) is more challenging. To address this, we split our sample into two groups. For the general estimation of stellar masses and SFRs across the entire sample, we rely solely on Subaru observations, as they provide complete coverage across all fields. For galaxies within Spitzer coverage (referred to as the ``Spizter Observed Sample''), we include the IRAC channel 1 (3.6 $\mu$m) observations and conduct a separate SED fitting to obtain rest-frame $UVJ$ information. This subsample includes 48 galaxies, and the discussion of quiescent galaxies is limited to this group. The additional process of matching IRAC channel 1 with our Subaru observations is described in detail in Section~\ref{sec:photometry}. The choice to use only IRAC channel 1 is motivated by its ability to provide sufficient wavelength coverage to estimate the rest-frame \textit{J-}band magnitude. lthough longer wavelengths from other IRAC channels and MIPS are also available, the decreasing spatial resolution in IRAC channels 3 and 4 raises concerns regarding reliable source matching. While the spatial resolution of IRAC channel 2 is acceptable, its relatively shallow depth significantly limits detections. Including both channels 1 and 2 would reduce our usable sample to just 39 galaxies. Given that IRAC channel 1 alone already allows us to robustly estimate the rest-frame \textit{J}-band magnitude, we restrict our analysis to this band for consistency and sample completeness. Archival CFHT $K$-band imaging is also available in the central region, but its relatively shallow depth (5$\sigma$ $\sim$ 22 mag; \citealt{Tomczak17}) compared to our Subaru observations limits its utility. Therefore, we only incorporate IRAC channel 1 where available, in addition to Subaru observations.


\subsubsection{Keck, Subaru and Gemini Spectroscopy}\label{sec:spec}
The spectroscopic redshift data utilized in this study were compiled from available KECK/DEIMOS+LRIS, Subaru/FOCAS, and Gemini-N/GMOS observations. The DEIMOS \citep{Faber03} and LRIS \citep{Oke95} observations on the Keck II 10 m telescope, focused on the original supercluster structure, were conducted between May 2003 and June 2010 \citep{Gal04, Gal08, Lemaux10, Lemaux12}, yielding secure redshifts for 525 objects within the adopted redshift range of $0.84 < z < 0.96$. Following the discovery of new structures extending to the northern and southern regions through HSC-SSP data, \citet{Hayashi19} conducted follow-up observations of these newly identified regions with Subaru/FOCAS \citep{Kashikawa02} and Gemini-N/GMOS \citep{Hook04}. These observations, conducted through two programs (S18A-125 for the FOCAS run, PI: T. Kodama, and GN-2018A-FT-107 for the GMOS run, PI: M. Hayashi), confirmed 137 galaxies at $z = 0.8$ -- $1.1$ among the 164 galaxies observed. The spectroscopically confirmed members are shown in Figure \ref{fig:spatial}.


\section{Analysis}\label{sec:analysis}
\subsection{Photometry}\label{sec:photometry}
With all the reduced images, we proceed with object detection and photometry on the final combined image using \texttt{SExtractor} \citep{bertin1996}. To ensure consistent measurements across bands, we match the point spread functions (PSFs) of the optical and NIR Subaru images using \texttt{IRAF} \citep{Tody1986}, as the PSF sizes vary with wavelength. The final matched images have a common PSF full width at FWHM of $0.\!''79$. Following a methodology similar to \citet{Asano20}, we use the dual-image mode, with the narrow-band images as the detection images. We adopt MAG\underline{\hspace{0.5em}}APER with an aperture diameter set to twice the PSF size for measuring the \textit{J - NB1244} and \textit{z - NB921} colors. For individual magnitudes across all bands, we use MAG\underline{\hspace{0.5em}}AUTO, based on the Kron aperture method \citep{kron1980}.

The photometric matching between Subaru and Spitzer is performed independently for this subsample. We retrieve the catalog from the Spitzer Heritage Archive, including only sources with \texttt{i1\_fluxtype = 1}, which indicates a detection at a significance level above 3$\sigma$ in IRAC channel 1. To match the spatial resolution, we smooth our Subaru \textit{J}-band images to match the FWHM of IRAC channel 1 ($1.\!''95$). After smoothing, we conduct aperture photometry on the Subaru image using a circular aperture with a diameter of $3.\!''8$ (denoted as FLUX\underline{\hspace{0.5em}}3.8). We then calculate the ratio between the $3.\!''8$ aperture flux (FLUX\underline{\hspace{0.5em}}3.8) and the original Subaru photometry (FLUX\underline{\hspace{0.5em}}AUTO), as described in Section \ref{sec:photometry}. This ratio quantifies the effect of instrumental differences, such as contamination from nearby sources that cannot be deblended at the lower resolution. Finally, we divide the Spitzer $3.\!''8$ aperture flux (\texttt{i1\_f\_ap1}) by this ratio to obtain an IRAC flux comparable to our original Subaru photometry. This corrected IRAC channel 1 flux is then adopted for our SED fitting to estimate the rest-frame \textit{U}, \textit{V}, and \textit{J} magnitudes.

\subsection{Sample Selection}\label{sec:sample}

\begin{figure}[ht]
    \centering
    \includegraphics[width=0.46\textwidth]{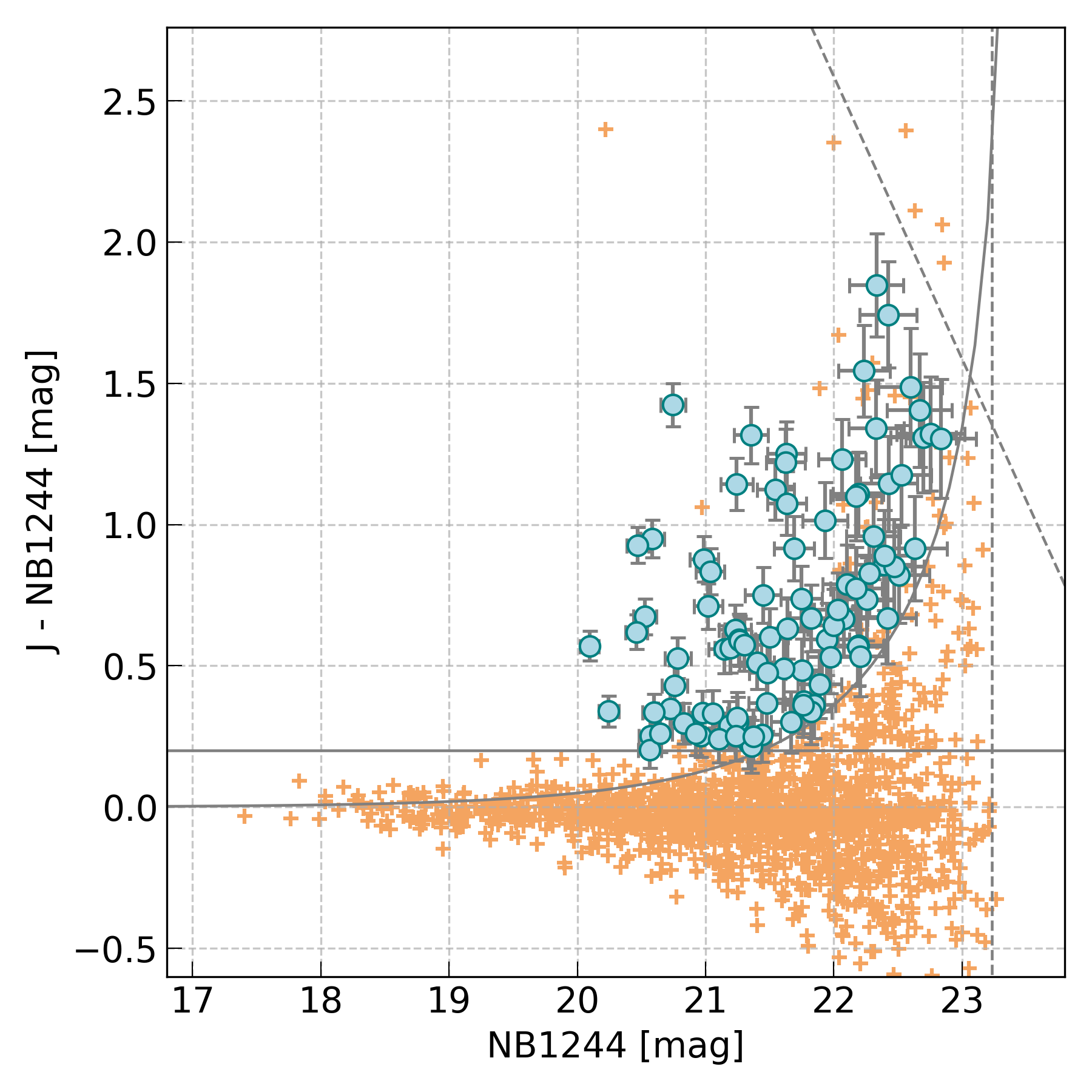}  
    \caption{Narrow–broad band color–magnitude diagram used to select \ha\ emitters. The color excess is calculated using MAG\_APER. The gray dashed lines represent the 5$\sigma$ limiting magnitudes in the \textit{J} and \textit{NB1244} bands. The gray solid curve and the line indicate the 3$\sigma$ excess in \textit{J - NB1244} colors and the \textit{J - NB1244} = 0.2 cut, respectively. Orange crosses represent galaxies brighter than the narrow-band limiting magnitude. Galaxies meeting our selection criteria (see Section \ref{sec:sample}) are shown as blue points with error bars representing magnitude/color uncertainties.}
    \label{fig:emitter}
\end{figure}

We perform sample selection based on \textit{NB1244} detection, which is designed to select emission-line galaxies through their color excess. When an emission line falls within both the narrow-band and broad-band filters, the flux density in the narrow-band will be greater than that in the broad-band. In other words, \textit{J - NB1244} colors of emitters will appear redder compared to those of non-emitters. Our candidate selection is based on this color excess in the \textit{J - NB1244} color. We define \textit{NB1244} emitters as galaxies that meet the following criteria: (1) Both \textit{J} and \textit{NB1244} magnitudes are brighter than the 5$\sigma$ limiting magnitude, with the \textit{J}-band magnitude also required to be fainter than 19.5 to avoid photometric scatter at the bright end. (2) \textit{J - NB1244} $\textgreater$ 3$\sigma$, where $\sigma$ represents the 1$\sigma$ photometric error in the \textit{J - NB1244} color \citep{Bunker95}. (3) \textit{J - NB1244} $\textgreater$ 0.2. A total of 118 emitters satisfy this selection criteria, all of which are detected in both the HSC-SSP catalog and the \textit{NB921} catalog. The color–magnitude diagram used to select \ha\ emitters are shown in Figure \ref{fig:emitter}.

While the flux excess in \textit{NB1244} can be attributed to \ha\ emission at z = 0.9, it could also be due to [O\,\textsc{iii}] emission at $z = 1.5$ or [O\,\textsc{ii}] emission at $z = 2.4$. To mitigate potential contamination from [O\,\textsc{iii}] or [O\,\textsc{ii}], we refer to the photometric redshift data from HSC-SSP PDR2 \citep{Aihara18, Aihara19} and exclude galaxies with photometric redshifts greater than $z = 1.3$. We adopt the Extended Photometric Redshift (EPHOR) values from the catalog, which were calculated using CModel fluxes across all five bands \citep{Tanaka18}. The photometric redshifts achieve an accuracy of $\sigma\left( \frac{\Delta z_{\mathrm{phot}}}{1 + z_{\mathrm{phot}}} \right) \sim 0.05$. Next, we cross-match the \textit{NB1244} emitter catalog with the \textit{NB921} photometric catalog with the matching radius set to $1.\!''0$. While the \ha\ emitters are selected based on the \textit{J$-$NB1244} excess (\textgreater3$\sigma$ and \textgreater0.2~mag), we do not require the same selection criteria for \hb\ emitters, as the \hb\ line is significantly fainter and may not produce a strong narrow-band excess. In this step, we simply cross-match and retain sources with NB921 and z-band magnitudes brighter than the detection limits. After cross-matching with the \hb\ emitter catalog and removing those sources, we are left with 102 emitters.

\begin{figure*}
  \centering
  \includegraphics[height=0.32\textheight]{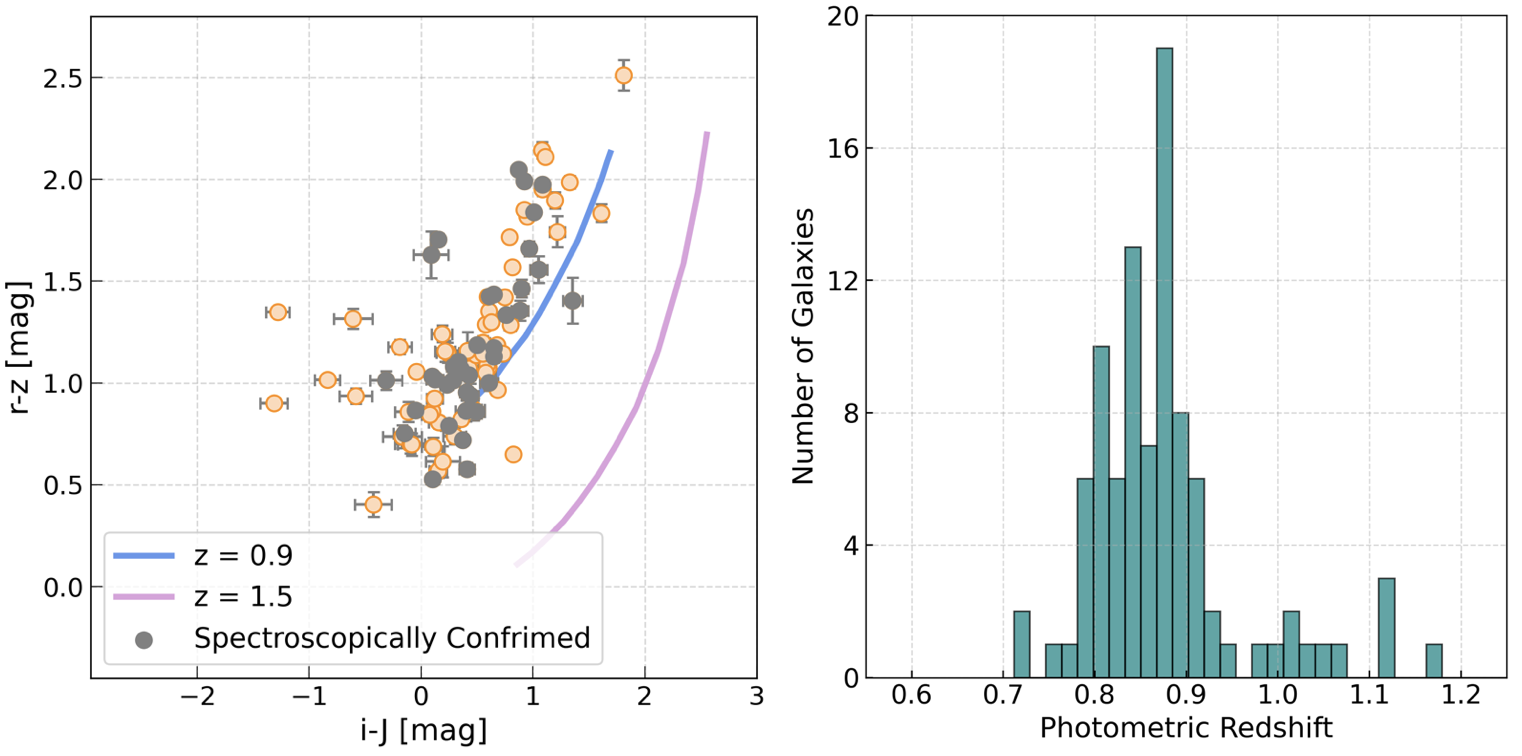}
  \caption{(Left) Predicted color tracks at $z = 0.9$  and $z = 1.5$ based on the \citet{Kodama1999} model, illustrating variations in the bulge-to-total ratio from 1 to 0 along the curves. The final sample of 94 galaxies is shown as filled circles, with orange circles representing narrow-band selected emitters and gray circles representing narrow-band selected emitters confirmed with spectroscopically redshifts. Error bars indicate magnitude uncertainties. (Right) Photometric redshift distribution of the final sample.}
  \label{fig:photoz}
\end{figure*}

\begin{figure}[ht]
    \centering
    \includegraphics[width=0.46\textwidth]{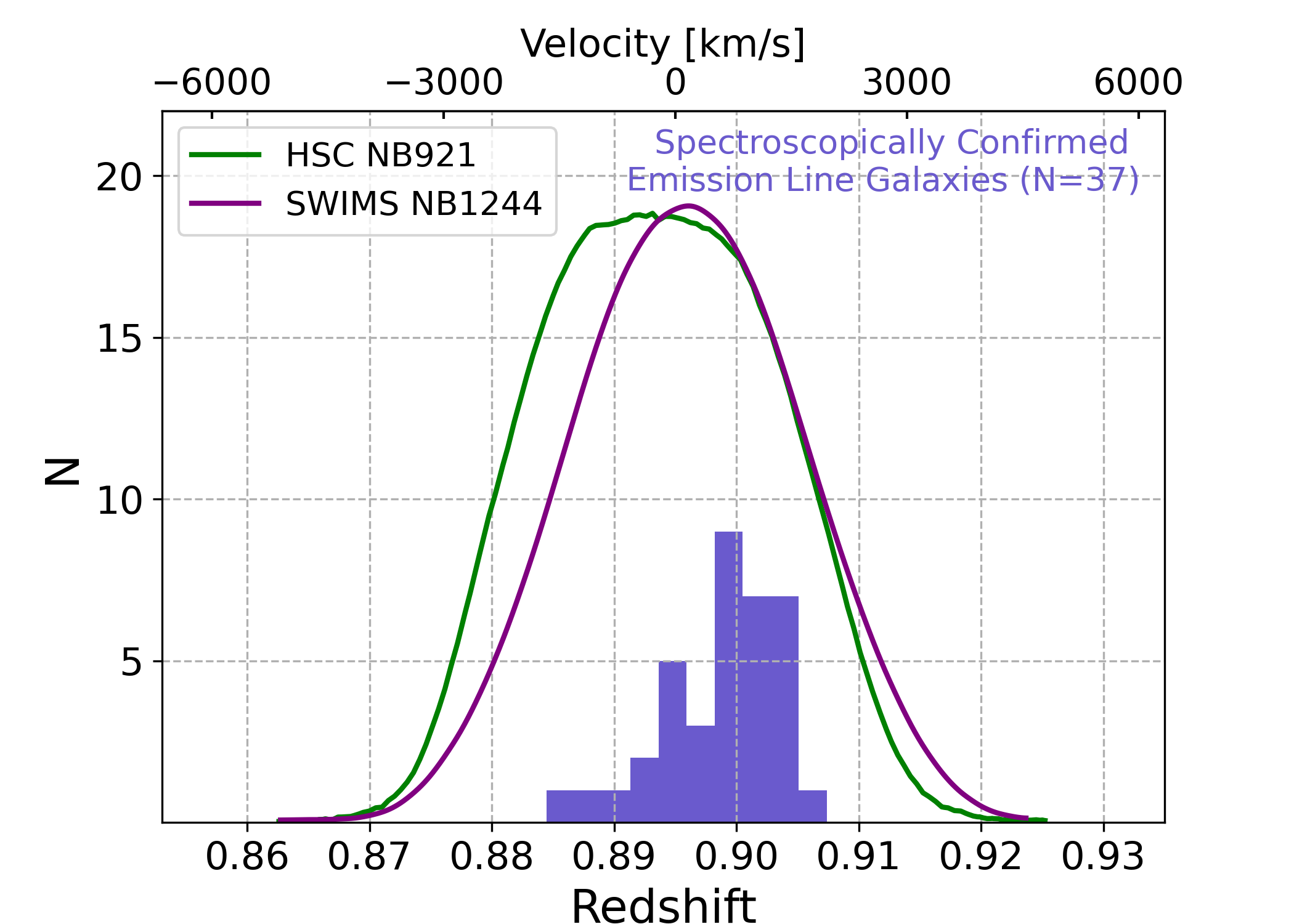}
    \caption{Redshift distribution of 37 spectroscopically confirmed emitters among 94 photometrically selected emitters. The transmission curves of the \textit{NB921} and \textit{NB1244} narrow-band filters are also shown.}
    \label{fig:filter_spec_z}
\end{figure}

We then inspect the emitter images in \textit{NB1244}, \textit{J}, \textit{NB921}, and \textit{z} bands to avoid galaxies that fall near the edges of the detector. For example, if an emitter has full \textit{NB1244} coverage but only partial \textit{J-}band coverage --- which may occur for galaxies near the image FoV edge --- it may result in an overestimated color excess in the \textit{J-NB1244} color, leading to a false emitter detection. We carefully inspect their images and remove eight galaxies in this step that have incomplete coverage or satellite contamination. Our final sample consists of 94 galaxies. The photometric redshift distribution and the color-color track of the final sample are shown in Figure \ref{fig:photoz}. Among these 94 galaxies, 37 have spectroscopic confirmation, with their redshift distribution presented in Figure \ref{fig:filter_spec_z}.

\subsection{Emission Line Flux Measurements and Filter Response Corrections}\label{sec:line flux}

Our study leverage the double narrow-band techinique, therefore could obtain the \ha\ and \hb\ emission line flux to estimate the dust extinction \citep{Bunker95}. We note that, because of their close proximity in wavelength, \ha\ and \nii\ lines come into the \textit{NB1244} filter simultaneously for most of our member galaxies in the supercluster under concern. Therefore we first compute the combined \ha\ + \nii\ flux and then apply a correction to subtract the \nii\ contribution.
We calculate the \ha\ + \nii\ and \hb\ fluxes using the narrow-band and broad-band magnitudes, following:

\begin{equation}\label{eq1}
    f_{BB}\times\Delta_{BB}  =  F_{\rm{{line}}}¥ + f_{\rm{c}}\times\Delta_{BB} 
\end{equation}
\begin{equation}\label{eq2}
    f_{NB}\times\Delta_{NB}  = F_{\rm{{line}}}+ f_{\rm{c}}\times\Delta_{NB} 
\end{equation}

where $f_{NB}$ and $f_{BB}$ represent the flux densities in the narrow-band and the corresponding broad-band, respectively. $\Delta_{NB}$ and $\Delta_{BB}$ denote the FWHMs of the narrow and broad bands. The emission line flux, $F_{\rm{line}}$, corresponds to the emission lines captured by the two narrow-band filters employed in this study. Specifically, the \textit{NB1244} filter captures the \ha\ + \nii\ emission, while the \textit{NB921} filter captures the \hb\ emission, $f_{\rm{c}}$ represents the continuum flux density. To isolate the \ha\ flux from the combined \ha\ + \nii\ emission line flux, we apply the \nii/\ha\ ratios as a function of stellar mass at \(z \sim 0.9\), following the calibration by the \(\text{KMOS}^{3D}\) program \citep{Wuyts_kmos}. However, the scatter in the \nii/\ha\ ratio introduces uncertainty in the derived \ha\ flux. At the highest stellar mass bin, this scatter can lead to a $\sim$10\% variation in the estimated \ha\ flux, which corresponds to an uncertainty of $\sim$ 0.27 mag in A(\ha). Our analysis is subject to this level of uncertainty.

Although narrow-band filter is effective tool for capturing emission lines by detecting flux excess within their redshift coverage, one significant uncertainty associated with narrow-band imaging is the underestimation of emission line fluxes caused by the varying filter response function across its wavelength range. The filter's transmission efficiency is not uniform; it peaks at a central wavelength and decreases toward the edges of the filter's bandwidth (Figure \ref{fig:filter}). Galaxies whose emission lines fall near the peak of the filter's response curve benefit from maximum transmission efficiency, resulting in accurate flux measurements. In contrast, galaxies whose emission lines are shifted away from the peak --- due to slight differences in redshift --- suffer reduced transmission efficiency. This leads to an underestimation of their observed flux.

This effect is particularly important when studying a sample of galaxies spread over a range of redshifts within the filter's coverage. The varying transmission can introduce systematic biases in the measured emission line fluxes, affecting derived properties such as SFRs and emission line ratios. To mitigate this issue, it is essential to account for the filter transmission function when calculating fluxes. However, not all of the emitters we select have spectroscopic redshifts that allow us to correct for this effect. Therefore, for those galaxies with accurate spectroscopic redshifts, we correct for the filter response curve; for those without, we mark their flux values as lower limits in the following analysis. We correct for the filter response effect using the equation below:

\begin{equation} F_{\text{int}} = F_{\text{obs}} \times \left( \frac{T_{\text{peak}}}{T_{\lambda}} \right) \end{equation}

where $F_{\text{int}}$ is the intrinsic emission line flux, $F_{\text{obs}}$ is the observed flux measured from the narrow-band imaging, $T_{\text{peak}}$ is the peak transmission of the filter, $T_{\lambda}$ is the filter transmission at the wavelength corresponding to the galaxy's emission line, determined by its spectroscopic redshift. For galaxies without a secure spectroscopic redshift, we assume $T_{\lambda} = T_{\text{peak}}$.

By applying this correction, we adjust the observed fluxes to more accurately reflect the intrinsic emission line fluxes, reducing systematic biases in our derived galaxy properties. Although \ha\ and \hb\ emission line fluxes, thus SFRs, tend to be underestimated for a significant fraction of galaxies, the \ha/\hb\ ratio is kept almost unchanged for the majority of galaxies because of similarity of the two NB filter response functions at the redshift range where most of our galaxies under concern actually fall in (Figure \ref{fig:filter}). Therefore the dust extinction value derived from this line ratio, which we will focus on in this paper, will be largely unaffected. However, the SFRs for these sources should be interpreted as lower limits. Based on the spectroscopic sample, we find that the median correction factor is $\sim$1.083 for \ha\ and $\sim$1.073 for \hb. Thus, the derived SFRs should be interpreted as lower limits.

\subsection{Dust Extinction, Stellar Mass and SFRs}\label{physical properties}
Using the \ha\ and \hb\ emission line fluxes calculated in Section \ref{sec:line flux}, we derive the dust extinction by comparing the observed \ha/\hb\ ratio to theoretical predictions. In this study, we adopt Case B recombination conditions, assuming ${\rm T_e = 10^{4}}$ K and ${\rm n_e = 10^{2}}$ ${\rm cm^{-3}}$, based on photoionization models from CLOUDY version 17.02 \citep{Ferland17}. Under these conditions, the intrinsic \ha/\hb\ intensity ratio is predicted to be 2.79. To calculate $E(B-V)_{\text{neb}}$ for our sample, we apply the reddening curve from \citet{Cardelli89} ($R_{\rm{v}} = 3.1$) and compare the observed \ha/\hb\ flux ratios to the expected Case B values following:
 
\[ E(B-V)_{neb} = \frac{2.5}{k(H\beta) - k(H\alpha)} \log_{10}\left( \frac{H\alpha / H\beta}{2.79} \right)\]

Here, $k(\lambda)$ represents the extinction coefficient at wavelength \( \lambda \) as given by \citet{Cardelli89}. We then derive the extinction values A(\ha) using:

\[ A(\lambda) = k(\lambda) \cdot E(B-V) \]

We then perform SED fitting using the broad-band photometry including HSC (\textit{g,r,i,z,y}) and SWIMS/MOIRCS (\textit{J, H}) to estimate stellar masses for our sample galaxies. The emission-line contributions in the \textit{z}-, \textit{J}-, and \textit{H}-bands have been removed based on the corresponding narrow-band measurements. Our SED fitting is carried out with the Code Investigating GALaxy Emission ({\tt{CIGALE}}, \citealt{Boquien19}), which is optimized for data spanning a wide range of wavelengths, from X-ray to radio. We adopt a delayed exponential star-formation history ({\tt sfhdelayed}) with the functional form ${\rm SFR}(t) \propto t \exp({-t}/{\tau})$. The stellar population synthesis model is based on \citet{Bruzual03} with solar metallicity. The dust extinction values in SED fitting are fixed based on the results from \ha/\hb\ flux. However, dust extinction in nebular lines is typically higher than in the stellar continuum because line-emitting regions trace very young stars that are still embedded in dusty birth clouds \citep[e.g.,][]{Calzetti2000, Wuyts11, Kashino13, Reddy15}. The conversion between nebular and stellar color excess was first quantified by \citet{Calzetti2000} in local starburst galaxies, where $E(B-V)_{\mathrm{s}} \approx 0.44\,E(B-V)_{\mathrm{neb}}$. At higher redshift, the ratio varies; for instance, \citet{Kashino13} reported $E(B-V)_{\mathrm{stars}}/E(B-V)_{\mathrm{neb}} = 0.69\text{--}0.83$ for star-forming galaxies at $1.4<z<1.7$. In this work, we infer $E(B-V)_{\mathrm{neb}}$ from the \ha/\hb\ flux ratio and, adopting the conversion factor of 0.44, apply the \citet{Calzetti2000} attenuation law to model the stellar continuum dust component in our SED fitting. For galaxies with IRAC coverage (see Section~\ref{sec:spitzer}), we include IRAC channel 1 photometry in the SED fitting to predict rest-frame $UVJ$ colors. These colors are used to identify quiescent galaxies, following the widely used $UVJ$ diagram method for distinguishing between star-forming and quiescent populations \citep{williams2009}. Although SED fitting with IRAC photometry is performed for this subsample, we use only the resulting rest-frame $UVJ$ colors for classification purposes, in order to avoid introducing biases when comparing to the full sample. The criteria used for classification follow the definitions in \citet{Mao22}. Other physical parameters, such as stellar mass, are consistently derived from the initial SED fitting based solely on Subaru broad-band photometry.


The SFRs are estimated using the intrinsic \ha\ emission line flux, after correcting for dust extinction. With the dust-corrected \ha\ flux, we follow the SFR-$L_{\mathrm{H\alpha}}$ relation given by \citet{kennicutt98_2}:

\begin{equation}\label{eq3}
 \frac{\mathrm{SFR}}{\mathrm{M_{\odot}}\,\mathrm{yr^{-1}}} = 7.9 \times 10^{-42}\,\,\mathrm{L}_{\mathrm{H\alpha}}(\mathrm{erg\,s}^{-1})
\end{equation}

We then rescale the SFR by dividing by a factor of 1.53 \citep{Driver13} to align with the \citet{Chabrier03} IMF.

\subsection{Environment Quantification}\label{environment}

To examine the environmental effects on galaxy properties, we need to define the environment in which a galaxy is located. Various statistical methods have been employed to estimate the richness and structure of large-scale cosmic environments, such as the nearest-neighbor density method \citep[e.g.,][]{Balogh04, Tanaka05, Koyama08}, friends-of-friends algorithms \citep[e.g.,][]{Crook07, Lavaux11}, the CAMIRA cluster finding algorithm \citep{Oguri14} and the Monte Carlo Voronoi method \citep{Darvish15, Lemaux17, Lemaux18}.

In the case of CL1604, a complex system extending both transversely and along the line of sight (Figure \ref{fig:spatial}), defining the environment becomes more challenging. Methods based on photometric redshifts or red-sequence galaxies can be effective, particularly given the high-quality photometric redshifts and extensive spectroscopic coverage available in this field \citep[e.g.,][]{Hung20}. However, projection effects can still pose difficulties, especially in the outer regions, where both Spitzer photometry and sufficient spectroscopic confirmation are lacking. Therefore, although using only spectroscopically confirmed sources or narrowband-selected emitters provides clean samples, such approaches introduce biases toward star-forming systems and may not fully capture environmental diversity, particularly beyond the core of the supercluster. To avoid the complexity induced by this issue, in this study, we assess environmental effects in the known cluster A and group G solely based on the distance from the overdensity centers. The center of each structure is defined by \citet{Lemaux12} based on the \textit{i}- or \textit{z}-band weighted positions of spectral members, as shown by the yellow cross in Figure \ref{fig:spatial}.

In addition to the cluster and group with well-defined centers where we can study galaxy properties as a function of distance from the cluster centroid, we are also interested in the rest of the Subaru pointings, specifically C1, C2, N1, and S3. These fields either lack both photometric and spectroscopic coverage (N1 and S3) or consist of filamentary structures where a center does not exist (C1 and C2). Therefore, we discuss the environmental dependence of galaxies in these regions based on the \ha\ emitter number density and split the sample into two groups: galaxies in the rich cluster (A) and galaxies in groups or filamentary structures (G, C1, C2, N1, S3).

\begin{figure}[ht]
    \centering
    \includegraphics[width=0.46\textwidth]{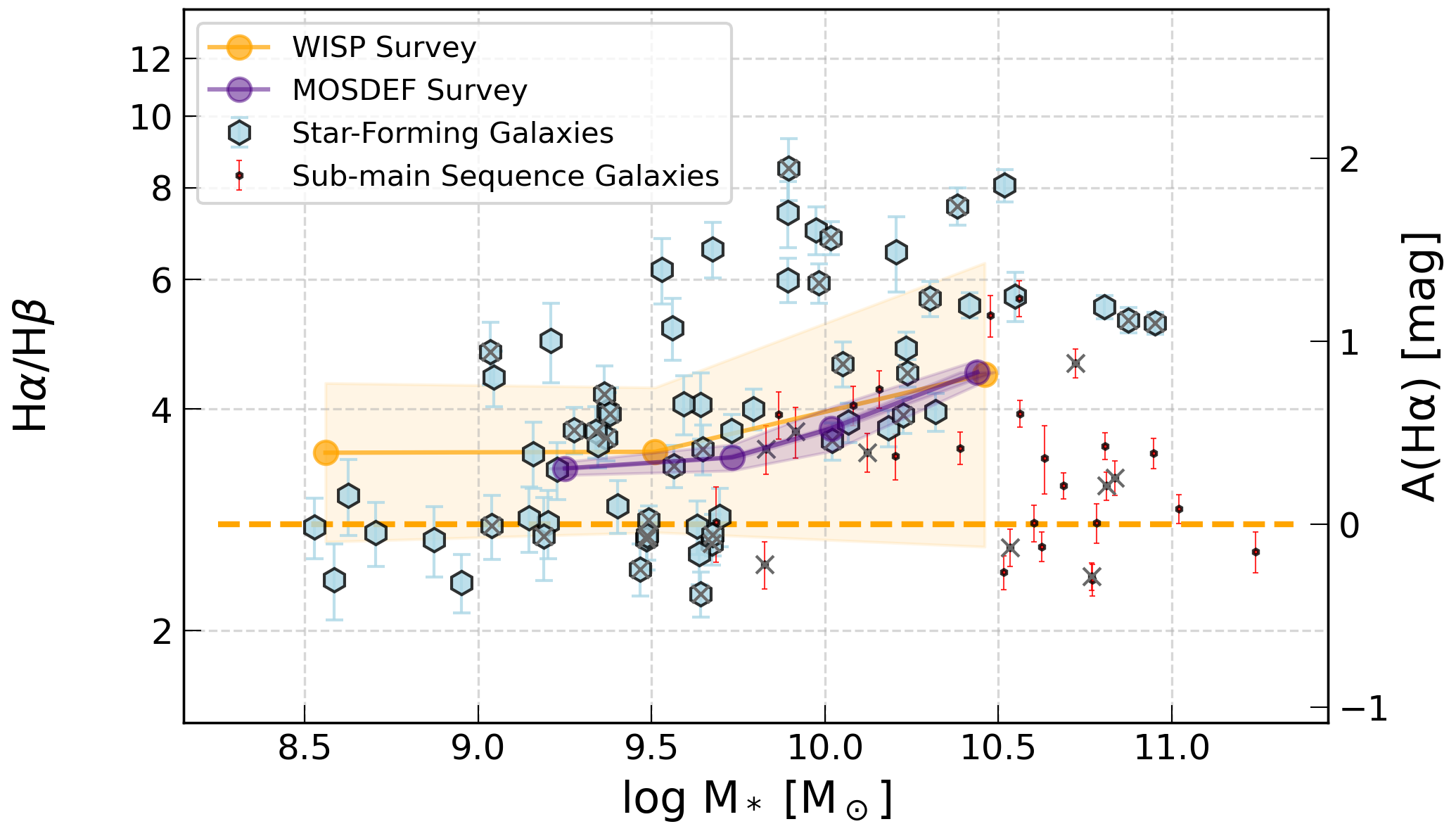}  
    \caption{Stellar mass versus extinction diagram. Blue hexagons represent star-forming galaxies, while small red hexagons denote sub-main-sequence galaxies, defined as those lying more than 0.3 dex below the star-forming main sequence. Black crosses mark spectroscopically confirmed emitters with filter response corrections applied to both \ha\ and \hb\ fluxes. The orange and indigo lines represent the \ha/\hb\ ratios as a function of stellar mass derived from the HST WISP survey ($0.75 \leq z \leq 1.5$; \citealt{2013ApJ...763..145D}) and the Keck MOSDEF survey ($z \sim 2.3$; \citealt{Shapley22}), respectively. The shaded regions around these lines indicate the corresponding uncertainties. The dashed line indicates the intrinsic \ha/\hb\ ratio in the dust-free case under the Case B recombination assumption.}
    \label{fig:extinction}
\end{figure}

\begin{figure}
  \centering
  \includegraphics[width=0.46\textwidth]{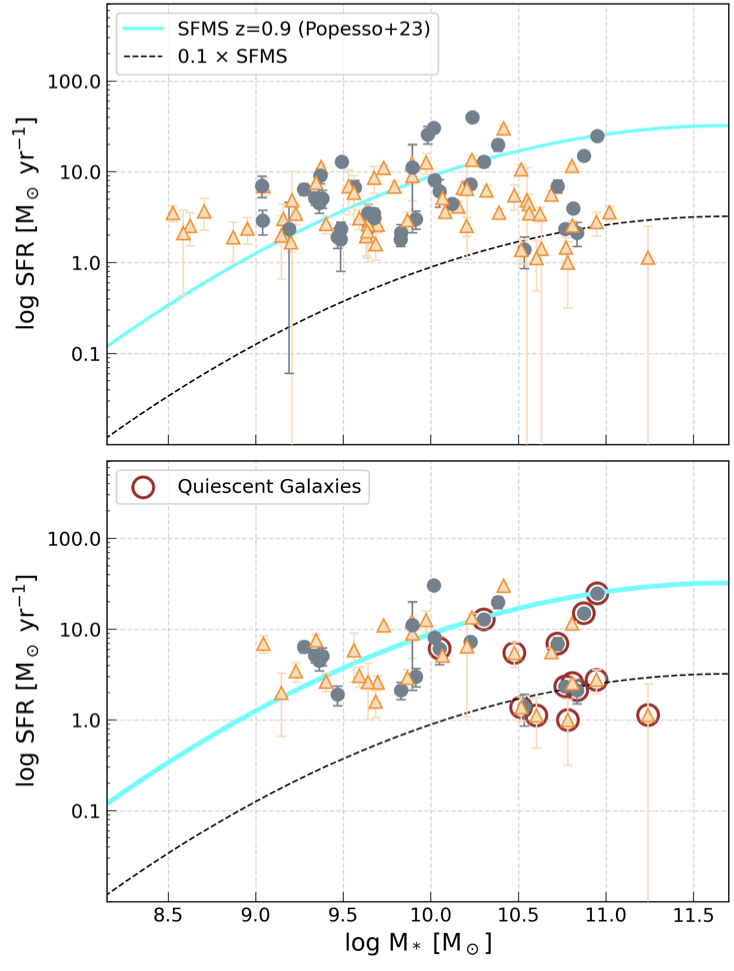}
  \caption{(Top) Stellar mass versus SFR plot for our final sample of 94 emitters. The colors of the data points are the same as those in Figure~\ref{fig:photoz}, with orange triangles representing narrow-band selected emitters and gray circles representing those confirmed with spectroscopic redshifts. The SFRs are calculated based on the H$\alpha$ emission line flux after removing the \nii\ contribution and correcting for dust extinction. Error bars are propagated from uncertainties in the H$\alpha$ flux and extinction, derived from the H$\alpha$/H$\beta$ ratio. The cyan line indicates the star-forming main sequence at $z=0.9$ from \textcite{Popesso23}, and the black dashed line marks 0.1 times the main sequence level. (Bottom) Same as the top panel but for the subsample with IRAC coverage, where the rest-frame $UVJ$ diagram is used to classify quiescent galaxies.}
  \label{fig:ms cl1604}
\end{figure}


\begin{figure}[ht]
    \centering
    \includegraphics[width=0.46\textwidth]{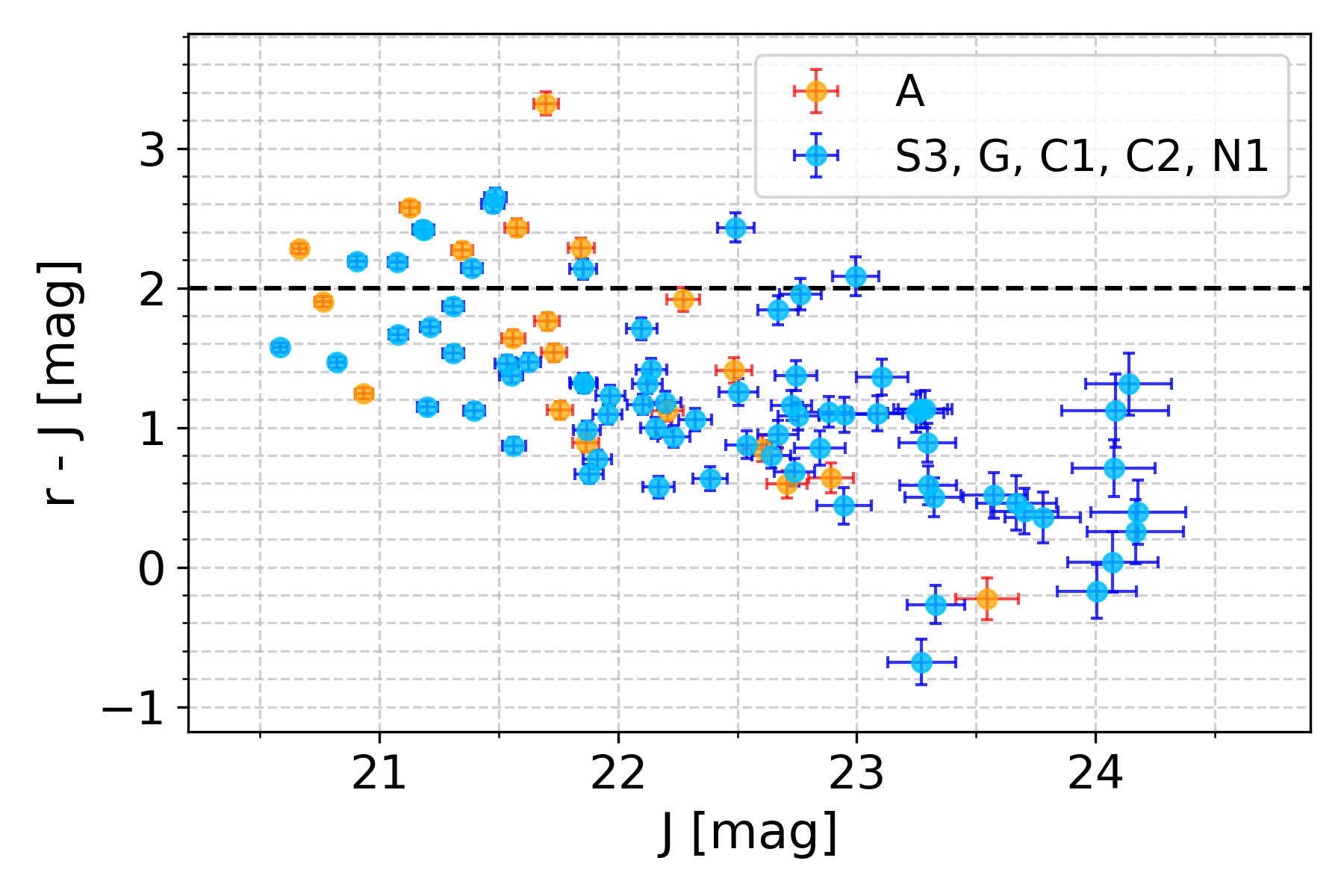} 
    \caption{Color–magnitude diagram for selected emission-line galaxies across all six fields. Orange data points indicate galaxies in cluster A (overdense regions), while blue data points represent galaxies in the remaining fields (moderate-density regions).}
    \label{fig:color}
\end{figure}

\begin{figure*}[ht!]
    \centering
    \includegraphics[width=0.89\textwidth]{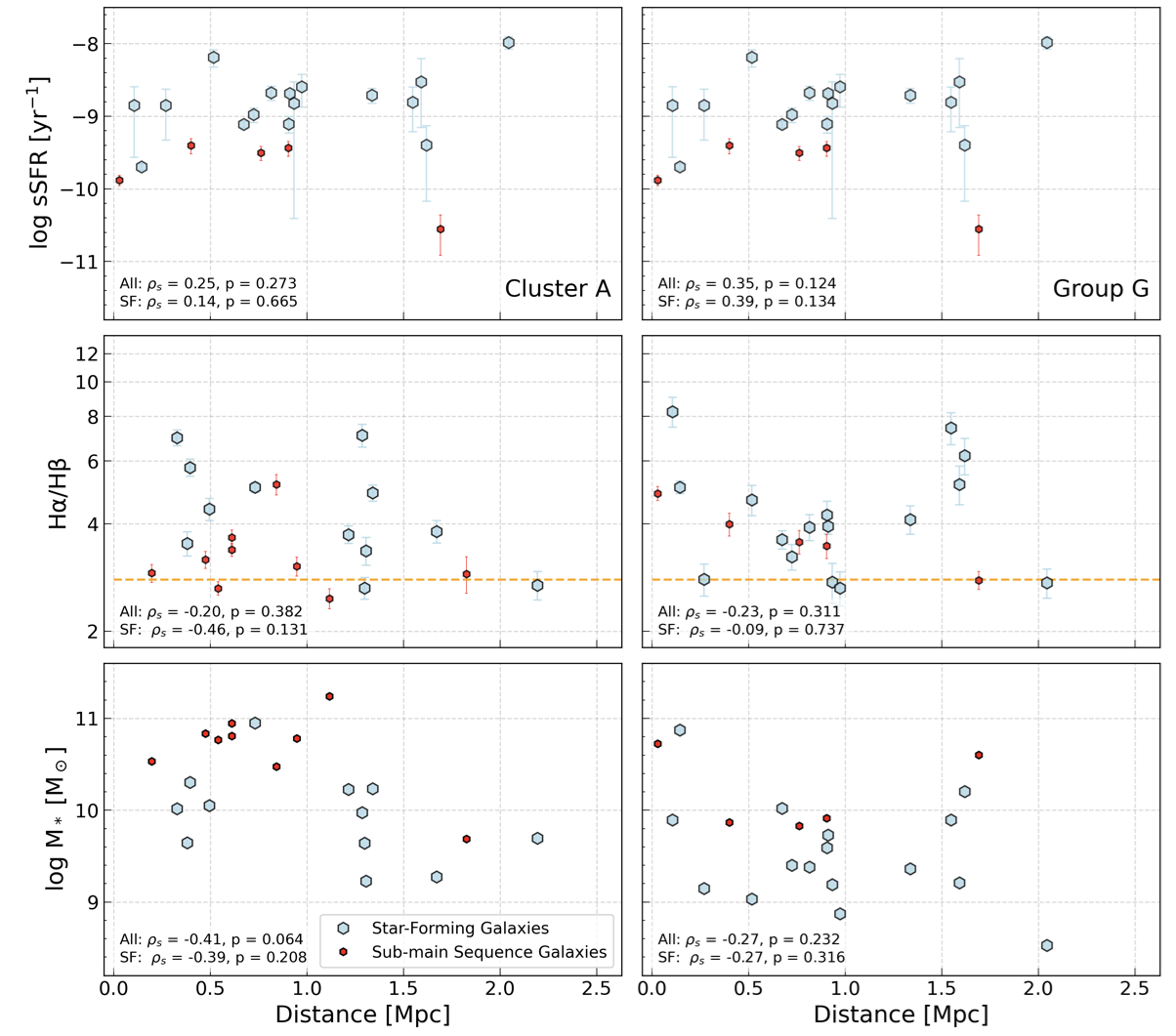} 
    \caption{
        Clustocentric radius versus physical properties of galaxies for cluster A (left) and group G (right). 
        Symbols for star-forming and sub-main-sequence galaxies are consistent with those defined in Figure~\ref{fig:extinction}. The Spearman rank correlation coefficients ($\rho_s$) and associated $p$-values for both the full population 
        and the star-forming subsample are indicated within each panel.
    }
    \label{fig:clusters}
\end{figure*}



\begin{figure*}
  \centering
  \includegraphics[height=0.55\textheight]{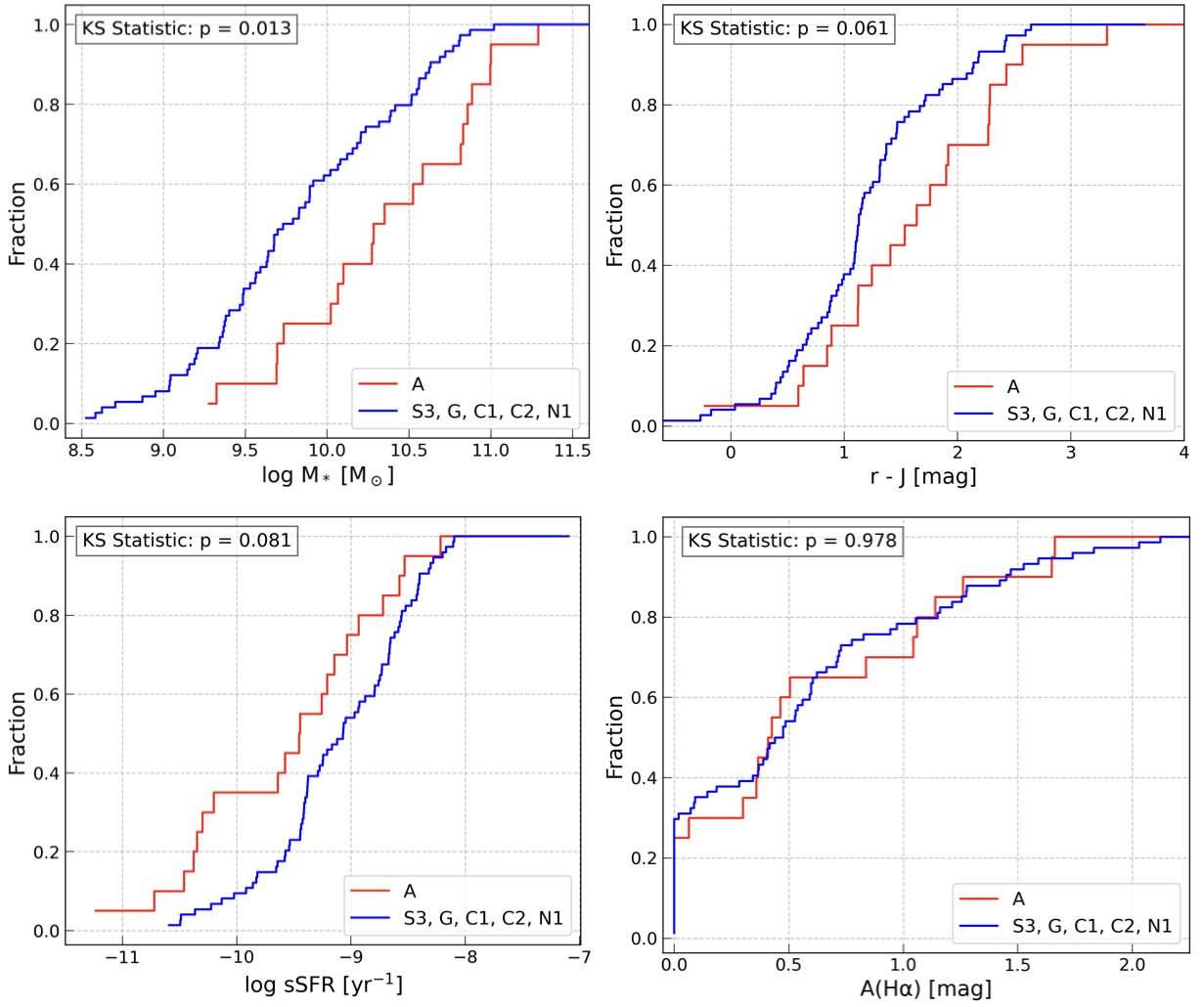}
  \caption{Cumulative distribution functions and p-value derived from the KS test comparing galaxies in cluster A (overdense regions) and the rest of the fields (moderate-density regions).
  }
  \label{fig:ks cl1604}
\end{figure*}

\section{RESULTS}\label{result}

In this section, we present the results of our analyses. First, we explore the correlation between dust extinction and stellar mass. We then investigate the star formation activities of galaxies across the entire structure. Lastly, we examine the potential correlations between environments and these physical quantities, including stellar mass, specific star formation rates (sSFRs), and dust extinction.

\subsection{Dust Extinction and Stellar Mass}\label{dust_extinction_result}

In Section \ref{sec:data}, we introduced our narrow-band observations and the relevant data reduction procedures used to study the H$\alpha$ and H$\beta$ emission lines. Subsequently, in Section \ref{physical properties}, we described the calculation processes for stellar mass and dust extinction. In Figure \ref{fig:extinction}, we present our results, where we divide galaxies into two populations: star-forming galaxies and sub-main-sequence galaxies (those more than 0.3 dex below the main sequence). This division is based on the fact that intense star formation generates dust through processes associated with the stellar life cycle, leading to higher extinction. Therefore, it is appropriate to compare dust extinction within galaxy populations that have similar star formation activities.

We observe an increasing trend of dust extinction with increasing stellar mass for star-forming galaxies. In contrast, less massive galaxies (i.e., $M_* < 10^{9.0}\,M_\odot$) and sub-main-sequence galaxies exhibit relatively low or negligible dust extinction compared to the more massive, star-forming galaxies. We also note that there are a fraction of galaxies which show negative extinction values. We will discuss these results later in Section \ref{discussion:dust extionction}.

\subsection{Star Formation Activities}\label{result:sf activities}

With accurate dust extinction measurements and H$\alpha$ emission line fluxes, we are able to constrain the galaxy SFRs with high precision (see Section \ref{physical properties}). Figure \ref{fig:ms cl1604} presents our results, where we study the galaxy SFR as a function of stellar mass. We notice that the majority of our galaxies scatter around the star-forming main sequence at this redshift, as derived by \citet{Popesso23}, thus exhibiting similar properties to field galaxies in terms of star formation activity. However, we also observe a drop-off from the main sequence at the massive end (i.e., $M_{*} > 10^{10.0}\,M_\odot$), where galaxies fall below by 0.3 dex, or even one dex, compared to the star-forming main sequence.

The subsample with IRAC coverage is plotted in the bottom panel of Figure \ref{fig:ms cl1604}, thanks to its 3.6 $\mu$m observed frame coverage, we are able to model the rest-frame $UVJ$ color more precisely, and thus discuss the potential quiescent population. We find that most of the galaxies at the massive end are classified as quiescent based on the rest-frame $UVJ$ diagram. Although they are still detected with \ha\ emission lines, indicating their star formation is not entirely ceased, their positions on both the main sequence diagram and the $UVJ$ diagram suggest that these galaxies are in the process of quenching or, at the very least, experiencing suppressed star formation. In fact, the H$\alpha$ detections in these massive galaxies may partially originate from blended \nii\ emission, which could be associated with AGN or LINER-type processes rather than star formation in H{\footnotesize II} regions \citep{Lemaux17}. Since our narrow-band technique cannot deblend H$\alpha$ and \nii, and our correction for \nii\ contamination is based on calibrations from star-forming galaxies (see Section \ref{sec:line flux}), this may lead to an overestimation of the SFRs in these galaxies.

\subsection{Galaxy transition in Color, Stellar Mass and Star Formation Activities Across Environments}

Finally, we examine the correlation between key physical properties and the environment. As detailed in Section \ref{environment}, we investigate galaxy properties in two ways: (1) For cluster A and group G, which have well-defined centers from the literature, we study the physical properties as a function of clustocentric radius. (2) We define two classes of galaxies based on the \ha\ number density and compare the distribution of properties between these classes to understand the statistical differences between galaxies in different environments.

We first examine the colors of our \ha\ emitters. Figure \ref{fig:color} presents the $r-J$ vs. $J$ color–magnitude diagram. Based on our number-density-based environmental classification, region A (the previously known cluster) is considered a high-density region, while G, C1, C2, S3, and N1 are classified as less-dense regions. Our analysis reveals that \ha\ emitters in the high-density region A exhibit redder colors compared to the overall color distribution across the six fields. To further investigate their spatial distribution, we display the distribution of \ha\ emitters on the density map in the left panel of Figure \ref{fig:spatial}, with emitters color-coded according to their $r-J$ colors. The right panel illustrates the spectroscopically confirmed galaxies and the coverage of our narrow-band filters. We find that quiescent galaxies generally have red $r-J$ colors. As discussed in Section \ref{dust_extinction_result}, most sub-main-sequence galaxies exhibit low dust extinction, indicating that the red colors of quiescent galaxies are primarily due to older stellar populations rather than increased dust content.

We then examine other physical properties as a function of clustocentric radius in cluster A and group G, as shown in Figure \ref{fig:clusters}. In both systems, galaxies near the center are on average more massive and exhibit lower sSFR than their counterparts in the outskirts, with the contrast being strongest in cluster A. Dust extinction, by contrast, shows no statistically significant radial gradient.

To explore whether this result extends beyond bound structures, we compare galaxies in dense versus less-dense field regions. The Kolmogorov–Smirnov test (Figure \ref{fig:ks cl1604}) confirms that their dust-extinction distributions are indistinguishable, although galaxies in denser fields are systematically more massive, display marginally lower sSFRs, and have slightly redder $r-J$ colours. Because dust extinction correlates with both stellar mass and star-formation activity, attributing its behaviour solely to environment can be misleading. This interplay is clearest in cluster A: the inner and outer zones host populations with markedly different masses and SFRs. For the star-forming galaxies alone, the Balmer decrement appears to increase toward the cluster centre, although the correlation is not statistically significant.

We note that our selection, which is based on \ha\ flux excess in NB1244, introduces biases against fainter or low–equivalent-width emitters. This may in turn affect the interpretation of environmental trends, for instance, dusty or merger-driven star formation could be missed if \hb\ counterparts are undetected, and the apparent lack of low-mass emitters in Cluster A may partly reflect such selection effects in addition to intrinsic cluster growth.

\section{DISCUSSION}\label{sec:discussion}

\subsection{Dust Extinction: Interplay of Stellar Mass, Star Formation, and Environment}\label{discussion:dust extionction}
One of our primary interests is to investigate the dust extinction within CL1604. As a key component of the ISM in galaxies, interstellar dust plays a crucial role in star-forming regions by shielding gas from UV radiation. This shielding allows the gas to cool and form molecules, ensuring that molecular clouds remain cold and dense to support ongoing star formation \citep{Calzetti2000, Dwek11}. Therefore, a precise measurement of dust extinction is not only important for recovering the intrinsic properties of galaxies before they are obscured, but the dust extinction itself is an important property to study, as its presence is directly connected to the mode of star formation.

Our analysis reveals that dust extinction, as measured by the Balmer decrement, increases with stellar mass in star-forming galaxies. This is consistent with previous studies that used Balmer or Paschen line ratios to assess dust extinction \citep[e.g.,][]{Garn10, Zahid13, Reddy15, Fudamoto20, shapley23, Liu24, Jose24}. This trend is expected, as dust is produced by mass loss from asymptotic giant branch (AGB) stars and supernova ejecta \citep[e.g.,][]{Nozawa03, Gall11, Boyer12}. In more massive galaxies, shorter timescales of star formation and deeper gravitational potentials allow them to proceed further in chemical enrichment \citep[e.g.,][]{Issa90, Matthieu15, Suzuki21}. As a result, more massive galaxies tend to be more metal-rich, providing a larger reservoir of elements necessary for dust formation.

Based on our analysis of the correlation between dust extinction and environment, we do not find a clear environmental dependence of galaxy dust extinction, whether we define environment by cluster-centric distance (Figure \ref{fig:clusters}) or by a KS-test approach (Figure \ref{fig:ks cl1604}). This result is similar to what has been observed in the Spiderweb protocluster at $z=2.16$, as reported by \citet{Jose24}, who derived dust extinction using Pa$\beta$ and \ha\ observed with JWST and Subaru narrow-band filters. Both \citet{Jose24} and this work seem to indicate that the mode of star formation does not dramatically change across different environments.

The apparent absence of a dust-extinction–environment correlation in the full sample could be explained by the inclusion of quenching galaxies, which have intrinsically lower dust content and may wash out any underlying trend. When we focus on normal star-forming galaxies (i.e., excluding sub-main-sequence galaxies), we see a marginal environmental dependence of dust extinction in cluster A. As shown in the left middle panel of Figure \ref{fig:clusters}, star-forming galaxies located within the central 1 Mpc exhibit higher dust extinction than those in the outskirts. However, this correlation is not statistically significant and is largely driven by the higher stellar masses of galaxies in the cluster center. If present, such a trend could also be linked to gas accretion toward the cluster core, triggered by interactions or mergers, as is commonly expected in overdense environments \citep[e.g.,][]{Sancisi08, Voort11, Hayashi18, Umehata19}. 

If more mergers and interactions were prevalent in overdense regions, driving intense starbursts and facilitating enhanced dust production, we would expect to see a stronger environmental signature. Such signatures are indeed observed in some higher-redshift protoclusters, where starbursting cores are reported --- often associated with submillimeter galaxies (SMGs) --- and are thought to form brightest cluster galaxies (BCGs) through rapid accretion and star formation \citep[e.g.,][]{Oteo18, Long20, Calvi21, Hill22, Zhou24}. We argue that the discrepancy may stem from differing evolutionary states and intracluster medium (ICM) conditions. In younger, non-virialized protoclusters, ample cold-gas accretion fuels vigorous star formation, accelerating mass build-up and dust production. However, as clusters mature, the ICM becomes hot and X-ray luminous, and star formation in their cores declines. At this stage, the cluster begins to grow predominantly through the accretion of galaxies in the outskirts, where there are still sufficient cold gas supplies for star formation.

Although central star formation may have largely ceased in these more evolved systems, past episodes of star formation and starbursts have already deposited some dust into massive galaxies. Consequently, even if the environment no longer actively enhances star formation and dust production at the core, the residual dust content from previous activity may persist. This scenario could explain why no significant environmental gradient in dust extinction is observed in our analysis. Indeed, as shown in Figure \ref{fig:clusters}, the central region of cluster A includes both star-forming and quiescent galaxies. While the presence of quiescent, low-dust systems suggests a maturing cluster core, their coexistence with dustier, star-forming galaxies dilutes any clear environmental trend. Thus, any potential environmental dependence of dust extinction may be suppressed by the interplay between dust content, stellar mass, and star formation activity.

There are also some unexpected observations in our study, namely that the observed \ha/\hb\ ratios are occasionally lower than the Case B predicted dust-free minimum of 2.79, implying (unphysically) negative extinction. While underlying Balmer absorption arising primarily from the atmospheres of A--type stars can introduce additional uncertainties in the measured line fluxes and reduce the apparent Balmer emission line strengths, particularly the weaker \hb\ line \citep{Groves12}, previous studies suggest this effect is modest (typical corrections of $\sim$3\% for \ha\ and up to $\sim$10\% for \hb) \citep[e.g.,][]{Reddy15}, and is small compared to other sources of uncertainty.

In particular, narrow-band imaging lacks the precision of high-resolution spectroscopy for deblending the \ha\ and \nii\ complex, and variations in the filter response across redshift introduce additional uncertainties, especially for galaxies without spectroscopic redshifts. As shown in Figure \ref{fig:filter_spec_z}, galaxies at $z<0.89$ experience significantly reduced sensitivity in the \ha\ narrow-band relative to the \hb\ narrow-band, leading to an underestimation of the \ha/\hb\ ratio and derived dust extinction -- for instance, at $z=0.885$, the \ha/\hb\ ratio may be underestimated by $\sim$40\%.

While Balmer absorption can slightly overestimate the derived \ha/\hb\ ratios and E(B–V) estimates, its impact is much smaller than that from line blending or redshift-dependent filter effects, which tend to underestimate the ratios. Given these limitations, we treat galaxies with observed \ha/\hb\ ratios below 2.79 as having negligible extinction ($A_V=0$) when applying dust corrections. We note, however, this approach may influence our interpretation of the mass and environmental dependence of dust attenuation.

\subsection{Suppressed Star Formation in Massive Galaxies at the Cluster Core}

Dedicated \ha\ mapping across the entire structure enables us to investigate the star formation activities within the CL1604 supercluster. By analyzing the \ha\ emission and correcting for dust extinction using the Balmer Decrement, we can precisely estimate the SFRs. As shown in Figure ~\ref{fig:ms cl1604}, the majority of our galaxies scatter around the star-forming main sequence, indicating that star formation is still actively ongoing within the structure.

Interestingly, we also observe the presence of inactive \ha\ emitters with red colors, as defined by their $r - J$ color, which predominantly appear at the massive end (i.e., $M_* > 10^{10.0}\,M_\odot$). A similar color transition from blue to red at higher stellar masses has been reported in previous studies of cluster galaxies at $z < 2$ \citep[e.g.,][]{Kodama01, Tanaka05, Cassata07, Koyama10, Cooke16, Lemaux19}. The majority of these red galaxies show a drop-off from the main sequence, suggesting potential suppressed star formation or an ongoing transition to quenching. Nevertheless, since all of them are detected with the \ha\ emission line, there is still ongoing star formation associated with detectable H{\footnotesize II} regions. 

The observation of suppressed star formation in massive, red \ha\ emitters is intriguing, although the exact mechanisms responsible for their lower SFRs compared to the star-forming main sequence remain unclear. Various processes have been proposed to drive galaxies from an active star-forming phase to a quiescent state, including mass-driven quenching and environmental quenching \citep[e.g.,][]{Peng10, Muzzin12, Kawinwanichakij17, Tomczak19, Chartab20, Mao22, jose25}. 

As galaxies attain significant mass, they can undergo internal processes that halt their star formation and drive them into quiescence. One such process is the development of a substantial bulge that stabilizes the galaxy and suppresses disk instabilities, preventing new star formation and leading to the formation of red elliptical galaxies \citep[e.g.,][]{Martig09, Bluck14, Bruce14, Morishita15, Dimauro22, Liu24b}. AGN feedback is another mechanism often cited in massive galaxies, where active galactic nuclei can heat the surrounding gas or expel it from the galaxy, reducing the availability of cold gas necessary for star formation \citep[e.g.,][]{Voit05, Gaspari11, Teyssier11, Donnari21, Kubo22}. Although AGN feedback is known to play an important role in driving massive galaxies onto the red sequence by suppressing star formation, most red galaxies with reduced star formation in our sample are not classified as AGNs, based on Chandra X-ray observations combined with optical spectroscopy from HST/ACS and Keck/DEIMOS \citep{Kocevski09}. The detected AGNs in these data have moderate luminosities, ranging from $L_X = 0.78$–$4.47 \times 10^{43}\ \mathrm{erg\ s^{-1}}$ (0.5–8 keV), but fainter and/ or more heavily obscured AGNs may remain undetected.



Galaxies in overdense environments undergo more complex quenching mechanisms beyond mass-driven quenching, known as environmental quenching. When galaxies interact with their surrounding environments, several physical processes are responsible for reducing star formation and driving galaxies toward quiescence. For instance, interactions between galaxies and the ICM can strip away or heat the galaxy's halo gas, halting star formation through the process known as starvation (or strangulation; \citealt{Larson80}). Additionally, gas stripping processes within clusters that directly remove the cold molecular gas reservoir from a galaxy's disk are commonly observed mechanisms that suppress galaxy formation. These processes include ram pressure stripping \citep{Gunn1972, Balogh2000} and viscous stripping \citep{Nulsen82}. Furthermore, in high-density environments, galaxies have a higher chance of interacting with other galaxies, which can induce harassment \citep{Moore96, Moore98, Bialas15}. Harassment introduces disk instabilities that can drive gas toward the central regions, potentially triggering a central starburst followed by quenching. 


The high stellar mass of red galaxies in our sample makes them susceptible to internal quenching mechanisms, such as bulge stabilization and morphological quenching. At the same time, their location within the cluster core subjects them to environmental effects including gas stripping by the ICM and gravitational interactions. These combined internal and external processes deplete the cold gas reservoirs needed for star formation, driving these galaxies toward the red sequence and causing them to fall below the main sequence. Such complexity in quenching processes within overdense environments has also been reported in other studies. For example, \citet{jose25} examined a protocluster at $z=2.16$ and discussed the roles of ICM growth, AGN activity and heating, gas overconsumption, and stripping --- highlighting how multiple mechanisms likely operate in tandem. These processes, together with internal factors such as bulge formation, may act at different evolutionary phases of both the cluster and its constituent galaxies. However, the dominant quenching driver at each stage remains unclear.

This evolutionary trend --- where more mature clusters host massive galaxies with older stellar populations and suppressed star formation --- is consistent with previous findings \citep[e.g.,][]{Hayashi10, Muzzin12, Maier19}. To unambiguously identify the primary quenching mechanisms, whether internal or environmental, future observations with high-resolution emission line mapping or kiloparsec-scale molecular gas studies will be critical.

\section{Summary}
\label{sec:summary}

This paper presents a novel methodology for investigating dust extinction and star formation activities in the  $z = 0.9$  supercluster CL1604 using pure narrow-band imaging with the Subaru Telescope. The unique capabilities of Subaru's filters enabled us to map the physical properties of emission-line galaxies across a large-scale structure. The key findings of this study are summarized as follows:
\begin{itemize}
    \item Our work represents the first attempt to measure the \ha/\hb\ ratio, or Balmer Decrement, solely from narrow-band imaging. By employing two distinct narrow-band filters available only on the Subaru Telescope, we surveyed the CL1604 supercluster over a degree-scale area. Under the Case B recombination assumption, we derived dust extinction, measured dust-corrected SFRs and estimated stellar masses through SED fitting. This comprehensive approach enabled the characterization of 94 emission-line galaxies within the structure.
    \item The majority of emitters lie near the star-forming main sequence at this redshift. However, galaxies at the massive end ($M_{*} > 10^{10.0}\,M_\odot$) deviate from this relation. By examining their colors and spatial distribution, we find that these massive galaxies are redder and preferentially located closer to the cluster center. This finding supports an inside-out growth and quenching scenario in overdense regions, where galaxies rapidly assemble their mass at earlier times and experience quenching through a combination of internal processes and environmental effects earlier than their field counterparts.
    \item Most of the galaxies in our sample exhibit dust extinction in the range $0 < A_{\mathrm{H}\alpha} < 3$ mag, which is typical for galaxies at this redshift. However, a subset of our sample shows \ha/\hb\ ratios lower than the Case B prediction, implying negative extinction values. These deviations may be partly attributed to observational uncertainties, including limited data depth, variations in narrow-band filter transmission, and [NII] contamination. Additionally, physical effects such as variations in electron density and temperature in the ionized regions could also influence the observed \ha/\hb\ ratios.
    
    \item Dust extinction shows a positive correlation with stellar mass, and star-forming galaxies exhibit higher extinction values than quiescent ones. We find no clear environmental dependence of galaxy dustiness. When sub-main-sequence galaxies are excluded (those lying more than 0.3 dex below the star-forming main sequence), however, there is a weak trend in cluster A: galaxies within a 1 Mpc radius show higher dust extinction than those in the outskirts. Because dust extinction is also related to stellar mass and star formation mode, any potential environmental dependence may be masked by the interplay of these factors. Moreover, given the current observational uncertainties and limited sample size, we cannot yet draw definitive conclusions.
\end{itemize}

Our results underscore the importance of accurately correcting for dust extinction when studying star formation and early galaxy properties within distant large-scale structures, such as galaxy clusters. Wide-field facilities are essential for these studies, and the Subaru Telescope's large field of view combined with its unique set of colorful narrow-band filters provides an exceptional opportunity to observe and analyze these distant galaxies effectively. This study demonstrates the effectiveness of using paired narrow-band filters to efficiently study galaxy evolution at high redshifts. Looking ahead, the upcoming PFS onboard Subaru will uniquely enable such studies by offering unprecedented spectroscopic capabilities over wide fields, facilitating more detailed investigations into galaxy formation and evolution in the distant universe.

We would like to express our gratitude to the anonymous referee and editor for their insightful comments that enhanced the manuscript. This paper is based on data collected at the Subaru Telescope and retrieved from the HSC data archive system, which is operated by the Subaru Telescope and Astronomy Data Center (ADC) at National Astronomical Observatory of Japan. Data analysis was in part carried out with the cooperation of Center for Computational Astrophysics (CfCA), National Astronomical Observatory of Japan. The Subaru Telescope is honored and grateful for the opportunity of observing the Universe from Maunakea, which has the cultural, historical and natural significance in Hawaii. Operation of SWIMS at Subaru telescope is supported by MEXT/JSPS KAKENHI grant 20H00171. The Hyper Suprime-Cam (HSC) collaboration includes the astronomical communities of Japan and Taiwan, and Princeton University. The HSC instrumentation and software were developed by the National Astronomical Observatory of Japan (NAOJ), the Kavli Institute for the Physics and Mathematics of the Universe (Kavli IPMU), the University of Tokyo, the High Energy Accelerator Research Organization (KEK), the Academia Sinica Institute for Astronomy and Astrophysics in Taiwan (ASIAA), and Princeton University. Funding was contributed by the FIRST program from the Japanese Cabinet Office, the Ministry of Education, Culture, Sports, Science and Technology (MEXT), the Japan Society for the Promotion of Science (JSPS), Japan Science and Technology Agency (JST), the Toray Science Foundation, NAOJ, Kavli IPMU, KEK, ASIAA, and Princeton University. The \textit{NB921} filter was supported by KAKENHI (23244025) Grant-in-Aid for Scientific Research (A) through the Japan Society for the Promotion of Science (JSPS). This paper makes use of software developed for the Large Synoptic Survey Telescope. We thank the LSST Project for making their code available as free software at  http://dm.lsst.org. Data analysis was carried out on the Multi-wavelength Data Analysis System operated by the Astronomy Data Center (ADC), National Astronomical Observatory of Japan. ZL acknowledges support from JSPS KAKENHI Grant Number 24KJ0394. This work is supported by JSPS KAKENHI Grant Numbers 24H00002 (Specially Promoted Research by T. Kodama et al.). BCL acknowledges support by the international Gemini Observatory, a program of NSF NOIRLab, which is managed by the Association of Universities for Research in Astronomy (AURA) under a cooperative agreement with the U.S. National Science Foundation, on behalf of the Gemini partnership of Argentina, Brazil, Canada, Chile, the Republic of Korea, and the United States of America. JMPM acknowledges support from the Agencia Estatal de Investigación del Ministerio de Ciencia, Innovación y Universidades (MCIU/AEI) under grant (Construcción de cúmulos de galaxias en formación a través de la formación estelar oscurecida por el polvo) and the European Regional Development Fund (ERDF) with reference (PID2022-143243NB-I00/10.13039/501100011033), as well as funding from the European Unions Horizon-Europe research and innovation program under the Marie Sklodowska-Curie grant agreement No. 101106626.

\appendix
\label{appendix}

\section{Definition of Quiescent Galaxies and Sub-main-sequence Galaxies}

In this paper, we define two types of galaxy populations with reduced star formation activity: quiescent galaxies, identified using the rest-frame $UVJ$ diagram, and sub-main-sequence galaxies, identified based on their SFRs. This is due to the non-uniform IRAC coverage, which limits our ability to reliably constrain the rest-frame $J$-band color across the entire sample.

\begin{figure}
\centering
\includegraphics[width=0.5\textwidth]{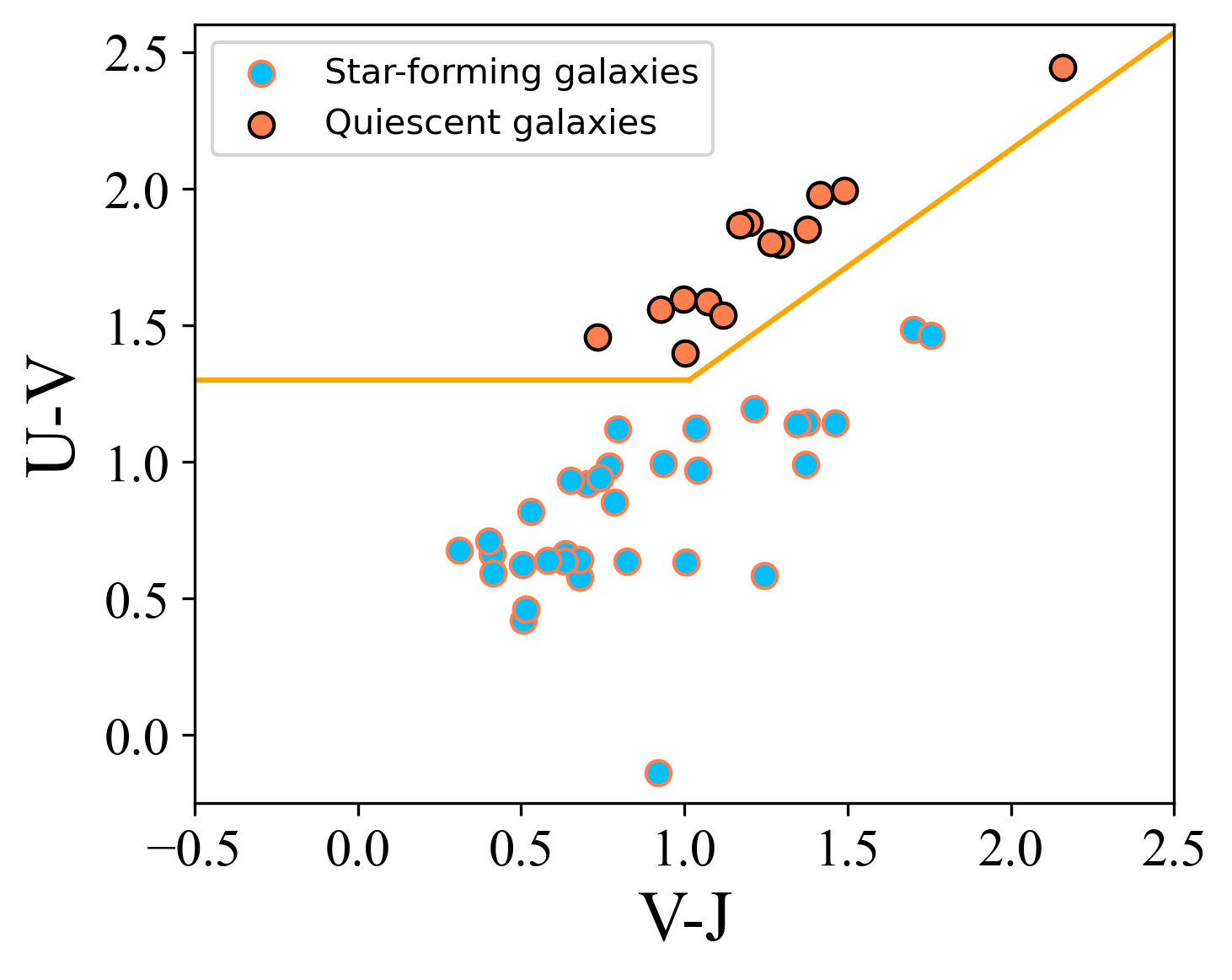} 
\caption{Rest-frame $UVJ$ diagram used to classify galaxy populations, colors are derived from SED fitting including Subaru and IRAC observations as described in Section \ref{sec:photometry}. The orange solid lines represent the quiescent galaxy boundaries defined in \citet{Mao22}. Blue filled circles represent star-forming galaxies, and red filled circles represent quiescent galaxies identified by the $UVJ$ diagram.}
\label{fig:uvj}
\end{figure}

For the galaxies with IRAC coverage, we classify quiescent galaxies based on their location in the rest-frame $UVJ$ color-color diagram, following the quiescence criteria defined in \citet{Mao22}. Figure~\ref{fig:uvj} shows the $UVJ$-diagram for our sample. However, many galaxies in our sample lack IRAC coverage and therefore do not have reliable rest-frame $J$-band colors. To enable the discussion of dust extinction among star-forming galaxies, an additional classification scheme is needed. We thus define another population, namely sub-main-sequence galaxies, based on their SFRs. sub-main-sequence galaxies are defined as those lying more than 0.3 dex below the star-forming main sequence. It is important to ensure the robustness of our SFR measurements, particularly given the potential uncertainties introduced by redshift-dependent dust extinction effects and the underestimation of \nii\ contributions, as discussed in Section~\ref{result:sf activities}. These uncertainties could lead to underestimated dust-corrected SFRs, potentially causing star-forming galaxies to be misclassified as sub-main-sequence galaxies. Nevertheless, we find that the majority of sub-main-sequence galaxies with IRAC coverage, for which rest-frame $UVJ$ colors can be reliably characterized, are classified as $UVJ$-quiescent galaxies (see Figure~\ref{fig:ms cl1604}). This consistency supports the robustness of our selection of sub-main-sequence galaxies based on SFRs.

\ref{fig:ms cl1604}



\software{\texttt{Astropy} \citep{Astropy13,Astropy18},
          \texttt{CIGALE} \citep{Boquien19}, \texttt{SExtractor} \citep{bertin1996}}

\bibliography{sample631}{}

\begin{thebibliography}{}
\expandafter\ifx\csname natexlab\endcsname\relax\def\natexlab#1{#1}\fi
\providecommand{\url}[1]{\href{#1}{#1}}
\providecommand{\dodoi}[1]{doi:~\href{http://doi.org/#1}{\nolinkurl{#1}}}
\providecommand{\doeprint}[1]{\href{http://ascl.net/#1}{\nolinkurl{http://ascl.net/#1}}}
\providecommand{\doarXiv}[1]{\href{https://arxiv.org/abs/#1}{\nolinkurl{https://arxiv.org/abs/#1}}}

\bibitem[{{Abadi} {et~al.}(1999){Abadi}, {Moore}, \& {Bower}}]{Abadi99}
{Abadi}, M.~G., {Moore}, B., \& {Bower}, R.~G. 1999, \mnras, 308, 947, \dodoi{10.1046/j.1365-8711.1999.02715.x}

\bibitem[{{Aihara} {et~al.}(2018){Aihara}, {Arimoto}, {Armstrong}, {Arnouts}, {Bahcall}, {Bickerton}, {Bosch}, {Bundy}, {Capak}, {Chan}, {Chiba}, {Coupon}, {Egami}, {Enoki}, {Finet}, {Fujimori}, {Fujimoto}, {Furusawa}, {Furusawa}, {Goto}, {Goulding}, {Greco}, {Greene}, {Gunn}, {Hamana}, {Harikane}, {Hashimoto}, {Hattori}, {Hayashi}, {Hayashi}, {He{\l}miniak}, {Higuchi}, {Hikage}, {Ho}, {Hsieh}, {Huang}, {Huang}, {Ikeda}, {Imanishi}, {Inoue}, {Iwasawa}, {Iwata}, {Jaelani}, {Jian}, {Kamata}, {Karoji}, {Kashikawa}, {Katayama}, {Kawanomoto}, {Kayo}, {Koda}, {Koike}, {Kojima}, {Komiyama}, {Konno}, {Koshida}, {Koyama}, {Kusakabe}, {Leauthaud}, {Lee}, {Lin}, {Lin}, {Lupton}, {Mandelbaum}, {Matsuoka}, {Medezinski}, {Mineo}, {Miyama}, {Miyatake}, {Miyazaki}, {Momose}, {More}, {More}, {Moritani}, {Moriya}, {Morokuma}, {Mukae}, {Murata}, {Murayama}, {Nagao}, {Nakata}, {Niida}, {Niikura}, {Nishizawa}, {Obuchi}, {Oguri}, {Oishi}, {Okabe}, {Okamoto}, {Okura}, {Ono}, {Onodera}, {Onoue}, {Osato}, {Ouchi}, {Price}, {Pyo},
  {Sako}, {Sawicki}, {Shibuya}, {Shimasaku}, {Shimono}, {Shirasaki}, {Silverman}, {Simet}, {Speagle}, {Spergel}, {Strauss}, {Sugahara}, {Sugiyama}, {Suto}, {Suyu}, {Suzuki}, {Tait}, {Takada}, {Takata}, {Tamura}, {Tanaka}, {Tanaka}, {Tanaka}, {Tanaka}, {Terai}, {Terashima}, {Toba}, {Tominaga}, {Toshikawa}, {Turner}, {Uchida}, {Uchiyama}, {Umetsu}, {Uraguchi}, {Urata}, {Usuda}, {Utsumi}, {Wang}, {Wang}, {Wong}, {Yabe}, {Yamada}, {Yamanoi}, {Yasuda}, {Yeh}, {Yonehara}, \& {Yuma}}]{Aihara18}
{Aihara}, H., {Arimoto}, N., {Armstrong}, R., {et~al.} 2018, \pasj, 70, S4, \dodoi{10.1093/pasj/psx066}

\bibitem[{{Aihara} {et~al.}(2019){Aihara}, {AlSayyad}, {Ando}, {Armstrong}, {Bosch}, {Egami}, {Furusawa}, {Furusawa}, {Goulding}, {Harikane}, {Hikage}, {Ho}, {Hsieh}, {Huang}, {Ikeda}, {Imanishi}, {Ito}, {Iwata}, {Jaelani}, {Kakuma}, {Kawana}, {Kikuta}, {Kobayashi}, {Koike}, {Komiyama}, {Li}, {Liang}, {Lin}, {Luo}, {Lupton}, {Lust}, {MacArthur}, {Matsuoka}, {Mineo}, {Miyatake}, {Miyazaki}, {More}, {Murata}, {Namiki}, {Nishizawa}, {Oguri}, {Okabe}, {Okamoto}, {Okura}, {Ono}, {Onodera}, {Onoue}, {Osato}, {Ouchi}, {Shibuya}, {Strauss}, {Sugiyama}, {Suto}, {Takada}, {Takagi}, {Takata}, {Takita}, {Tanaka}, {Terai}, {Toba}, {Uchiyama}, {Utsumi}, {Wang}, {Wang}, \& {Yamada}}]{Aihara19}
{Aihara}, H., {AlSayyad}, Y., {Ando}, M., {et~al.} 2019, \pasj, 71, 114, \dodoi{10.1093/pasj/psz103}

\bibitem[{{Asano} {et~al.}(2020){Asano}, {Kodama}, {Motohara}, {Lubin}, {Lemaux}, {Gal}, {Tomczak}, {Kocevski}, {Hayashi}, {Koyama}, {Tanaka}, {Suzuki}, {Yamamoto}, {Kimura}, {Konishi}, {Takahashi}, {Terao}, {Kushibiki}, {Kono}, {Yoshii}, \& {Swims Team}}]{Asano20}
{Asano}, T., {Kodama}, T., {Motohara}, K., {et~al.} 2020, \apj, 899, 64, \dodoi{10.3847/1538-4357/ab9dfb}

\bibitem[{{Astropy Collaboration} {et~al.}(2013){Astropy Collaboration}, {Robitaille}, {Tollerud}, {Greenfield}, {Droettboom}, {Bray}, {Aldcroft}, {Davis}, {Ginsburg}, {Price-Whelan}, {Kerzendorf}, {Conley}, {Crighton}, {Barbary}, {Muna}, {Ferguson}, {Grollier}, {Parikh}, {Nair}, {Unther}, {Deil}, {Woillez}, {Conseil}, {Kramer}, {Turner}, {Singer}, {Fox}, {Weaver}, {Zabalza}, {Edwards}, {Azalee Bostroem}, {Burke}, {Casey}, {Crawford}, {Dencheva}, {Ely}, {Jenness}, {Labrie}, {Lim}, {Pierfederici}, {Pontzen}, {Ptak}, {Refsdal}, {Servillat}, \& {Streicher}}]{Astropy13}
{Astropy Collaboration}, {Robitaille}, T.~P., {Tollerud}, E.~J., {et~al.} 2013, \aap, 558, A33, \dodoi{10.1051/0004-6361/201322068}

\bibitem[{{Astropy Collaboration} {et~al.}(2018){Astropy Collaboration}, {Price-Whelan}, {Sip{\H{o}}cz}, {G{\"u}nther}, {Lim}, {Crawford}, {Conseil}, {Shupe}, {Craig}, {Dencheva}, {Ginsburg}, {VanderPlas}, {Bradley}, {P{\'e}rez-Su{\'a}rez}, {de Val-Borro}, {Aldcroft}, {Cruz}, {Robitaille}, {Tollerud}, {Ardelean}, {Babej}, {Bach}, {Bachetti}, {Bakanov}, {Bamford}, {Barentsen}, {Barmby}, {Baumbach}, {Berry}, {Biscani}, {Boquien}, {Bostroem}, {Bouma}, {Brammer}, {Bray}, {Breytenbach}, {Buddelmeijer}, {Burke}, {Calderone}, {Cano Rodr{\'\i}guez}, {Cara}, {Cardoso}, {Cheedella}, {Copin}, {Corrales}, {Crichton}, {D'Avella}, {Deil}, {Depagne}, {Dietrich}, {Donath}, {Droettboom}, {Earl}, {Erben}, {Fabbro}, {Ferreira}, {Finethy}, {Fox}, {Garrison}, {Gibbons}, {Goldstein}, {Gommers}, {Greco}, {Greenfield}, {Groener}, {Grollier}, {Hagen}, {Hirst}, {Homeier}, {Horton}, {Hosseinzadeh}, {Hu}, {Hunkeler}, {Ivezi{\'c}}, {Jain}, {Jenness}, {Kanarek}, {Kendrew}, {Kern}, {Kerzendorf}, {Khvalko}, {King}, {Kirkby}, {Kulkarni},
  {Kumar}, {Lee}, {Lenz}, {Littlefair}, {Ma}, {Macleod}, {Mastropietro}, {McCully}, {Montagnac}, {Morris}, {Mueller}, {Mumford}, {Muna}, {Murphy}, {Nelson}, {Nguyen}, {Ninan}, {N{\"o}the}, {Ogaz}, {Oh}, {Parejko}, {Parley}, {Pascual}, {Patil}, {Patil}, {Plunkett}, {Prochaska}, {Rastogi}, {Reddy Janga}, {Sabater}, {Sakurikar}, {Seifert}, {Sherbert}, {Sherwood-Taylor}, {Shih}, {Sick}, {Silbiger}, {Singanamalla}, {Singer}, {Sladen}, {Sooley}, {Sornarajah}, {Streicher}, {Teuben}, {Thomas}, {Tremblay}, {Turner}, {Terr{\'o}n}, {van Kerkwijk}, {de la Vega}, {Watkins}, {Weaver}, {Whitmore}, {Woillez}, {Zabalza}, \& {Astropy Contributors}}]{Astropy18}
{Astropy Collaboration}, {Price-Whelan}, A.~M., {Sip{\H{o}}cz}, B.~M., {et~al.} 2018, \aj, 156, 123, \dodoi{10.3847/1538-3881/aabc4f}

\bibitem[{{Balogh} {et~al.}(2004){Balogh}, {Baldry}, {Nichol}, {Miller}, {Bower}, \& {Glazebrook}}]{Balogh04}
{Balogh}, M.~L., {Baldry}, I.~K., {Nichol}, R., {et~al.} 2004, \apjl, 615, L101, \dodoi{10.1086/426079}

\bibitem[{{Balogh} {et~al.}(2000){Balogh}, {Navarro}, \& {Morris}}]{Balogh2000}
{Balogh}, M.~L., {Navarro}, J.~F., \& {Morris}, S.~L. 2000, \apj, 540, 113, \dodoi{10.1086/309323}

\bibitem[{{Balogh} {et~al.}(2017){Balogh}, {Gilbank}, {Muzzin}, {Rudnick}, {Cooper}, {Lidman}, {Biviano}, {Demarco}, {McGee}, {Nantais}, {Noble}, {Old}, {Wilson}, {Yee}, {Bellhouse}, {Cerulo}, {Chan}, {Pintos-Castro}, {Simpson}, {van der Burg}, {Zaritsky}, {Ziparo}, {Alonso}, {Bower}, {De Lucia}, {Finoguenov}, {Lambas}, {Muriel}, {Parker}, {Rettura}, {Valotto}, \& {Wetzel}}]{Balogh17}
{Balogh}, M.~L., {Gilbank}, D.~G., {Muzzin}, A., {et~al.} 2017, \mnras, 470, 4168, \dodoi{10.1093/mnras/stx1370}

\bibitem[{{Bell} {et~al.}(2004){Bell}, {Wolf}, {Meisenheimer}, {Rix}, {Borch}, {Dye}, {Kleinheinrich}, {Wisotzki}, \& {McIntosh}}]{Bell04}
{Bell}, E.~F., {Wolf}, C., {Meisenheimer}, K., {et~al.} 2004, \apj, 608, 752, \dodoi{10.1086/420778}

\bibitem[{{Berta} {et~al.}(2016){Berta}, {Lutz}, {Genzel}, {F{\"o}rster-Schreiber}, \& {Tacconi}}]{Berta16}
{Berta}, S., {Lutz}, D., {Genzel}, R., {F{\"o}rster-Schreiber}, N.~M., \& {Tacconi}, L.~J. 2016, \aap, 587, A73, \dodoi{10.1051/0004-6361/201527746}

\bibitem[{{Bertin} \& {Arnouts}(1996)}]{bertin1996}
{Bertin}, E., \& {Arnouts}, S. 1996, \aaps, 117, 393, \dodoi{10.1051/aas:1996164}

\bibitem[{{B{\'e}thermin} {et~al.}(2015){B{\'e}thermin}, {Daddi}, {Magdis}, {Lagos}, {Sargent}, {Albrecht}, {Aussel}, {Bertoldi}, {Buat}, {Galametz}, {Heinis}, {Ilbert}, {Karim}, {Koekemoer}, {Lacey}, {Le Floc'h}, {Navarrete}, {Pannella}, {Schreiber}, {Smol{\v{c}}i{\'c}}, {Symeonidis}, \& {Viero}}]{Matthieu15}
{B{\'e}thermin}, M., {Daddi}, E., {Magdis}, G., {et~al.} 2015, \aap, 573, A113, \dodoi{10.1051/0004-6361/201425031}

\bibitem[{{Bialas} {et~al.}(2015){Bialas}, {Lisker}, {Olczak}, {Spurzem}, \& {Kotulla}}]{Bialas15}
{Bialas}, D., {Lisker}, T., {Olczak}, C., {Spurzem}, R., \& {Kotulla}, R. 2015, \aap, 576, A103, \dodoi{10.1051/0004-6361/201425235}

\bibitem[{{Bluck} {et~al.}(2014){Bluck}, {Mendel}, {Ellison}, {Moreno}, {Simard}, {Patton}, \& {Starkenburg}}]{Bluck14}
{Bluck}, A. F.~L., {Mendel}, J.~T., {Ellison}, S.~L., {et~al.} 2014, \mnras, 441, 599, \dodoi{10.1093/mnras/stu594}

\bibitem[{{Bond} {et~al.}(1996){Bond}, {Kofman}, \& {Pogosyan}}]{Bond96}
{Bond}, J.~R., {Kofman}, L., \& {Pogosyan}, D. 1996, \nat, 380, 603, \dodoi{10.1038/380603a0}

\bibitem[{{Boquien} {et~al.}(2019){Boquien}, {Burgarella}, {Roehlly}, {Buat}, {Ciesla}, {Corre}, {Inoue}, \& {Salas}}]{Boquien19}
{Boquien}, M., {Burgarella}, D., {Roehlly}, Y., {et~al.} 2019, \aap, 622, A103, \dodoi{10.1051/0004-6361/201834156}

\bibitem[{{Bosch} {et~al.}(2018){Bosch}, {Armstrong}, {Bickerton}, {Furusawa}, {Ikeda}, {Koike}, {Lupton}, {Mineo}, {Price}, {Takata}, {Tanaka}, {Yasuda}, {AlSayyad}, {Becker}, {Coulton}, {Coupon}, {Garmilla}, {Huang}, {Krughoff}, {Lang}, {Leauthaud}, {Lim}, {Lust}, {MacArthur}, {Mandelbaum}, {Miyatake}, {Miyazaki}, {Murata}, {More}, {Okura}, {Owen}, {Swinbank}, {Strauss}, {Yamada}, \& {Yamanoi}}]{Bosch18}
{Bosch}, J., {Armstrong}, R., {Bickerton}, S., {et~al.} 2018, \pasj, 70, S5, \dodoi{10.1093/pasj/psx080}

\bibitem[{{Boselli} \& {Gavazzi}(2006)}]{Boselli06}
{Boselli}, A., \& {Gavazzi}, G. 2006, \pasp, 118, 517, \dodoi{10.1086/500691}

\bibitem[{{Bower} {et~al.}(1992){Bower}, {Lucey}, \& {Ellis}}]{Bower92}
{Bower}, R.~G., {Lucey}, J.~R., \& {Ellis}, R.~S. 1992, \mnras, 254, 601, \dodoi{10.1093/mnras/254.4.601}

\bibitem[{{Boyer} {et~al.}(2012){Boyer}, {Srinivasan}, {Riebel}, {McDonald}, {van Loon}, {Clayton}, {Gordon}, {Meixner}, {Sargent}, \& {Sloan}}]{Boyer12}
{Boyer}, M.~L., {Srinivasan}, S., {Riebel}, D., {et~al.} 2012, \apj, 748, 40, \dodoi{10.1088/0004-637X/748/1/40}

\bibitem[{{Bruce} {et~al.}(2014){Bruce}, {Dunlop}, {McLure}, {Cirasuolo}, {Buitrago}, {Bowler}, {Targett}, {Bell}, {McIntosh}, {Dekel}, {Faber}, {Ferguson}, {Grogin}, {Hartley}, {Kocevski}, {Koekemoer}, {Koo}, \& {McGrath}}]{Bruce14}
{Bruce}, V.~A., {Dunlop}, J.~S., {McLure}, R.~J., {et~al.} 2014, \mnras, 444, 1001, \dodoi{10.1093/mnras/stu1478}

\bibitem[{{Bruzual} \& {Charlot}(2003)}]{Bruzual03}
{Bruzual}, G., \& {Charlot}, S. 2003, \mnras, 344, 1000, \dodoi{10.1046/j.1365-8711.2003.06897.x}

\bibitem[{{Buat} \& {Xu}(1996)}]{Buat96}
{Buat}, V., \& {Xu}, C. 1996, \aap, 306, 61

\bibitem[{{Buat} {et~al.}(2012){Buat}, {Noll}, {Burgarella}, {Giovannoli}, {Charmandaris}, {Pannella}, {Hwang}, {Elbaz}, {Dickinson}, {Magdis}, {Reddy}, \& {Murphy}}]{Buat12}
{Buat}, V., {Noll}, S., {Burgarella}, D., {et~al.} 2012, \aap, 545, A141, \dodoi{10.1051/0004-6361/201219405}

\bibitem[{{Bunker} {et~al.}(1995){Bunker}, {Warren}, {Hewett}, \& {Clements}}]{Bunker95}
{Bunker}, A.~J., {Warren}, S.~J., {Hewett}, P.~C., \& {Clements}, D.~L. 1995, \mnras, 273, 513, \dodoi{10.1093/mnras/273.2.513}

\bibitem[{{Butcher} \& {Oemler}(1978)}]{Butcher78}
{Butcher}, H., \& {Oemler}, Jr., A. 1978, \apj, 219, 18, \dodoi{10.1086/155751}

\bibitem[{{Butcher} \& {Oemler}(1984)}]{Butcher84}
---. 1984, \apj, 285, 426, \dodoi{10.1086/162519}

\bibitem[{{Calvi} {et~al.}(2021){Calvi}, {Dannerbauer}, {Arrabal Haro}, {Rodr{\'\i}guez Espinosa}, {Mu{\~n}oz-Tu{\~n}{\'o}n}, {P{\'e}rez Gonz{\'a}lez}, \& {Geier}}]{Calvi21}
{Calvi}, R., {Dannerbauer}, H., {Arrabal Haro}, P., {et~al.} 2021, \mnras, 502, 4558, \dodoi{10.1093/mnras/staa4037}

\bibitem[{{Calzetti} {et~al.}(2000){Calzetti}, {Armus}, {Bohlin}, {Kinney}, {Koornneef}, \& {Storchi-Bergmann}}]{Calzetti2000}
{Calzetti}, D., {Armus}, L., {Bohlin}, R.~C., {et~al.} 2000, \apj, 533, 682, \dodoi{10.1086/308692}

\bibitem[{{Cardelli} {et~al.}(1989){Cardelli}, {Clayton}, \& {.Mathis}}]{Cardelli89}
{Cardelli}, J.~A., {Clayton}, G.~C., \& {.Mathis}, J.~S. 1989, \apj, 345, 245, \dodoi{10.1086/167900}

\bibitem[{{Cassata} {et~al.}(2007){Cassata}, {Guzzo}, {Franceschini}, {Scoville}, {Capak}, {Ellis}, {Koekemoer}, {McCracken}, {Mobasher}, {Renzini}, {Ricciardelli}, {Scodeggio}, {Taniguchi}, \& {Thompson}}]{Cassata07}
{Cassata}, P., {Guzzo}, L., {Franceschini}, A., {et~al.} 2007, \apjs, 172, 270, \dodoi{10.1086/516591}

\bibitem[{{Chabrier}(2003)}]{Chabrier03}
{Chabrier}, G. 2003, \pasp, 115, 763, \dodoi{10.1086/376392}

\bibitem[{{Chambers} {et~al.}(2016){Chambers}, {Magnier}, {Metcalfe}, {Flewelling}, {Huber}, {Waters}, {Denneau}, {Draper}, {Farrow}, {Finkbeiner}, {Holmberg}, {Koppenhoefer}, {Price}, {Rest}, {Saglia}, {Schlafly}, {Smartt}, {Sweeney}, {Wainscoat}, {Burgett}, {Chastel}, {Grav}, {Heasley}, {Hodapp}, {Jedicke}, {Kaiser}, {Kudritzki}, {Luppino}, {Lupton}, {Monet}, {Morgan}, {Onaka}, {Shiao}, {Stubbs}, {Tonry}, {White}, {Ba{\~n}ados}, {Bell}, {Bender}, {Bernard}, {Boegner}, {Boffi}, {Botticella}, {Calamida}, {Casertano}, {Chen}, {Chen}, {Cole}, {Deacon}, {Frenk}, {Fitzsimmons}, {Gezari}, {Gibbs}, {Goessl}, {Goggia}, {Gourgue}, {Goldman}, {Grant}, {Grebel}, {Hambly}, {Hasinger}, {Heavens}, {Heckman}, {Henderson}, {Henning}, {Holman}, {Hopp}, {Ip}, {Isani}, {Jackson}, {Keyes}, {Koekemoer}, {Kotak}, {Le}, {Liska}, {Long}, {Lucey}, {Liu}, {Martin}, {Masci}, {McLean}, {Mindel}, {Misra}, {Morganson}, {Murphy}, {Obaika}, {Narayan}, {Nieto-Santisteban}, {Norberg}, {Peacock}, {Pier}, {Postman}, {Primak}, {Rae}, {Rai},
  {Riess}, {Riffeser}, {Rix}, {R{\"o}ser}, {Russel}, {Rutz}, {Schilbach}, {Schultz}, {Scolnic}, {Strolger}, {Szalay}, {Seitz}, {Small}, {Smith}, {Soderblom}, {Taylor}, {Thomson}, {Taylor}, {Thakar}, {Thiel}, {Thilker}, {Unger}, {Urata}, {Valenti}, {Wagner}, {Walder}, {Walter}, {Watters}, {Werner}, {Wood-Vasey}, \& {Wyse}}]{Chambers16}
{Chambers}, K.~C., {Magnier}, E.~A., {Metcalfe}, N., {et~al.} 2016, arXiv e-prints, arXiv:1612.05560, \dodoi{10.48550/arXiv.1612.05560}

\bibitem[{{Chartab} {et~al.}(2020){Chartab}, {Mobasher}, {Darvish}, {Finkelstein}, {Guo}, {Kodra}, {Lee}, {Newman}, {Pacifici}, {Papovich}, {Sattari}, {Shahidi}, {Dickinson}, {Faber}, {Ferguson}, {Giavalisco}, \& {Jafariyazani}}]{Chartab20}
{Chartab}, N., {Mobasher}, B., {Darvish}, B., {et~al.} 2020, \apj, 890, 7, \dodoi{10.3847/1538-4357/ab61fd}

\bibitem[{{Chiang} {et~al.}(2017){Chiang}, {Overzier}, {Gebhardt}, \& {Henriques}}]{Chiang17}
{Chiang}, Y.-K., {Overzier}, R.~A., {Gebhardt}, K., \& {Henriques}, B. 2017, \apjl, 844, L23, \dodoi{10.3847/2041-8213/aa7e7b}

\bibitem[{{Cooke} {et~al.}(2016){Cooke}, {Hatch}, {Stern}, {Rettura}, {Brodwin}, {Galametz}, {Wylezalek}, {Bridge}, {Conselice}, {De Breuck}, {Gonzalez}, \& {Jarvis}}]{Cooke16}
{Cooke}, E.~A., {Hatch}, N.~A., {Stern}, D., {et~al.} 2016, \apj, 816, 83, \dodoi{10.3847/0004-637X/816/2/83}

\bibitem[{{Crook} {et~al.}(2007){Crook}, {Huchra}, {Martimbeau}, {Masters}, {Jarrett}, \& {Macri}}]{Crook07}
{Crook}, A.~C., {Huchra}, J.~P., {Martimbeau}, N., {et~al.} 2007, \apj, 655, 790, \dodoi{10.1086/510201}

\bibitem[{{da Cunha} {et~al.}(2015){da Cunha}, {Walter}, {Smail}, {Swinbank}, {Simpson}, {Decarli}, {Hodge}, {Weiss}, {van der Werf}, {Bertoldi}, {Chapman}, {Cox}, {Danielson}, {Dannerbauer}, {Greve}, {Ivison}, {Karim}, \& {Thomson}}]{Cunha15}
{da Cunha}, E., {Walter}, F., {Smail}, I.~R., {et~al.} 2015, \apj, 806, 110, \dodoi{10.1088/0004-637X/806/1/110}

\bibitem[{{Dannerbauer} {et~al.}(2014){Dannerbauer}, {Kurk}, {De Breuck}, {Wylezalek}, {Santos}, {Koyama}, {Seymour}, {Tanaka}, {Hatch}, {Altieri}, {Coia}, {Galametz}, {Kodama}, {Miley}, {R{\"o}ttgering}, {Sanchez-Portal}, {Valtchanov}, {Venemans}, \& {Ziegler}}]{Dannerbauer14}
{Dannerbauer}, H., {Kurk}, J.~D., {De Breuck}, C., {et~al.} 2014, \aap, 570, A55, \dodoi{10.1051/0004-6361/201423771}

\bibitem[{{Darvish} {et~al.}(2015){Darvish}, {Mobasher}, {Sobral}, {Scoville}, \& {Aragon-Calvo}}]{Darvish15}
{Darvish}, B., {Mobasher}, B., {Sobral}, D., {Scoville}, N., \& {Aragon-Calvo}, M. 2015, \apj, 805, 121, \dodoi{10.1088/0004-637X/805/2/121}

\bibitem[{{Dimauro} {et~al.}(2022){Dimauro}, {Daddi}, {Shankar}, {Cattaneo}, {Huertas-Company}, {Bernardi}, {Caro}, {Dupke}, {H{\"a}u{\ss}ler}, {Johnston}, {Cortesi}, {Mei}, \& {Peletier}}]{Dimauro22}
{Dimauro}, P., {Daddi}, E., {Shankar}, F., {et~al.} 2022, \mnras, 513, 256, \dodoi{10.1093/mnras/stac884}

\bibitem[{{Dom{\'\i}nguez} {et~al.}(2013){Dom{\'\i}nguez}, {Siana}, {Henry}, {Scarlata}, {Bedregal}, {Malkan}, {Atek}, {Ross}, {Colbert}, {Teplitz}, {Rafelski}, {McCarthy}, {Bunker}, {Hathi}, {Dressler}, {Martin}, \& {Masters}}]{2013ApJ...763..145D}
{Dom{\'\i}nguez}, A., {Siana}, B., {Henry}, A.~L., {et~al.} 2013, \apj, 763, 145, \dodoi{10.1088/0004-637X/763/2/145}

\bibitem[{{Dong} {et~al.}(2008){Dong}, {Wang}, {Wang}, {Yuan}, {Zhou}, {Dai}, \& {Zhang}}]{Dong08}
{Dong}, X., {Wang}, T., {Wang}, J., {et~al.} 2008, \mnras, 383, 581, \dodoi{10.1111/j.1365-2966.2007.12560.x}

\bibitem[{{Donnari} {et~al.}(2021){Donnari}, {Pillepich}, {Joshi}, {Nelson}, {Genel}, {Marinacci}, {Rodriguez-Gomez}, {Pakmor}, {Torrey}, {Vogelsberger}, \& {Hernquist}}]{Donnari21}
{Donnari}, M., {Pillepich}, A., {Joshi}, G.~D., {et~al.} 2021, \mnras, 500, 4004, \dodoi{10.1093/mnras/staa3006}

\bibitem[{{Dressler}(1980)}]{Dressler80}
{Dressler}, A. 1980, \apj, 236, 351, \dodoi{10.1086/157753}

\bibitem[{{Dressler} {et~al.}(1997){Dressler}, {Oemler}, {Couch}, {Smail}, {Ellis}, {Barger}, {Butcher}, {Poggianti}, \& {Sharples}}]{Dressler97}
{Dressler}, A., {Oemler}, Augustus, J., {Couch}, W.~J., {et~al.} 1997, \apj, 490, 577, \dodoi{10.1086/304890}

\bibitem[{{Driver} {et~al.}(2013){Driver}, {Robotham}, {Bland-Hawthorn}, {Brown}, {Hopkins}, {Liske}, {Phillipps}, \& {Wilkins}}]{Driver13}
{Driver}, S.~P., {Robotham}, A.~S.~G., {Bland-Hawthorn}, J., {et~al.} 2013, \mnras, 430, 2622, \dodoi{10.1093/mnras/sts717}

\bibitem[{{Dwek} \& {Cherchneff}(2011)}]{Dwek11}
{Dwek}, E., \& {Cherchneff}, I. 2011, \apj, 727, 63, \dodoi{10.1088/0004-637X/727/2/63}

\bibitem[{{Faber} {et~al.}(2003){Faber}, {Phillips}, {Kibrick}, {Alcott}, {Allen}, {Burrous}, {Cantrall}, {Clarke}, {Coil}, {Cowley}, {Davis}, {Deich}, {Dietsch}, {Gilmore}, {Harper}, {Hilyard}, {Lewis}, {McVeigh}, {Newman}, {Osborne}, {Schiavon}, {Stover}, {Tucker}, {Wallace}, {Wei}, {Wirth}, \& {Wright}}]{Faber03}
{Faber}, S.~M., {Phillips}, A.~C., {Kibrick}, R.~I., {et~al.} 2003, in Society of Photo-Optical Instrumentation Engineers (SPIE) Conference Series, Vol. 4841, Instrument Design and Performance for Optical/Infrared Ground-based Telescopes, ed. M.~{Iye} \& A.~F.~M. {Moorwood}, 1657--1669, \dodoi{10.1117/12.460346}

\bibitem[{{Faisst} {et~al.}(2020){Faisst}, {Fudamoto}, {Oesch}, {Scoville}, {Riechers}, {Pavesi}, \& {Capak}}]{Faisst20}
{Faisst}, A.~L., {Fudamoto}, Y., {Oesch}, P.~A., {et~al.} 2020, \mnras, 498, 4192, \dodoi{10.1093/mnras/staa2545}

\bibitem[{{Ferland} {et~al.}(2017){Ferland}, {Chatzikos}, {Guzm{\'a}n}, {Lykins}, {van Hoof}, {Williams}, {Abel}, {Badnell}, {Keenan}, {Porter}, \& {Stancil}}]{Ferland17}
{Ferland}, G.~J., {Chatzikos}, M., {Guzm{\'a}n}, F., {et~al.} 2017, \rmxaa, 53, 385, \dodoi{10.48550/arXiv.1705.10877}

\bibitem[{{F{\"o}rster Schreiber} {et~al.}(2009){F{\"o}rster Schreiber}, {Genzel}, {Bouch{\'e}}, {Cresci}, {Davies}, {Buschkamp}, {Shapiro}, {Tacconi}, {Hicks}, {Genel}, {Shapley}, {Erb}, {Steidel}, {Lutz}, {Eisenhauer}, {Gillessen}, {Sternberg}, {Renzini}, {Cimatti}, {Daddi}, {Kurk}, {Lilly}, {Kong}, {Lehnert}, {Nesvadba}, {Verma}, {McCracken}, {Arimoto}, {Mignoli}, \& {Onodera}}]{Schreiber09}
{F{\"o}rster Schreiber}, N.~M., {Genzel}, R., {Bouch{\'e}}, N., {et~al.} 2009, \apj, 706, 1364, \dodoi{10.1088/0004-637X/706/2/1364}

\bibitem[{{Fudamoto} {et~al.}(2017){Fudamoto}, {Oesch}, {Schinnerer}, {Groves}, {Karim}, {Magnelli}, {Sargent}, {Cassata}, {Lang}, {Liu}, {Le F{\`e}vre}, {Leslie}, {Smol{\v{c}}i{\'c}}, \& {Tasca}}]{Fudamoto17}
{Fudamoto}, Y., {Oesch}, P.~A., {Schinnerer}, E., {et~al.} 2017, \mnras, 472, 483, \dodoi{10.1093/mnras/stx1948}

\bibitem[{{Fudamoto} {et~al.}(2020){Fudamoto}, {Oesch}, {Magnelli}, {Schinnerer}, {Liu}, {Lang}, {Jim{\'e}nez-Andrade}, {Groves}, {Leslie}, \& {Sargent}}]{Fudamoto20}
{Fudamoto}, Y., {Oesch}, P.~A., {Magnelli}, B., {et~al.} 2020, \mnras, 491, 4724, \dodoi{10.1093/mnras/stz3248}

\bibitem[{{Fukugita} {et~al.}(1996){Fukugita}, {Ichikawa}, {Gunn}, {Doi}, {Shimasaku}, \& {Schneider}}]{Fukugita96}
{Fukugita}, M., {Ichikawa}, T., {Gunn}, J.~E., {et~al.} 1996, \aj, 111, 1748, \dodoi{10.1086/117915}

\bibitem[{{Gal} {et~al.}(2008){Gal}, {Lemaux}, {Lubin}, {Kocevski}, \& {Squires}}]{Gal08}
{Gal}, R.~R., {Lemaux}, B.~C., {Lubin}, L.~M., {Kocevski}, D., \& {Squires}, G.~K. 2008, \apj, 684, 933, \dodoi{10.1086/590416}

\bibitem[{{Gal} \& {Lubin}(2004)}]{Gal04}
{Gal}, R.~R., \& {Lubin}, L.~M. 2004, \apjl, 607, L1, \dodoi{10.1086/421463}

\bibitem[{{Gall} {et~al.}(2011){Gall}, {Hjorth}, \& {Andersen}}]{Gall11}
{Gall}, C., {Hjorth}, J., \& {Andersen}, A.~C. 2011, \aapr, 19, 43, \dodoi{10.1007/s00159-011-0043-7}

\bibitem[{{Garn} \& {Best}(2010)}]{Garn10}
{Garn}, T., \& {Best}, P.~N. 2010, \mnras, 409, 421, \dodoi{10.1111/j.1365-2966.2010.17321.x}

\bibitem[{{Gaspari} {et~al.}(2011){Gaspari}, {Brighenti}, {D'Ercole}, \& {Melioli}}]{Gaspari11}
{Gaspari}, M., {Brighenti}, F., {D'Ercole}, A., \& {Melioli}, C. 2011, \mnras, 415, 1549, \dodoi{10.1111/j.1365-2966.2011.18806.x}

\bibitem[{{Gilbank} {et~al.}(2011){Gilbank}, {Gladders}, {Yee}, \& {Hsieh}}]{Gilbank11}
{Gilbank}, D.~G., {Gladders}, M.~D., {Yee}, H.~K.~C., \& {Hsieh}, B.~C. 2011, \aj, 141, 94, \dodoi{10.1088/0004-6256/141/3/94}

\bibitem[{{Gladders} \& {Yee}(2005)}]{Gladders05}
{Gladders}, M.~D., \& {Yee}, H.~K.~C. 2005, \apjs, 157, 1, \dodoi{10.1086/427327}

\bibitem[{{Groves} {et~al.}(2012){Groves}, {Brinchmann}, \& {Walcher}}]{Groves12}
{Groves}, B., {Brinchmann}, J., \& {Walcher}, C.~J. 2012, \mnras, 419, 1402, \dodoi{10.1111/j.1365-2966.2011.19796.x}

\bibitem[{{Gunn} \& {Gott}(1972)}]{Gunn1972}
{Gunn}, J.~E., \& {Gott}, J.~Richard, I. 1972, \apj, 176, 1, \dodoi{10.1086/151605}

\bibitem[{{Gunn} {et~al.}(1986){Gunn}, {Hoessel}, \& {Oke}}]{Gunn1986}
{Gunn}, J.~E., {Hoessel}, J.~G., \& {Oke}, J.~B. 1986, \apj, 306, 30, \dodoi{10.1086/164317}

\bibitem[{{Hayashi} {et~al.}(2010){Hayashi}, {Kodama}, {Koyama}, {Tanaka}, {Shimasaku}, \& {Okamura}}]{Hayashi10}
{Hayashi}, M., {Kodama}, T., {Koyama}, Y., {et~al.} 2010, \mnras, 402, 1980, \dodoi{10.1111/j.1365-2966.2009.16026.x}

\bibitem[{{Hayashi} {et~al.}(2012){Hayashi}, {Kodama}, {Tadaki}, {Koyama}, \& {Tanaka}}]{Hayashi12}
{Hayashi}, M., {Kodama}, T., {Tadaki}, K.-i., {Koyama}, Y., \& {Tanaka}, I. 2012, \apj, 757, 15, \dodoi{10.1088/0004-637X/757/1/15}

\bibitem[{{Hayashi} {et~al.}(2018){Hayashi}, {Tadaki}, {Kodama}, {Kohno}, {Yamaguchi}, {Hatsukade}, {Koyama}, {Shimakawa}, {Tamura}, \& {Suzuki}}]{Hayashi18}
{Hayashi}, M., {Tadaki}, K.-i., {Kodama}, T., {et~al.} 2018, \apj, 856, 118, \dodoi{10.3847/1538-4357/aab3e7}

\bibitem[{{Hayashi} {et~al.}(2019){Hayashi}, {Koyama}, {Kodama}, {Komiyama}, {Lin}, {Miyazaki}, {Shimakawa}, {Suzuki}, {Tanaka}, {Yamamoto}, \& {Yamamoto}}]{Hayashi19}
{Hayashi}, M., {Koyama}, Y., {Kodama}, T., {et~al.} 2019, \pasj, 71, 112, \dodoi{10.1093/pasj/psz097}

\bibitem[{{Hill} {et~al.}(2022){Hill}, {Chapman}, {Phadke}, {Aravena}, {Archipley}, {Ashby}, {B{\'e}thermin}, {Canning}, {Gonzalez}, {Greve}, {Gururajan}, {Hayward}, {Hezaveh}, {Jarugula}, {MacIntyre}, {Marrone}, {Miller}, {Rennehan}, {Reuter}, {Rotermund}, {Scott}, {Spilker}, {Vieira}, {Wang}, \& {Wei{\ss}}}]{Hill22}
{Hill}, R., {Chapman}, S., {Phadke}, K.~A., {et~al.} 2022, \mnras, 512, 4352, \dodoi{10.1093/mnras/stab3539}

\bibitem[{{Hirashita} {et~al.}(2017){Hirashita}, {Burgarella}, \& {Bouwens}}]{Hirashita17}
{Hirashita}, H., {Burgarella}, D., \& {Bouwens}, R.~J. 2017, \mnras, 472, 4587, \dodoi{10.1093/mnras/stx2349}

\bibitem[{{Hook} {et~al.}(2004){Hook}, {J{\o}rgensen}, {Allington-Smith}, {Davies}, {Metcalfe}, {Murowinski}, \& {Crampton}}]{Hook04}
{Hook}, I.~M., {J{\o}rgensen}, I., {Allington-Smith}, J.~R., {et~al.} 2004, \pasp, 116, 425, \dodoi{10.1086/383624}

\bibitem[{{Hung} {et~al.}(2020){Hung}, {Lemaux}, {Gal}, {Tomczak}, {Lubin}, {Cucciati}, {Pelliccia}, {Shen}, {Le F{\`e}vre}, {Wu}, {Kocevski}, {Mei}, \& {Squires}}]{Hung20}
{Hung}, D., {Lemaux}, B.~C., {Gal}, R.~R., {et~al.} 2020, \mnras, 491, 5524, \dodoi{10.1093/mnras/stz3164}

\bibitem[{{Ichikawa} {et~al.}(2006){Ichikawa}, {Suzuki}, {Tokoku}, {Uchimoto}, {Konishi}, {Yoshikawa}, {Yamada}, {Tanaka}, {Omata}, \& {Nishimura}}]{Ichikawa06}
{Ichikawa}, T., {Suzuki}, R., {Tokoku}, C., {et~al.} 2006, in Society of Photo-Optical Instrumentation Engineers (SPIE) Conference Series, Vol. 6269, Ground-based and Airborne Instrumentation for Astronomy, ed. I.~S. {McLean} \& M.~{Iye}, 626916, \dodoi{10.1117/12.670078}

\bibitem[{{Issa} {et~al.}(1990){Issa}, {MacLaren}, \& {Wolfendale}}]{Issa90}
{Issa}, M.~R., {MacLaren}, I., \& {Wolfendale}, A.~W. 1990, \aap, 236, 237

\bibitem[{{Kashikawa} {et~al.}(2002){Kashikawa}, {Aoki}, {Asai}, {Ebizuka}, {Inata}, {Iye}, {Kawabata}, {Kosugi}, {Ohyama}, {Okita}, {Ozawa}, {Saito}, {Sasaki}, {Sekiguchi}, {Shimizu}, {Taguchi}, {Takata}, {Yadoumaru}, \& {Yoshida}}]{Kashikawa02}
{Kashikawa}, N., {Aoki}, K., {Asai}, R., {et~al.} 2002, \pasj, 54, 819, \dodoi{10.1093/pasj/54.6.819}

\bibitem[{{Kashino} {et~al.}(2013){Kashino}, {Silverman}, {Rodighiero}, {Renzini}, {Arimoto}, {Daddi}, {Lilly}, {Sanders}, {Kartaltepe}, {Zahid}, {Nagao}, {Sugiyama}, {Capak}, {Carollo}, {Chu}, {Hasinger}, {Ilbert}, {Kajisawa}, {Kewley}, {Koekemoer}, {Kova{\v{c}}}, {Le F{\`e}vre}, {Masters}, {McCracken}, {Onodera}, {Scoville}, {Strazzullo}, {Symeonidis}, \& {Taniguchi}}]{Kashino13}
{Kashino}, D., {Silverman}, J.~D., {Rodighiero}, G., {et~al.} 2013, \apjl, 777, L8, \dodoi{10.1088/2041-8205/777/1/L8}

\bibitem[{{Kawinwanichakij} {et~al.}(2017){Kawinwanichakij}, {Papovich}, {Quadri}, {Glazebrook}, {Kacprzak}, {Allen}, {Bell}, {Croton}, {Dekel}, {Ferguson}, {Forrest}, {Grogin}, {Guo}, {Kocevski}, {Koekemoer}, {Labb{\'e}}, {Lucas}, {Nanayakkara}, {Spitler}, {Straatman}, {Tran}, {Tomczak}, \& {van Dokkum}}]{Kawinwanichakij17}
{Kawinwanichakij}, L., {Papovich}, C., {Quadri}, R.~F., {et~al.} 2017, \apj, 847, 134, \dodoi{10.3847/1538-4357/aa8b75}

\bibitem[{{Kennicutt}(1998)}]{kennicutt98_2}
{Kennicutt}, Robert~C., J. 1998, \araa, 36, 189, \dodoi{10.1146/annurev.astro.36.1.189}

\bibitem[{{Kennicutt} {et~al.}(1994){Kennicutt}, {Tamblyn}, \& {Congdon}}]{Kennicutt94}
{Kennicutt}, Robert~C., J., {Tamblyn}, P., \& {Congdon}, C.~E. 1994, \apj, 435, 22, \dodoi{10.1086/174790}

\bibitem[{{Kewley} \& {Dopita}(2002)}]{Kewley02}
{Kewley}, L.~J., \& {Dopita}, M.~A. 2002, \apjs, 142, 35, \dodoi{10.1086/341326}

\bibitem[{{Kocevski} {et~al.}(2009){Kocevski}, {Lubin}, {Lemaux}, {Gal}, {Fassnacht}, {Lin}, \& {Squires}}]{Kocevski09}
{Kocevski}, D.~D., {Lubin}, L.~M., {Lemaux}, B.~C., {et~al.} 2009, \apj, 700, 901, \dodoi{10.1088/0004-637X/700/2/901}

\bibitem[{{Kocevski} {et~al.}(2011){Kocevski}, {Lemaux}, {Lubin}, {Gal}, {McGrath}, {Fassnacht}, {Squires}, {Surace}, \& {Lacy}}]{Kocevski11}
{Kocevski}, D.~D., {Lemaux}, B.~C., {Lubin}, L.~M., {et~al.} 2011, \apj, 736, 38, \dodoi{10.1088/0004-637X/736/1/38}

\bibitem[{{Kodama} {et~al.}(2004){Kodama}, {Balogh}, {Smail}, {Bower}, \& {Nakata}}]{Kodama04}
{Kodama}, T., {Balogh}, M.~L., {Smail}, I., {Bower}, R.~G., \& {Nakata}, F. 2004, \mnras, 354, 1103, \dodoi{10.1111/j.1365-2966.2004.08271.x}

\bibitem[{{Kodama} {et~al.}(1999){Kodama}, {Bell}, \& {Bower}}]{Kodama1999}
{Kodama}, T., {Bell}, E.~F., \& {Bower}, R.~G. 1999, \mnras, 302, 152, \dodoi{10.1046/j.1365-8711.1999.02184.x}

\bibitem[{{Kodama} {et~al.}(2001){Kodama}, {Smail}, {Nakata}, {Okamura}, \& {Bower}}]{Kodama01}
{Kodama}, T., {Smail}, I., {Nakata}, F., {Okamura}, S., \& {Bower}, R.~G. 2001, \apjl, 562, L9, \dodoi{10.1086/338100}

\bibitem[{{Kodama} {et~al.}(2005){Kodama}, {Tanaka}, {Tamura}, {Yahagi}, {Nagashima}, {Tanaka}, {Arimoto}, {Futamase}, {Iye}, {Karasawa}, {Kashikawa}, {Kawasaki}, {Kitayama}, {Matsuhara}, {Nakata}, {Ohashi}, {Ohta}, {Okamoto}, {Okamura}, {Shimasaku}, {Suto}, {Tamura}, {Umetsu}, \& {Yamada}}]{Kodama05}
{Kodama}, T., {Tanaka}, M., {Tamura}, T., {et~al.} 2005, \pasj, 57, 309, \dodoi{10.1093/pasj/57.2.309}

\bibitem[{{Konishi} {et~al.}(2018){Konishi}, {Motohara}, {Takahashi}, {Kato}, {Terao}, {Ohashi}, {Kono}, {Kushibiki}, {Kodama}, {Hayashi}, {Tanaka}, {Tadaki}, {Toshikawa}, {Koyama}, {Shimakawa}, {Suzuki}, {Tateuchi}, {Kitagawa}, {Kobayakawa}, {Todo}, {Aoki}, {Doi}, {Hatsukade}, {Kamizuka}, {Kohno}, {Minezaki}, {Miyata}, {Morokuma}, {Sako}, {Soyano}, {Tanab{\'e}}, {Tanaka}, {Tarusawa}, {Tamura}, {Koshida}, {Ohsawa}, {Uchiyama}, {Mori}, {Yamaguchi}, {Yoshida}, \& {Yoshii}}]{konishi18}
{Konishi}, M., {Motohara}, K., {Takahashi}, H., {et~al.} 2018, in Society of Photo-Optical Instrumentation Engineers (SPIE) Conference Series, Vol. 10702, Ground-based and Airborne Instrumentation for Astronomy VII, ed. C.~J. {Evans}, L.~{Simard}, \& H.~{Takami}, 1070226, \dodoi{10.1117/12.2310060}

\bibitem[{{Koyama} {et~al.}(2010){Koyama}, {Kodama}, {Shimasaku}, {Hayashi}, {Okamura}, {Tanaka}, \& {Tokoku}}]{Koyama10}
{Koyama}, Y., {Kodama}, T., {Shimasaku}, K., {et~al.} 2010, \mnras, 403, 1611, \dodoi{10.1111/j.1365-2966.2009.16226.x}

\bibitem[{{Koyama} {et~al.}(2019){Koyama}, {Shimakawa}, {Yamamura}, {Kodama}, \& {Hayashi}}]{Koyama19}
{Koyama}, Y., {Shimakawa}, R., {Yamamura}, I., {Kodama}, T., \& {Hayashi}, M. 2019, \pasj, 71, 8, \dodoi{10.1093/pasj/psy113}

\bibitem[{{Koyama} {et~al.}(2008){Koyama}, {Kodama}, {Shimasaku}, {Okamura}, {Tanaka}, {Lee}, {Im}, {Matsuhara}, {Takagi}, {Wada}, \& {Oyabu}}]{Koyama08}
{Koyama}, Y., {Kodama}, T., {Shimasaku}, K., {et~al.} 2008, \mnras, 391, 1758, \dodoi{10.1111/j.1365-2966.2008.13931.x}

\bibitem[{{Kron}(1980)}]{kron1980}
{Kron}, R.~G. 1980, \apjs, 43, 305, \dodoi{10.1086/190669}

\bibitem[{{Kubo} {et~al.}(2022){Kubo}, {Umehata}, {Matsuda}, {Kajisawa}, {Steidel}, {Yamada}, {Tanaka}, {Hatsukade}, {Tamura}, {Nakanishi}, {Kohno}, {Lee}, {Matsuda}, {Ao}, {Nagao}, \& {Yun}}]{Kubo22}
{Kubo}, M., {Umehata}, H., {Matsuda}, Y., {et~al.} 2022, \apj, 935, 89, \dodoi{10.3847/1538-4357/ac7f2d}

\bibitem[{{Laishram} {et~al.}(2024){Laishram}, {Kodama}, {Morishita}, {Faisst}, {Koyama}, \& {Yamamoto}}]{Laishram24}
{Laishram}, R., {Kodama}, T., {Morishita}, T., {et~al.} 2024, \apjl, 964, L33, \dodoi{10.3847/2041-8213/ad3238}

\bibitem[{{Larson} {et~al.}(1980){Larson}, {Tinsley}, \& {Caldwell}}]{Larson80}
{Larson}, R.~B., {Tinsley}, B.~M., \& {Caldwell}, C.~N. 1980, \apj, 237, 692, \dodoi{10.1086/157917}

\bibitem[{{Lavaux} \& {Hudson}(2011)}]{Lavaux11}
{Lavaux}, G., \& {Hudson}, M.~J. 2011, \mnras, 416, 2840, \dodoi{10.1111/j.1365-2966.2011.19233.x}

\bibitem[{{Lemaux} {et~al.}(2010){Lemaux}, {Lubin}, {Shapley}, {Kocevski}, {Gal}, \& {Squires}}]{Lemaux10}
{Lemaux}, B.~C., {Lubin}, L.~M., {Shapley}, A., {et~al.} 2010, \apj, 716, 970, \dodoi{10.1088/0004-637X/716/2/970}

\bibitem[{{Lemaux} {et~al.}(2017){Lemaux}, {Tomczak}, {Lubin}, {Wu}, {Gal}, {Rumbaugh}, {Kocevski}, \& {Squires}}]{Lemaux17}
{Lemaux}, B.~C., {Tomczak}, A.~R., {Lubin}, L.~M., {et~al.} 2017, \mnras, 472, 419, \dodoi{10.1093/mnras/stx1579}

\bibitem[{{Lemaux} {et~al.}(2012){Lemaux}, {Gal}, {Lubin}, {Kocevski}, {Fassnacht}, {McGrath}, {Squires}, {Surace}, \& {Lacy}}]{Lemaux12}
{Lemaux}, B.~C., {Gal}, R.~R., {Lubin}, L.~M., {et~al.} 2012, \apj, 745, 106, \dodoi{10.1088/0004-637X/745/2/106}

\bibitem[{{Lemaux} {et~al.}(2018){Lemaux}, {Le F{\`e}vre}, {Cucciati}, {Ribeiro}, {Tasca}, {Zamorani}, {Ilbert}, {Thomas}, {Bardelli}, {Cassata}, {Hathi}, {Pforr}, {Smol{\v{c}}i{\'c}}, {Delvecchio}, {Novak}, {Berta}, {McCracken}, {Koekemoer}, {Amor{\'\i}n}, {Garilli}, {Maccagni}, {Schaerer}, \& {Zucca}}]{Lemaux18}
{Lemaux}, B.~C., {Le F{\`e}vre}, O., {Cucciati}, O., {et~al.} 2018, \aap, 615, A77, \dodoi{10.1051/0004-6361/201730870}

\bibitem[{{Lemaux} {et~al.}(2019){Lemaux}, {Tomczak}, {Lubin}, {Gal}, {Shen}, {Pelliccia}, {Wu}, {Hung}, {Mei}, {Le F{\`e}vre}, {Rumbaugh}, {Kocevski}, \& {Squires}}]{Lemaux19}
{Lemaux}, B.~C., {Tomczak}, A.~R., {Lubin}, L.~M., {et~al.} 2019, \mnras, 490, 1231, \dodoi{10.1093/mnras/stz2661}

\bibitem[{{Lewis} {et~al.}(2002){Lewis}, {Balogh}, {De Propris}, {Couch}, {Bower}, {Offer}, {Bland-Hawthorn}, {Baldry}, {Baugh}, {Bridges}, {Cannon}, {Cole}, {Colless}, {Collins}, {Cross}, {Dalton}, {Driver}, {Efstathiou}, {Ellis}, {Frenk}, {Glazebrook}, {Hawkins}, {Jackson}, {Lahav}, {Lumsden}, {Maddox}, {Madgwick}, {Norberg}, {Peacock}, {Percival}, {Peterson}, {Sutherland}, \& {Taylor}}]{Lewis02}
{Lewis}, I., {Balogh}, M., {De Propris}, R., {et~al.} 2002, \mnras, 334, 673, \dodoi{10.1046/j.1365-8711.2002.05558.x}

\bibitem[{{Liu} {et~al.}(2024){Liu}, {Morishita}, \& {Kodama}}]{Liu24}
{Liu}, Z., {Morishita}, T., \& {Kodama}, T. 2024, arXiv e-prints, arXiv:2406.11188, \dodoi{10.48550/arXiv.2406.11188}

\bibitem[{{Liu} {et~al.}(2025){Liu}, {Kodama}, {Morishita}, {Lee}, {Sun}, {Kubo}, {Cai}, {Wu}, \& {Li}}]{Liu24b}
{Liu}, Z., {Kodama}, T., {Morishita}, T., {et~al.} 2025, \apj, 980, 69, \dodoi{10.3847/1538-4357/ada937}

\bibitem[{{Long} {et~al.}(2020){Long}, {Cooray}, {Ma}, {Casey}, {Wardlow}, {Nayyeri}, {Ivison}, {Farrah}, \& {Dannerbauer}}]{Long20}
{Long}, A.~S., {Cooray}, A., {Ma}, J., {et~al.} 2020, \apj, 898, 133, \dodoi{10.3847/1538-4357/ab9d1f}

\bibitem[{{Lubin} {et~al.}(2009){Lubin}, {Gal}, {Lemaux}, {Kocevski}, \& {Squires}}]{Lubin09}
{Lubin}, L.~M., {Gal}, R.~R., {Lemaux}, B.~C., {Kocevski}, D.~D., \& {Squires}, G.~K. 2009, \aj, 137, 4867, \dodoi{10.1088/0004-6256/137/6/4867}

\bibitem[{{Magnier} {et~al.}(2013){Magnier}, {Schlafly}, {Finkbeiner}, {Juric}, {Tonry}, {Burgett}, {Chambers}, {Flewelling}, {Kaiser}, {Kudritzki}, {Morgan}, {Price}, {Sweeney}, \& {Stubbs}}]{Magnier13}
{Magnier}, E.~A., {Schlafly}, E., {Finkbeiner}, D., {et~al.} 2013, \apjs, 205, 20, \dodoi{10.1088/0067-0049/205/2/20}

\bibitem[{{Maier} {et~al.}(2019){Maier}, {Hayashi}, {Ziegler}, \& {Kodama}}]{Maier19}
{Maier}, C., {Hayashi}, M., {Ziegler}, B.~L., \& {Kodama}, T. 2019, \aap, 626, A14, \dodoi{10.1051/0004-6361/201935522}

\bibitem[{{Mao} {et~al.}(2022){Mao}, {Kodama}, {P{\'e}rez-Mart{\'\i}nez}, {Suzuki}, {Yamamoto}, \& {Adachi}}]{Mao22}
{Mao}, Z., {Kodama}, T., {P{\'e}rez-Mart{\'\i}nez}, J.~M., {et~al.} 2022, \aap, 666, A141, \dodoi{10.1051/0004-6361/202243733}

\bibitem[{{Martig} {et~al.}(2009){Martig}, {Bournaud}, {Teyssier}, \& {Dekel}}]{Martig09}
{Martig}, M., {Bournaud}, F., {Teyssier}, R., \& {Dekel}, A. 2009, \apj, 707, 250, \dodoi{10.1088/0004-637X/707/1/250}

\bibitem[{{Miller} {et~al.}(2018){Miller}, {Chapman}, {Aravena}, {Ashby}, {Hayward}, {Vieira}, {Wei{\ss}}, {Babul}, {B{\'e}thermin}, {Bradford}, {Brodwin}, {Carlstrom}, {Chen}, {Cunningham}, {De Breuck}, {Gonzalez}, {Greve}, {Harnett}, {Hezaveh}, {Lacaille}, {Litke}, {Ma}, {Malkan}, {Marrone}, {Morningstar}, {Murphy}, {Narayanan}, {Pass}, {Perry}, {Phadke}, {Rennehan}, {Rotermund}, {Simpson}, {Spilker}, {Sreevani}, {Stark}, {Strandet}, \& {Strom}}]{miller18}
{Miller}, T.~B., {Chapman}, S.~C., {Aravena}, M., {et~al.} 2018, \nat, 556, 469, \dodoi{10.1038/s41586-018-0025-2}

\bibitem[{{Miyazaki} {et~al.}(2012){Miyazaki}, {Komiyama}, {Nakaya}, {Kamata}, {Doi}, {Hamana}, {Karoji}, {Furusawa}, {Kawanomoto}, {Morokuma}, {Ishizuka}, {Nariai}, {Tanaka}, {Uraguchi}, {Utsumi}, {Obuchi}, {Okura}, {Oguri}, {Takata}, {Tomono}, {Kurakami}, {Namikawa}, {Usuda}, {Yamanoi}, {Terai}, {Uekiyo}, {Yamada}, {Koike}, {Aihara}, {Fujimori}, {Mineo}, {Miyatake}, {Yasuda}, {Nishizawa}, {Saito}, {Tanaka}, {Uchida}, {Katayama}, {Wang}, {Chen}, {Lupton}, {Loomis}, {Bickerton}, {Price}, {Gunn}, {Suzuki}, {Miyazaki}, {Muramatsu}, {Yamamoto}, {Endo}, {Ezaki}, {Itoh}, {Miwa}, {Yokota}, {Matsuda}, {Ebinuma}, \& {Takeshi}}]{Miyazaki12}
{Miyazaki}, S., {Komiyama}, Y., {Nakaya}, H., {et~al.} 2012, in Society of Photo-Optical Instrumentation Engineers (SPIE) Conference Series, Vol. 8446, Ground-based and Airborne Instrumentation for Astronomy IV, ed. I.~S. {McLean}, S.~K. {Ramsay}, \& H.~{Takami}, 84460Z, \dodoi{10.1117/12.926844}

\bibitem[{{Miyazaki} {et~al.}(2018){Miyazaki}, {Komiyama}, {Kawanomoto}, {Doi}, {Furusawa}, {Hamana}, {Hayashi}, {Ikeda}, {Kamata}, {Karoji}, {Koike}, {Kurakami}, {Miyama}, {Morokuma}, {Nakata}, {Namikawa}, {Nakaya}, {Nariai}, {Obuchi}, {Oishi}, {Okada}, {Okura}, {Tait}, {Takata}, {Tanaka}, {Tanaka}, {Terai}, {Tomono}, {Uraguchi}, {Usuda}, {Utsumi}, {Yamada}, {Yamanoi}, {Aihara}, {Fujimori}, {Mineo}, {Miyatake}, {Oguri}, {Uchida}, {Tanaka}, {Yasuda}, {Takada}, {Murayama}, {Nishizawa}, {Sugiyama}, {Chiba}, {Futamase}, {Wang}, {Chen}, {Ho}, {Liaw}, {Chiu}, {Ho}, {Lai}, {Lee}, {Jeng}, {Iwamura}, {Armstrong}, {Bickerton}, {Bosch}, {Gunn}, {Lupton}, {Loomis}, {Price}, {Smith}, {Strauss}, {Turner}, {Suzuki}, {Miyazaki}, {Muramatsu}, {Yamamoto}, {Endo}, {Ezaki}, {Ito}, {Kawaguchi}, {Sofuku}, {Taniike}, {Akutsu}, {Dojo}, {Kasumi}, {Matsuda}, {Imoto}, {Miwa}, {Suzuki}, {Takeshi}, \& {Yokota}}]{Miyazaki18}
{Miyazaki}, S., {Komiyama}, Y., {Kawanomoto}, S., {et~al.} 2018, \pasj, 70, S1, \dodoi{10.1093/pasj/psx063}

\bibitem[{{Moore} {et~al.}(1996){Moore}, {Katz}, {Lake}, {Dressler}, \& {Oemler}}]{Moore96}
{Moore}, B., {Katz}, N., {Lake}, G., {Dressler}, A., \& {Oemler}, A. 1996, \nat, 379, 613, \dodoi{10.1038/379613a0}

\bibitem[{{Moore} {et~al.}(1998){Moore}, {Lake}, \& {Katz}}]{Moore98}
{Moore}, B., {Lake}, G., \& {Katz}, N. 1998, \apj, 495, 139, \dodoi{10.1086/305264}

\bibitem[{{Mori} \& {Burkert}(2000)}]{Mori2000}
{Mori}, M., \& {Burkert}, A. 2000, \apj, 538, 559, \dodoi{10.1086/309140}

\bibitem[{{Morishita} {et~al.}(2015){Morishita}, {Ichikawa}, {Noguchi}, {Akiyama}, {Patel}, {Kajisawa}, \& {Obata}}]{Morishita15}
{Morishita}, T., {Ichikawa}, T., {Noguchi}, M., {et~al.} 2015, \apj, 805, 34, \dodoi{10.1088/0004-637X/805/1/34}

\bibitem[{{Morishita} {et~al.}(2024){Morishita}, {Liu}, {Stiavelli}, {Treu}, {Trenti}, {Chartab}, {Roberts-Borsani}, {Vulcani}, {Bergamini}, {Castellano}, \& {Grillo}}]{Morishita24b}
{Morishita}, T., {Liu}, Z., {Stiavelli}, M., {et~al.} 2024, arXiv e-prints, arXiv:2408.10980, \dodoi{10.48550/arXiv.2408.10980}

\bibitem[{{Morishita} {et~al.}(2025){Morishita}, {Stiavelli}, {Vanzella}, {Bergamini}, {Boyett}, {Chiaberge}, {Grillo}, {Leethochawalit}, {Messa}, {Roberts-Borsani}, {Rosati}, \& {Shajib}}]{Morishita25}
{Morishita}, T., {Stiavelli}, M., {Vanzella}, E., {et~al.} 2025, arXiv e-prints, arXiv:2501.11879, \dodoi{10.48550/arXiv.2501.11879}

\bibitem[{{Motohara} {et~al.}(2016){Motohara}, {Konishi}, {Takahashi}, {Kato}, {Kitagawa}, {Kobayakawa}, {Terao}, {Ohashi}, {Aoki}, {Doi}, {Kamizuka}, {Kohno}, {Minezaki}, {Miyata}, {Morokuma}, {Mori}, {Ohsawa}, {Okada}, {Sako}, {Soyano}, {Tamura}, {Tanabe}, {Tanaka}, {Tarusawa}, {Uchiyama}, {Koshida}, {Asano}, {Tateuchi}, {Uchiyama}, {Todo}, \& {Yoshii}}]{motohara16}
{Motohara}, K., {Konishi}, M., {Takahashi}, H., {et~al.} 2016, in Society of Photo-Optical Instrumentation Engineers (SPIE) Conference Series, Vol. 9908, Ground-based and Airborne Instrumentation for Astronomy VI, ed. C.~J. {Evans}, L.~{Simard}, \& H.~{Takami}, 99083U, \dodoi{10.1117/12.2231386}

\bibitem[{{Muzzin} {et~al.}(2012){Muzzin}, {Wilson}, {Yee}, {Gilbank}, {Hoekstra}, {Demarco}, {Balogh}, {van Dokkum}, {Franx}, {Ellingson}, {Hicks}, {Nantais}, {Noble}, {Lacy}, {Lidman}, {Rettura}, {Surace}, \& {Webb}}]{Muzzin12}
{Muzzin}, A., {Wilson}, G., {Yee}, H.~K.~C., {et~al.} 2012, \apj, 746, 188, \dodoi{10.1088/0004-637X/746/2/188}

\bibitem[{{Nozawa} {et~al.}(2003){Nozawa}, {Kozasa}, {Umeda}, {Maeda}, \& {Nomoto}}]{Nozawa03}
{Nozawa}, T., {Kozasa}, T., {Umeda}, H., {Maeda}, K., \& {Nomoto}, K. 2003, \apj, 598, 785, \dodoi{10.1086/379011}

\bibitem[{{Nulsen}(1982)}]{Nulsen82}
{Nulsen}, P.~E.~J. 1982, \mnras, 198, 1007, \dodoi{10.1093/mnras/198.4.1007}

\bibitem[{{Oguri}(2014)}]{Oguri14}
{Oguri}, M. 2014, \mnras, 444, 147, \dodoi{10.1093/mnras/stu1446}

\bibitem[{{Oke} \& {Gunn}(1983)}]{oke83}
{Oke}, J.~B., \& {Gunn}, J.~E. 1983, \apj, 266, 713, \dodoi{10.1086/160817}

\bibitem[{{Oke} {et~al.}(1995){Oke}, {Cohen}, {Carr}, {Cromer}, {Dingizian}, {Harris}, {Labrecque}, {Lucinio}, {Schaal}, {Epps}, \& {Miller}}]{Oke95}
{Oke}, J.~B., {Cohen}, J.~G., {Carr}, M., {et~al.} 1995, \pasp, 107, 375, \dodoi{10.1086/133562}

\bibitem[{{Oteo} {et~al.}(2018){Oteo}, {Ivison}, {Dunne}, {Manilla-Robles}, {Maddox}, {Lewis}, {de Zotti}, {Bremer}, {Clements}, {Cooray}, {Dannerbauer}, {Eales}, {Greenslade}, {Omont}, {Perez{\textendash}Fourn{\'o}n}, {Riechers}, {Scott}, {van der Werf}, {Weiss}, \& {Zhang}}]{Oteo18}
{Oteo}, I., {Ivison}, R.~J., {Dunne}, L., {et~al.} 2018, \apj, 856, 72, \dodoi{10.3847/1538-4357/aaa1f1}

\bibitem[{{Pelliccia} {et~al.}(2019){Pelliccia}, {Lemaux}, {Tomczak}, {Lubin}, {Shen}, {Epinat}, {Wu}, {Gal}, {Rumbaugh}, {Kocevski}, {Tresse}, \& {Squires}}]{Pelliccia19}
{Pelliccia}, D., {Lemaux}, B.~C., {Tomczak}, A.~R., {et~al.} 2019, \mnras, 482, 3514, \dodoi{10.1093/mnras/sty2876}

\bibitem[{{Peng} {et~al.}(2010){Peng}, {Lilly}, {Kova{\v{c}}}, {Bolzonella}, {Pozzetti}, {Renzini}, {Zamorani}, {Ilbert}, {Knobel}, {Iovino}, {Maier}, {Cucciati}, {Tasca}, {Carollo}, {Silverman}, {Kampczyk}, {de Ravel}, {Sanders}, {Scoville}, {Contini}, {Mainieri}, {Scodeggio}, {Kneib}, {Le F{\`e}vre}, {Bardelli}, {Bongiorno}, {Caputi}, {Coppa}, {de la Torre}, {Franzetti}, {Garilli}, {Lamareille}, {Le Borgne}, {Le Brun}, {Mignoli}, {Perez Montero}, {Pello}, {Ricciardelli}, {Tanaka}, {Tresse}, {Vergani}, {Welikala}, {Zucca}, {Oesch}, {Abbas}, {Barnes}, {Bordoloi}, {Bottini}, {Cappi}, {Cassata}, {Cimatti}, {Fumana}, {Hasinger}, {Koekemoer}, {Leauthaud}, {Maccagni}, {Marinoni}, {McCracken}, {Memeo}, {Meneux}, {Nair}, {Porciani}, {Presotto}, \& {Scaramella}}]{Peng10}
{Peng}, Y.-j., {Lilly}, S.~J., {Kova{\v{c}}}, K., {et~al.} 2010, \apj, 721, 193, \dodoi{10.1088/0004-637X/721/1/193}

\bibitem[{{P{\'e}rez-Mart{\'\i}nez} {et~al.}(2024{\natexlab{a}}){P{\'e}rez-Mart{\'\i}nez}, {Kodama}, {Koyama}, {Shimakawa}, {Suzuki}, {Daikuhara}, {Adachi}, {Onodera}, \& {Tanaka}}]{jose24uss}
{P{\'e}rez-Mart{\'\i}nez}, J.~M., {Kodama}, T., {Koyama}, Y., {et~al.} 2024{\natexlab{a}}, \mnras, 527, 10221, \dodoi{10.1093/mnras/stad3805}

\bibitem[{{P{\'e}rez-Mart{\'\i}nez} {et~al.}(2024{\natexlab{b}}){P{\'e}rez-Mart{\'\i}nez}, {Dannerbauer}, {Koyama}, {P{\'e}rez-Gonz{\'a}lez}, {Shimakawa}, {Kodama}, {Zhang}, {Daikuhara}, {D'Eugenio}, \& {Naufal}}]{Jose24}
{P{\'e}rez-Mart{\'\i}nez}, J.~M., {Dannerbauer}, H., {Koyama}, Y., {et~al.} 2024{\natexlab{b}}, \apj, 977, 74, \dodoi{10.3847/1538-4357/ad8156}

\bibitem[{{P{\'e}rez-Mart{\'\i}nez} {et~al.}(2025){P{\'e}rez-Mart{\'\i}nez}, {Dannerbauer}, {Emonts}, {Allison}, {Champagne}, {Indermuehle}, {Norris}, {Serra}, {Seymour}, {Thomson}, {Casey}, {Chen}, {Daikuhara}, {De Breuck}, {D'Eugenio}, {Drouart}, {Hatch}, {Jin}, {Kodama}, {Koyama}, {Lehnert}, {Macgregor}, {Miley}, {Naufal}, {R{\"o}ttgering}, {S{\'a}nchez-Portal}, {Shimakawa}, {Zhang}, \& {Ziegler}}]{jose25}
{P{\'e}rez-Mart{\'\i}nez}, J.~M., {Dannerbauer}, H., {Emonts}, B.~H.~C., {et~al.} 2025, \aap, 696, A236, \dodoi{10.1051/0004-6361/202450785}

\bibitem[{{Poggianti} {et~al.}(2008){Poggianti}, {Desai}, {Finn}, {Bamford}, {De Lucia}, {Varela}, {Arag{\'o}n-Salamanca}, {Halliday}, {Noll}, {Saglia}, {Zaritsky}, {Best}, {Clowe}, {Milvang-Jensen}, {Jablonka}, {Pell{\'o}}, {Rudnick}, {Simard}, {von der Linden}, \& {White}}]{Poggianti08}
{Poggianti}, B.~M., {Desai}, V., {Finn}, R., {et~al.} 2008, \apj, 684, 888, \dodoi{10.1086/589936}

\bibitem[{{Popesso} {et~al.}(2023){Popesso}, {Concas}, {Cresci}, {Belli}, {Rodighiero}, {Inami}, {Dickinson}, {Ilbert}, {Pannella}, \& {Elbaz}}]{Popesso23}
{Popesso}, P., {Concas}, A., {Cresci}, G., {et~al.} 2023, \mnras, 519, 1526, \dodoi{10.1093/mnras/stac3214}

\bibitem[{{Postman} \& {Geller}(1984)}]{Postman84}
{Postman}, M., \& {Geller}, M.~J. 1984, \apj, 281, 95, \dodoi{10.1086/162078}

\bibitem[{{Postman} {et~al.}(1998){Postman}, {Lubin}, \& {Oke}}]{Postman98}
{Postman}, M., {Lubin}, L.~M., \& {Oke}, J.~B. 1998, \aj, 116, 560, \dodoi{10.1086/300463}

\bibitem[{{Pozzi} {et~al.}(2021){Pozzi}, {Calura}, {Fudamoto}, {Dessauges-Zavadsky}, {Gruppioni}, {Talia}, {Zamorani}, {Bethermin}, {Cimatti}, {Enia}, {Khusanova}, {Decarli}, {Le F{\`e}vre}, {Capak}, {Cassata}, {Faisst}, {Yan}, {Schaerer}, {Silverman}, {Bardelli}, {Boquien}, {Enia}, {Narayanan}, {Ginolfi}, {Hathi}, {Jones}, {Koekemoer}, {Lemaux}, {Loiacono}, {Maiolino}, {Riechers}, {Rodighiero}, {Romano}, {Vallini}, {Vergani}, \& {Zucca}}]{Pozzi21}
{Pozzi}, F., {Calura}, F., {Fudamoto}, Y., {et~al.} 2021, \aap, 653, A84, \dodoi{10.1051/0004-6361/202040258}

\bibitem[{{Reddy} {et~al.}(2023){Reddy}, {Topping}, {Sanders}, {Shapley}, \& {Brammer}}]{Reddy23}
{Reddy}, N.~A., {Topping}, M.~W., {Sanders}, R.~L., {Shapley}, A.~E., \& {Brammer}, G. 2023, \apj, 948, 83, \dodoi{10.3847/1538-4357/acc869}

\bibitem[{{Reddy} {et~al.}(2015){Reddy}, {Kriek}, {Shapley}, {Freeman}, {Siana}, {Coil}, {Mobasher}, {Price}, {Sanders}, \& {Shivaei}}]{Reddy15}
{Reddy}, N.~A., {Kriek}, M., {Shapley}, A.~E., {et~al.} 2015, \apj, 806, 259, \dodoi{10.1088/0004-637X/806/2/259}

\bibitem[{{Sancisi} {et~al.}(2008){Sancisi}, {Fraternali}, {Oosterloo}, \& {van der Hulst}}]{Sancisi08}
{Sancisi}, R., {Fraternali}, F., {Oosterloo}, T., \& {van der Hulst}, T. 2008, \aapr, 15, 189, \dodoi{10.1007/s00159-008-0010-0}

\bibitem[{{Schlafly} {et~al.}(2012){Schlafly}, {Finkbeiner}, {Juri{\'c}}, {Magnier}, {Burgett}, {Chambers}, {Grav}, {Hodapp}, {Kaiser}, {Kudritzki}, {Martin}, {Morgan}, {Price}, {Rix}, {Stubbs}, {Tonry}, \& {Wainscoat}}]{Schlafly12}
{Schlafly}, E.~F., {Finkbeiner}, D.~P., {Juri{\'c}}, M., {et~al.} 2012, \apj, 756, 158, \dodoi{10.1088/0004-637X/756/2/158}

\bibitem[{{Shapley} {et~al.}(2023){Shapley}, {Sanders}, {Reddy}, {Topping}, \& {Brammer}}]{shapley23}
{Shapley}, A.~E., {Sanders}, R.~L., {Reddy}, N.~A., {Topping}, M.~W., \& {Brammer}, G.~B. 2023, \apj, 954, 157, \dodoi{10.3847/1538-4357/acea5a}

\bibitem[{{Shapley} {et~al.}(2022){Shapley}, {Sanders}, {Salim}, {Reddy}, {Kriek}, {Mobasher}, {Coil}, {Siana}, {Price}, {Shivaei}, {Dunlop}, {McLure}, \& {Cullen}}]{Shapley22}
{Shapley}, A.~E., {Sanders}, R.~L., {Salim}, S., {et~al.} 2022, \apj, 926, 145, \dodoi{10.3847/1538-4357/ac4742}

\bibitem[{{Shen} {et~al.}(2017){Shen}, {Miller}, {Lemaux}, {Tomczak}, {Lubin}, {Rumbaugh}, {Fassnacht}, {Becker}, {Gal}, {Wu}, \& {Squires}}]{Shen17}
{Shen}, L., {Miller}, N.~A., {Lemaux}, B.~C., {et~al.} 2017, \mnras, 472, 998, \dodoi{10.1093/mnras/stx1984}

\bibitem[{{Shimakawa} {et~al.}(2018){Shimakawa}, {Kodama}, {Hayashi}, {Prochaska}, {Tanaka}, {Cai}, {Suzuki}, {Tadaki}, \& {Koyama}}]{Shimakawa18}
{Shimakawa}, R., {Kodama}, T., {Hayashi}, M., {et~al.} 2018, \mnras, 473, 1977, \dodoi{10.1093/mnras/stx2494}

\bibitem[{{Sobral} {et~al.}(2012){Sobral}, {Best}, {Matsuda}, {Smail}, {Geach}, \& {Cirasuolo}}]{sobral12}
{Sobral}, D., {Best}, P.~N., {Matsuda}, Y., {et~al.} 2012, \mnras, 420, 1926, \dodoi{10.1111/j.1365-2966.2011.19977.x}

\bibitem[{{Sobral} {et~al.}(2011){Sobral}, {Best}, {Smail}, {Geach}, {Cirasuolo}, {Garn}, \& {Dalton}}]{Sobral11}
{Sobral}, D., {Best}, P.~N., {Smail}, I., {et~al.} 2011, \mnras, 411, 675, \dodoi{10.1111/j.1365-2966.2010.17707.x}

\bibitem[{{Springel} {et~al.}(2005){Springel}, {White}, {Jenkins}, {Frenk}, {Yoshida}, {Gao}, {Navarro}, {Thacker}, {Croton}, {Helly}, {Peacock}, {Cole}, {Thomas}, {Couchman}, {Evrard}, {Colberg}, \& {Pearce}}]{Springel05}
{Springel}, V., {White}, S. D.~M., {Jenkins}, A., {et~al.} 2005, \nat, 435, 629, \dodoi{10.1038/nature03597}

\bibitem[{{Steidel} {et~al.}(2014){Steidel}, {Rudie}, {Strom}, {Pettini}, {Reddy}, {Shapley}, {Trainor}, {Erb}, {Turner}, {Konidaris}, {Kulas}, {Mace}, {Matthews}, \& {McLean}}]{Steidel14}
{Steidel}, C.~C., {Rudie}, G.~C., {Strom}, A.~L., {et~al.} 2014, \apj, 795, 165, \dodoi{10.1088/0004-637X/795/2/165}

\bibitem[{{Suzuki} {et~al.}(2008){Suzuki}, {Tokoku}, {Ichikawa}, {Uchimoto}, {Konishi}, {Yoshikawa}, {Tanaka}, {Yamada}, {Omata}, \& {Nishimura}}]{Suzuki08}
{Suzuki}, R., {Tokoku}, C., {Ichikawa}, T., {et~al.} 2008, \pasj, 60, 1347, \dodoi{10.1093/pasj/60.6.1347}

\bibitem[{{Suzuki} {et~al.}(2021){Suzuki}, {Onodera}, {Kodama}, {Daddi}, {Hayashi}, {Koyama}, {Shimakawa}, {Smail}, {Sobral}, {Tacchella}, \& {Tanaka}}]{Suzuki21}
{Suzuki}, T.~L., {Onodera}, M., {Kodama}, T., {et~al.} 2021, \apj, 908, 15, \dodoi{10.3847/1538-4357/abd4e7}

\bibitem[{{Takagi} {et~al.}(1999){Takagi}, {Arimoto}, \& {Vansevi{\v{c}}ius}}]{Takagi99}
{Takagi}, T., {Arimoto}, N., \& {Vansevi{\v{c}}ius}, V. 1999, \apj, 523, 107, \dodoi{10.1086/307706}

\bibitem[{{Tanaka} {et~al.}(2011){Tanaka}, {De Breuck}, {Kurk}, {Taniguchi}, {Kodama}, {Matsuda}, {Packham}, {Zirm}, {Kajisawa}, {Ichikawa}, {Seymour}, {Stern}, {Stockton}, {Venemans}, \& {Vernet}}]{Tanaka11}
{Tanaka}, I., {De Breuck}, C., {Kurk}, J.~D., {et~al.} 2011, \pasj, 63, 415, \dodoi{10.1093/pasj/63.sp2.S415}

\bibitem[{{Tanaka} {et~al.}(2005){Tanaka}, {Kodama}, {Arimoto}, {Okamura}, {Umetsu}, {Shimasaku}, {Tanaka}, \& {Yamada}}]{Tanaka05}
{Tanaka}, M., {Kodama}, T., {Arimoto}, N., {et~al.} 2005, \mnras, 362, 268, \dodoi{10.1111/j.1365-2966.2005.09300.x}

\bibitem[{{Tanaka} {et~al.}(2018){Tanaka}, {Coupon}, {Hsieh}, {Mineo}, {Nishizawa}, {Speagle}, {Furusawa}, {Miyazaki}, \& {Murayama}}]{Tanaka18}
{Tanaka}, M., {Coupon}, J., {Hsieh}, B.-C., {et~al.} 2018, \pasj, 70, S9, \dodoi{10.1093/pasj/psx077}

\bibitem[{{Teyssier} {et~al.}(2011){Teyssier}, {Moore}, {Martizzi}, {Dubois}, \& {Mayer}}]{Teyssier11}
{Teyssier}, R., {Moore}, B., {Martizzi}, D., {Dubois}, Y., \& {Mayer}, L. 2011, \mnras, 414, 195, \dodoi{10.1111/j.1365-2966.2011.18399.x}

\bibitem[{{Thomas} {et~al.}(2005){Thomas}, {Maraston}, {Bender}, \& {Mendes de Oliveira}}]{thomas05}
{Thomas}, D., {Maraston}, C., {Bender}, R., \& {Mendes de Oliveira}, C. 2005, \apj, 621, 673, \dodoi{10.1086/426932}

\bibitem[{{Tody}(1986)}]{Tody1986}
{Tody}, D. 1986, in Society of Photo-Optical Instrumentation Engineers (SPIE) Conference Series, Vol. 627, Instrumentation in astronomy VI, ed. D.~L. {Crawford}, 733, \dodoi{10.1117/12.968154}

\bibitem[{{Tomczak} {et~al.}(2017){Tomczak}, {Lemaux}, {Lubin}, {Gal}, {Wu}, {Holden}, {Kocevski}, {Mei}, {Pelliccia}, {Rumbaugh}, \& {Shen}}]{Tomczak17}
{Tomczak}, A.~R., {Lemaux}, B.~C., {Lubin}, L.~M., {et~al.} 2017, \mnras, 472, 3512, \dodoi{10.1093/mnras/stx2245}

\bibitem[{{Tomczak} {et~al.}(2019){Tomczak}, {Lemaux}, {Lubin}, {Pelliccia}, {Shen}, {Gal}, {Hung}, {Kocevski}, {Le F{\`e}vre}, {Mei}, {Rumbaugh}, {Squires}, \& {Wu}}]{Tomczak19}
---. 2019, \mnras, 484, 4695, \dodoi{10.1093/mnras/stz342}

\bibitem[{{Tonry} {et~al.}(2012){Tonry}, {Stubbs}, {Lykke}, {Doherty}, {Shivvers}, {Burgett}, {Chambers}, {Hodapp}, {Kaiser}, {Kudritzki}, {Magnier}, {Morgan}, {Price}, \& {Wainscoat}}]{Tonry12}
{Tonry}, J.~L., {Stubbs}, C.~W., {Lykke}, K.~R., {et~al.} 2012, \apj, 750, 99, \dodoi{10.1088/0004-637X/750/2/99}

\bibitem[{{Umehata} {et~al.}(2015){Umehata}, {Tamura}, {Kohno}, {Ivison}, {Alexander}, {Geach}, {Hatsukade}, {Hughes}, {Ikarashi}, {Kato}, {Izumi}, {Kawabe}, {Kubo}, {Lee}, {Lehmer}, {Makiya}, {Matsuda}, {Nakanishi}, {Saito}, {Smail}, {Yamada}, {Yamaguchi}, \& {Yun}}]{Umehata15}
{Umehata}, H., {Tamura}, Y., {Kohno}, K., {et~al.} 2015, \apjl, 815, L8, \dodoi{10.1088/2041-8205/815/1/L8}

\bibitem[{{Umehata} {et~al.}(2019){Umehata}, {Fumagalli}, {Smail}, {Matsuda}, {Swinbank}, {Cantalupo}, {Sykes}, {Ivison}, {Steidel}, {Shapley}, {Vernet}, {Yamada}, {Tamura}, {Kubo}, {Nakanishi}, {Kajisawa}, {Hatsukade}, \& {Kohno}}]{Umehata19}
{Umehata}, H., {Fumagalli}, M., {Smail}, I., {et~al.} 2019, Science, 366, 97, \dodoi{10.1126/science.aaw5949}

\bibitem[{{van de Voort} {et~al.}(2011){van de Voort}, {Schaye}, {Booth}, {Haas}, \& {Dalla Vecchia}}]{Voort11}
{van de Voort}, F., {Schaye}, J., {Booth}, C.~M., {Haas}, M.~R., \& {Dalla Vecchia}, C. 2011, \mnras, 414, 2458, \dodoi{10.1111/j.1365-2966.2011.18565.x}

\bibitem[{{Voit} \& {Donahue}(2005)}]{Voit05}
{Voit}, G.~M., \& {Donahue}, M. 2005, \apj, 634, 955, \dodoi{10.1086/497063}

\bibitem[{{Williams} {et~al.}(2009){Williams}, {Quadri}, {Franx}, {van Dokkum}, \& {Labb{\'e}}}]{williams2009}
{Williams}, R.~J., {Quadri}, R.~F., {Franx}, M., {van Dokkum}, P., \& {Labb{\'e}}, I. 2009, \apj, 691, 1879, \dodoi{10.1088/0004-637X/691/2/1879}

\bibitem[{{Wuyts} {et~al.}(2016){Wuyts}, {Wisnioski}, {Fossati}, {F{\"o}rster Schreiber}, {Genzel}, {Davies}, {Mendel}, {Naab}, {R{\"o}ttgers}, {Wilman}, {Wuyts}, {Bandara}, {Beifiori}, {Belli}, {Bender}, {Brammer}, {Burkert}, {Chan}, {Galametz}, {Kulkarni}, {Lang}, {Lutz}, {Momcheva}, {Nelson}, {Rosario}, {Saglia}, {Seitz}, {Tacconi}, {Tadaki}, {{\"U}bler}, \& {van Dokkum}}]{Wuyts_kmos}
{Wuyts}, E., {Wisnioski}, E., {Fossati}, M., {et~al.} 2016, \apj, 827, 74, \dodoi{10.3847/0004-637X/827/1/74}

\bibitem[{{Wuyts} {et~al.}(2011){Wuyts}, {F{\"o}rster Schreiber}, {Lutz}, {Nordon}, {Berta}, {Altieri}, {Andreani}, {Aussel}, {Bongiovanni}, {Cepa}, {Cimatti}, {Daddi}, {Elbaz}, {Genzel}, {Koekemoer}, {Magnelli}, {Maiolino}, {McGrath}, {P{\'e}rez Garc{\'\i}a}, {Poglitsch}, {Popesso}, {Pozzi}, {Sanchez-Portal}, {Sturm}, {Tacconi}, \& {Valtchanov}}]{Wuyts11}
{Wuyts}, S., {F{\"o}rster Schreiber}, N.~M., {Lutz}, D., {et~al.} 2011, \apj, 738, 106, \dodoi{10.1088/0004-637X/738/1/106}

\bibitem[{{Yoshii} {et~al.}(2016){Yoshii}, {Doi}, {Kohno}, {Miyata}, {Motohara}, {Kawara}, {Tanaka}, {Minezaki}, {Sako}, {Morokuma}, {Tamura}, {Tanabe}, {Takahashi}, {Konishi}, {Kamizuka}, {Kato}, {Aoki}, {Soyano}, {Tarusawa}, {Handa}, {Koshida}, {Bronfman}, {Ruiz}, {Hamuy}, \& {Garay}}]{Yoshii16}
{Yoshii}, Y., {Doi}, M., {Kohno}, K., {et~al.} 2016, in Society of Photo-Optical Instrumentation Engineers (SPIE) Conference Series, Vol. 9906, Ground-based and Airborne Telescopes VI, ed. H.~J. {Hall}, R.~{Gilmozzi}, \& H.~K. {Marshall}, 99060R, \dodoi{10.1117/12.2231391}

\bibitem[{{Zahid} {et~al.}(2013){Zahid}, {Yates}, {Kewley}, \& {Kudritzki}}]{Zahid13}
{Zahid}, H.~J., {Yates}, R.~M., {Kewley}, L.~J., \& {Kudritzki}, R.~P. 2013, \apj, 763, 92, \dodoi{10.1088/0004-637X/763/2/92}

\bibitem[{{Zhou} {et~al.}(2024){Zhou}, {Wang}, {Daddi}, {Coogan}, {Sun}, {Xu}, {Arumugam}, {Jin}, {Liu}, {Lu}, {Sillassen}, {Wang}, {Shi}, {Zhang}, {Tan}, {Gu}, {Elbaz}, {Le Bail}, {Magnelli}, {G{\'o}mez-Guijarro}, {d'Eugenio}, {Magdis}, {Valentino}, {Ji}, {Gobat}, {Delvecchio}, {Xiao}, {Strazzullo}, {Finoguenov}, {Schinnerer}, {Rich}, {Huang}, {Dai}, {Chen}, {Gao}, {Yang}, \& {Hao}}]{Zhou24}
{Zhou}, L., {Wang}, T., {Daddi}, E., {et~al.} 2024, \aap, 684, A196, \dodoi{10.1051/0004-6361/202348351}

\end{thebibliography}
\bibliographystyle{aasjournal}



\end{document}